\documentclass[dvipsnames,12pt]{article}
\usepackage[letterpaper, total={19cm,25cm}]{geometry}
\usepackage{authblk}
\usepackage{graphicx}
\usepackage{amsmath,amssymb,amsthm,mathtools}
\usepackage{xcolor}
\usepackage[utf8]{inputenc}
\usepackage[T1]{fontenc}
\usepackage{url} 
\usepackage{wrapfig} 
\usepackage{float}
\usepackage{chemist} 
\usepackage{bm} 
\usepackage{caption} 

\newcommand{\dd}{\mathrm{d}}
\newcommand{\vct}[1]{\mathbf{#1}}
\newcommand{\IR}{\mathbb{R}}

\title{Lambda-ABF: Simplified, Portable, Accurate, and Cost-effective Alchemical Free Energy Computation}

\author[1,2,4]{Louis Lagardère,\thanks{Corresponding author: \texttt{louis.lagardere@sorbonne-universite.fr}}}
\author[3]{Lise Maurin}
\author[1]{Olivier Adjoua}
\author[4]{Krystel El Hage}
\author[1,3]{Pierre Monmarché,\thanks{Corresponding author: \texttt{pierre.monmarche@sorbonne-universite.fr}}}
\author[1,4]{Jean-Philip Piquemal,\thanks{Corresponding author: \texttt{jean-philip.piquemal@sorbonne-universite.fr}}}
\author[6]{Jérôme Hénin,\thanks{Corresponding author: \texttt{jerome.henin@cnrs.fr}}}
\date{}

\affil[1]{Sorbonne Université, Laboratoire de Chimie Théorique, UMR 7616 CNRS, 75005, Paris, France}
\affil[2]{Institut Parisien de Chimie Physique et Théorique, FR2622 CNRS, 75005, Paris, France}
\affil[3]{Sorbonne Université, Laboratoire Jacques-Louis Lions, UMR 7589 CNRS, 75005, Paris, France}
\affil[4]{Qubit Pharmaceuticals, 750014, Paris, France}
\affil[5]{Institut Universitaire de France}
\affil[6]{Laboratoire de Biochimie Théorique UPR 9080, CNRS, Université de Paris, 75005 Paris, France
Institut de Biologie Physico-Chimique-Fondation Edmond de Rothschild, PSL Research University, 75005 Paris, Paris, France}

\begin{document}

\maketitle
\newpage

\begin{abstract}
We introduce the lambda-ABF method for the computation of alchemical free energy differences. We propose a software implementation and showcase it on biomolecular systems. The method arises from coupling multiple-walker Adaptive Biasing Force (mwABF) with $\lambda$-dynamics. The sampling of the alchemical variable is continuous and converges towards a uniform distribution, making manual optimization of the $\lambda$ schedule unnecessary.
Contrary to most other approaches, alchemical free energy estimates are obtained immediately without any post-processing.
Free diffusion of $\lambda$  improves orthogonal relaxation compared to fixed-$\lambda$ Thermodynamic Integration or Free Energy Perturbation. Furthermore, multiple walkers provide generic orthogonal space coverage with minimal user input and negligible computational overhead. We show that our high-performance implementations coupling the Colvars library with NAMD and Tinker-HP can address real-world cases including ligand-receptor binding with both fixed-charge and polarizable models, with a demonstrably richer sampling than fixed-$\lambda$ methods. The implementation is fully open-source, publicly available, and readily usable by practitioners of current alchemical methods.
Thanks to the portable Colvars library, lambda-ABF presents a unified user interface regardless of the back-end (NAMD, Tinker-HP, or any software to be interfaced in the future), sparing users the effort of learning multiple interfaces.
Finally, the Colvars Dashboard extension of VMD provides an interactive monitoring and diagnostic tool for lambda-ABF simulations.
\end{abstract}

\section{Introduction}\label{sec:intro}

Free energy methods applied to biological and chemical systems have been the focus of active research and development for the last forty to fifty years\cite{jorgensen1985monte,chipot2007free,wang2015accurate,woo2005calculation,cournia2017relative}. This activity is powered by the ever-increasing compute capabilities available to practitioners (both in the academic and in the industrial context) as well as the introduction of new approaches\cite{kong1996lambda,zheng2008random,gapsys2021accurate,cruz2020combining}, the refinement of existing ones\cite{zheng2012practically,knight2011multisite,hayes2017adaptive}, and their efficient implementation in production codes, making them usable in conjunction with validated models in a practical context.

Among such methods, alchemical free energy methods, which rely on a non-physical alchemical pathway between two physical states of interest, have been shown to be both reliable and computationally efficient for the computation of relative and absolute solvation free energies as well as relative and absolute free energies of binding of small molecules\cite{chodera2011alchemical,song2020evolution}. Such techniques are built on sampling an ``alchemical pathway'' of intermediates between the two end-states, defined by interpolated Hamiltonians parameterized by a variable $\lambda$.
Most practical uses of these techniques rely on sampling discrete states at fixed, predefined values of $\lambda$. Free energy differences between these $\lambda$-states are then computed with a free energy estimator, such as the Zwanzig\cite{zwanzig1954high}, Bennett Acceptance Ratio (BAR)\cite{bennett1976efficient} or Multistate Bennett Acceptance Ratio (MBAR) estimators\cite{shirts2008statistically}.  Another widely used estimator is Thermodynamic Integration (TI), which consists of computing free energy derivatives with respect to the coupling parameter and then reconstructing the free energy differences through numerical integration\cite{straatsma1991multiconfiguration}. This approach has been shown to be more efficient in several situations\cite{cuendet2014free}. Recently, non-equilibrium approaches that introduce rapid alchemical change applied in several short simulations have also been introduced\cite{gapsys2012new,procacci2014fast}. The final free-energy estimate is then obtained with Jarzynski/Crooks relations and related approximations\cite{jarzynski1997nonequilibrium,crooks1999entropy,park2004calculating}.

Still, these strategies are always associated with a heavy computational cost and with practical complexity, limiting their adoption by non-expert users.
The necessary steps in state-of-the-art methods include the manual or automated optimization of the lambda schedule\cite{deRuiter2013, Zeng2023}), and a postprocessing step to obtain free energy differences.
Additionally, even in the simpler case of relative free energies, the need to sample slow orthogonal degrees of freedom such as solvent reorganization or binding site relaxation has pushed the development of dedicated techniques such as Hamiltonian Replica Exchange\cite{fukunishi2002hamiltonian,jang2003replica,jiang2010free} (HRE). In this case, sampling of such degrees of freedom is improved by exchanging configurations between different lambda states. But this comes at the cost of synchronization between the (a priori) independent trajectories and the improvement in accuracy of the final estimate is not systematic\cite{rizzi2018overview}.

Alternatively, Brooks and coworkers, inspired by the work from Tidor\cite{tidor1993simulated}, introduced another approach: $\lambda$-dynamics in which the coupling parameter is made dynamical through an extended Lagrangian/Hamiltonian scheme\cite{kong1996lambda}. It has been at first applied in the context of relative free energy differences and is systematically associated with both a biasing scheme in order to efficiently sample the end states as well as a free energy estimator. Brooks and coworkers started by using a fixed (non-adaptive bias) in $\lambda$-space to ensure proper sampling of the end states, and recently introduced an adaptive landscape flattening scheme to further refine this bias\cite{hayes2017adaptive}. In any case, they rely on an estimator of the ratio of probabilities of the end states (which require unbiasing) to get the free energy differences of interest. $\lambda$-dynamics has been shown to be more efficient than other methods in the context of relative free energies and has been extended to the simultaneous estimation of several of these differences through multi-site $\lambda$-dynamics\cite{knight2011multisite}. This efficiency can be related to the absence of a predefined (and thus unoptimized) lambda stratification strategy and to a better sampling of orthogonal space thanks to the relaxation of lambda in a single molecular dynamics trajectory\cite{hahn2020overcoming}.
$\lambda$-dynamics has also been used in conjunction with metadynamics, which yields at the same time an adaptive bias and a free energy estimator\cite{wu2011lambda} with some practical issues related to the use of reflective boundary conditions and the deposition of Gaussian potentials on the $\lambda$ reaction coordinate. Also, Tuckerman and collaborators studied how adiabatic free energy dynamics could be coupled with $\lambda$-dynamics\cite{abrams2006efficient} showing promising results, yet this combination has not been widely adopted by practitioners.

$\lambda$-dynamics has been further refined and rationalized by Hünenberger and coworkers with the introduction of the $\lambda$-LEUS scheme\cite{bieler2014local}. They systematically studied the impact of the parameters influencing the dynamics in $\lambda$, namely its mass and the thermostat parameters associated with $\lambda$ evolution in time, and showed that this mass needed to be large enough that the dynamics in $\lambda$ drives the others degrees of freedom but not so large that it slows down the exploration of $\lambda$-space. They also showed that the coupling with the thermostat needed to be strong enough to guarantee numerical stability and sampling efficiency. They resorted to an adaptive bias in $\lambda$ through the local elevation umbrella sampling method and to an unbiased probability estimator similar to the original work of Brooks and coworkers.
A thorough assessment of convergence showed improvements compared to standard Thermodynamic Integration. In subsequent studies, they detailed how their method outperforms state-of-the-art fixed-$\lambda$ techniques in the exploration of orthogonal space\cite{bieler2015orthogonal}. This sampling efficiency can be further augmented by their recently introduced Conveyor Belt Thermodynamic Integration scheme which enables the exchange of information between lambda values, at the cost of frequent global synchronization between trajectories\cite{hahn2019alchemical}.

Other techniques aim at improving the sampling in a specific additional direction. This is the case of the recent work of York and coworkers\cite{lee2023aces} as well as that of Shirts and collaborators which uses expanded ensemble simulations (with Metropolis-Hastings Monte-Carlo) and additional variables whose sampling is enhanced through metadynamics\cite{hsu2023alchemical}. OSRW (Orthogonal Space Random Walk) and DI-OST (Double Integration-Orthogonal Space Tempering)\cite{zheng2008random,zheng2012practically} follow another approach in which a generic additional coordinate, the derivative of the potential with respect to $\lambda$, is the focus (on top of the $\lambda$ variable) of an enhanced sampling scheme, namely metadynamics\cite{barducci2011metadynamics} or Adaptive Biasing Force\cite{Darve2001,darve2008adaptive,comer2015adaptive}. It has shown impressive sampling results but is not readily accessible by the community of users in a production code. All these families of methods may still suffer from slow relaxation in other degrees of freedom and from the intrinsic additional burden of having to estimate free energy differences in a higher dimensional space.

Despite the wide variety of existing $\lambda$-dynamics and more generally expanded-ensemble approaches, their adoption by the community of free energy practitioners is still lacking and, to the best of our knowledge, none of these have been applied to the more complex case of absolute protein-ligand binding free energy.

In this work, we propose a new $\lambda$-dynamics approach with the following features:
\begin{itemize}
    \item i) $\lambda$ propagation under strongly damped Langevin dynamics with a large mass;
    \item ii) Adaptively biased dynamics in $\lambda$ through an Adaptive Biasing Force scheme;
    \item iii) Efficient sampling (with minimal computational overhead) of orthogonal space with a multiple walker strategy;
    \item iv) Free energy calculation through the TI estimator (in its ABF version, as explained below).
\end{itemize}
The main benefits of this method lie in its user-friendliness and sampling efficiency.
The user protocol for obtaining free energy estimates is straightforward, largely on account of the intrinsic simplicity of the algorithm.
The method is supported by the open-source Colvars library,\cite{Fiorin2013} where the alchemical parameter is implemented as new extended variable.
The implementation arises from the interfacing of existing, flexible, and well-maintained software, with limited added complexity.
This also enables the practical manipulation of this coordinate with all the pre-existing functionalities of Colvars, including seamless biasing of this and other coordinates. This is illustrated below through the use of the recently introduced Distance to Bound Configuration (DBC) collective variable, which facilitates the definition of the bound state and its sampling in the context of absolute free energies of binding,\cite{salari2018} and through an example of custom $\lambda$-dependent restraint.

We begin by reviewing the theory of $\lambda$-dynamics and its combination with multiple-walker Adaptive Biasing Force. We then describe the implementation of the method in combination with two high-performance molecular dynamics codes: NAMD\cite{phillips2020scalable} and Tinker-HP\cite{lagardere2018tinker,adjoua2021tinker}. We then showcase applications ranging from simple solvation free energies to absolute free energies of ligand binding to proteins, described with either a fixed-charge (CHARMM\cite{mackerell2002charmm,vanommeslaeghe2010charmm}) or a polarizable (AMOEBA \cite{ren2003polarizable,ponder2010current,ren2011polarizable,shi2013polarizable,AMOEBAnucleic}) force field (FF), illustrating the robustness, portability, and efficiency of the method.
We focus in particular on the sampling efficiency of lambda-ABF compared to state-of-the-art fixed-$\lambda$ methods, by applying it to a few systems where we can monitor the relaxation of relevant slow degrees of freedom. Extensive benchmarks of the method on a large number of ligands and targets will be the focus of future work.

\section{Theory}

\subsection{Principle of $\lambda$-dynamics}\label{subsec:lambdadyn-framework}

In everything that follows, the alchemical coupling parameter $\lambda\in [0,1]$ corresponds to scaling by $\lambda$ of some bonded or non-bonded interactions between predefined groups of atoms. Well-defined physical Hamiltonians are then associated with $\lambda$ values $0$ and $1$, such as two chemical states of a portion of the system in the context of relative free energy simulation, or coupled/uncoupled non-bonded interactions in the context of absolute free energies of interaction. When $0 < \lambda < 1$, the associated Hamiltonian parameterized by $\lambda$ is an unphysical intermediate between the Hamiltonians describing the end-points. Evaluating the free energy change along this unphysical alchemical route is at the core of alchemical free energy methods.

The main idea behind $\lambda$-dynamics is to treat the variable $\lambda$ as a dynamical variable associated with a fictitious mass $\mathfrak{m}_{\lambda}>0$. A direct consequence is that one can now consider the extended microstate $(\vct{q},\vct{p}; \lambda, p_{\lambda}) \in \IR^{6N} \times [0,1] \times \IR$, where $p_{\lambda}=\mathfrak{m}_{\lambda} \dot{\lambda}$ is the fictitious particle's momentum and $\dot{\lambda}$ its time derivative. This extended system is assigned the following extended Hamiltonian:
\begin{equation}\label{eq:lambdadyn-Hamiltonian}
H(\vct{q},\vct{p};\lambda,p_\lambda)=E_{k}^{x}(\vct{p})+E_{k}^{\lambda}(p_\lambda) + V(\vct{q}; \lambda),
\end{equation}
where $E_{k}^{x}$ is the kinetic energy depending solely on the atomic momenta, $E_{k}^{\lambda}(p_\lambda)= p_\lambda^2 / (2\, \mathfrak{m}_{\lambda})$ is the kinetic energy of $\lambda$ and $V(\vct{q};\lambda)$ is the extended potential energy.
A convenient approach to sample the canonical ensemble at a given inverse temperature $\beta$ is to simulate Langevin dynamics for both the physical and the alchemical coordinates. This is the choice that we made for all the results that follow. The corresponding extended-system dynamics is given by:

\begin{equation}\label{eq:lambdadyn-Langevin-process}
\left\{ \begin{array}{l}
\dd \vct{q}_{t}= M^{-1} \vct{p}_{t} \dd t \\
\dd \vct{p}_{t}= -\nabla_{\vct{q}} V(\vct{q}_{t};\lambda_{t}) \dd t -\gamma M^{-1} \vct{p}_{t} \dd t + \sigma \dd W_{t}^{\vct{p}}\\
\dd \lambda_{t}= \mathfrak{m}_{\lambda}^{-1} p_{\lambda_{t}} \dd t \\
\dd p_{\lambda_{t}} = -\partial_{\lambda} V(\vct{q}_{t};\lambda_{t}) \dd t -\gamma_{\lambda} \mathfrak{m}_{\lambda}^{-1} p_{\lambda_{t}} \dd t + \sigma_{\lambda} \dd W_{t}^{\lambda},
\end{array}\right.
\end{equation}
where $(W_{t}^{\vct{p}})_{t \geq 0}$ (resp. $(W_{t}^{\lambda})_{t \geq 0}$) is a $3N$-dimensional (resp. $1$-dimensional) Brownian motion, $M$ is the mass matrix of $N$ original particles, and the constants $\gamma_{(\lambda)}, \sigma_{(\lambda)}>0$ satisfy the fluctuation-dissipation condition $\sigma \sigma^{\top}=\frac{2\gamma}{\beta}$.
Since $\lambda$ has no physical meaning outside the $[0,1]$ interval, we impose reflecting boundary conditions on $\lambda$ at these boundaries. Other choices have been made in the literature, including expressing $\lambda$ as a function of an auxiliary dynamical variable in order to tune the dynamics, \cite{knight2011applying,zheng2012practically} but we did not find it necessary in this study. This generalizes to constant-pressure simulations in a straightforward way by adding barostat terms to the Cartesian coordinate dynamics described by Equation~\ref{eq:lambdadyn-Langevin-process}, as implemented in the MD engines.

\subsection{The choice of the parameters influences the dynamics of $\lambda$}

Two tunable parameters influence the dynamics of the alchemical parameter $\lambda$: its mass $\mathfrak{m}_{\lambda}$ and the friction coefficient $\gamma_{\lambda}$. In principle, these parameters do not affect the distribution sampled by the dynamics, but in practice, as has been extensively documented \cite{bieler2015orthogonal}, it is critical for the efficiency of $\lambda$-dynamics-based methods. On the one hand, the mass has to be large enough so that geometric degrees of freedom can relax at a given value of $\lambda$, but not too large so that $\lambda$-space is still sampled in a reasonable time. Critically, however, no adiabatic separation between $\lambda$ and the Cartesian degrees of freedom is required. On the other hand, the friction should be large enough to ensure numerical stability. In practice, we resort to strongly damped dynamics of $\lambda$ with $\gamma_{\lambda}=1000$ ps$^{-1}$ and $\mathfrak{m}_{\lambda}=150000$ kcal/mol fs$^2$.\footnote{The extended mass does not have the dimension of a physical mass because $\lambda$ is dimensionless.} We found these parameters suited to all the systems we tested the method on,  from the simple solvation of ions or water molecules to protein-ligand interactions. We thus expect these to be appropriate for these broad classes of systems. A detailed sensitivity analysis showing that a wide range of these parameters lead to similar estimates of hydration free energies can be found in Supporting Information.

\subsection{Adaptive Biasing Force using $\lambda$ as a collective variable}

Unbiased $\lambda$-dynamics alone does not result in rapid sampling of alchemical space because of free energy barriers. This is why methods based on $\lambda$-dynamics leverage biases to ensure such a condition. The original $\lambda$-dynamics method used a fixed bias, which has recently been improved by the use of Adaptive Landscape Flattening\cite{hayes2017adaptive}, while other groups showed results using metadynamics or Wang-Landau as an adaptive bias on $\lambda$\cite{wu2011lambda,hsu2023alchemical}, Local Elevation Umbrella Sampling\cite{bieler2014local} or an Adaptive Biasing Force in conjunction with an additional reaction coordinate\cite{zheng2008random,zheng2012practically}. Here, we resort to an Adaptive Biasing Force applied directly onto the $\lambda$ collective variable, and name this method lambda-ABF.

The main idea of the Adaptive Biasing Force algorithm\cite{Darve2001,darve2008adaptive} is to adaptively compute the derivative of the free energy associated with $\lambda$ through the Thermodynamic Integration formula and to apply this estimate to the simulated system through the dynamics. One can show\cite{lelievre2008long} that the sampling of $\lambda$ then converges towards a uniform distribution, and thus facilitates free energy barrier crossing. Compared to other adaptive bias procedures such as metadynamics, one of its strengths is that only local information is necessary to estimate the mean force in a bin resulting from the discretization of collective variable space\cite{comer2015adaptive}.

Let us call $A(\lambda)$ the Helmholtz free energy (defined up to a constant) associated with the alchemical parameter:

\begin{equation}
    A(\lambda)=-\beta^{-1}\ln\int e^{-\beta V(\vct{q} ; \lambda)}  d\vct{q} .
\end{equation}

We note $A'$ its derivative, which can be estimated through the TI formula:
\begin{equation}
    A'(\lambda)=\left<\frac{\partial V}{\partial \lambda}\right>_{\lambda}
\end{equation}
and $A'_t(\lambda)$ its on-the-fly estimate at time t.
The Adaptive Biasing Force equations of motion are the same as Equation~\ref{eq:lambdadyn-Langevin-process}, except an additional term (highlighted in blue):
\begin{equation}\label{eq:lambda-abf}
\left\{ \begin{array}{l}
\dd \vct{q}_{t}= M^{-1} \vct{p}_{t} \dd t \\
\dd \vct{p}_{t}= -\nabla V(\vct{q}_{t};\lambda_{t}) \dd t -\gamma M^{-1} \vct{p}_{t} \dd t + \sigma \dd W_{t}^{\vct{p}}\\
\dd \lambda_{t}= \mathfrak{m}_{\lambda}^{-1} p_{\lambda_{t}} \dd t \\
\dd p_{\lambda_{t}} = \left[ -\partial_{\lambda} V(\vct{q}_{t};\lambda_{t}) \color{blue}{+ A'_t(\lambda)} \right] \dd t-\gamma_{\lambda} \mathfrak{m}_{\lambda}^{-1} p_{\lambda_{t}} \dd t + \sigma_{\lambda} \dd W_{t}^{\lambda}
\end{array}\right.
\end{equation}

In practice, the biasing force $A'_t$ is scaled by a smoothly varying term in the initial regime to limit its initial variance, as explained in detail below.
At convergence, the estimate $A'_t(\lambda)$ provides a basis for obtaining the associated free energy surface $A(\lambda)$ by integration, and the desired free energy difference $\Delta A = A(1)-A(0)$.

One-dimensional numerical integration is performed using a simple Riemann sum.
In the ABF case, the gradient is estimated as a bin average of samples covering the continuous $\lambda$ space, making it an estimate of the \emph{average gradient} over a bin, thereby greatly reducing discretization error.\cite{Henin2021}
This results from an important difference between conventional, fixed-$\lambda$ TI, and ABF, which started with the idea of ``unconstrained TI''\cite{Darve2001, Darve2002} then eliminated the calculation of a constraint force altogether.\cite{Henin2004}
In fixed-$\lambda$ TI (or constrained TI along collective variables), estimates of $A'(\lambda)$ contain no information on its values between the sampled points, which warrants higher-order quadrature methods to improve the interpolation.\cite{Bruckner2010,Bruckner2010a}

The Gibbs free energy difference is by definition $\Delta G = \Delta A + p \Delta V$, and the volume contribution is typically negligible, as is the case here, therefore we use directly the $\Delta A$ estimators.

\subsection{Orthogonal space sampling}

All the techniques aiming at getting a free energy profile along one or several collective variables may suffer from free energy barriers in the space orthogonal to the collective variable of interest. This general observation is well documented and several techniques have been developed to identify and tackle this intrinsic limitation\cite{hahn2020overcoming}. In the context of alchemical free energy simulations, the orthogonal degrees of freedom hindering convergence of sampling techniques only focusing on $\lambda$ may be the movement of part of the host system such as a side chain rearrangement, the solvent reorganization upon appearance/disappearance of the mutated system, or more involved high dimensional phenomena.

If such orthogonal degrees of freedom are well identified, then the most straightforward strategy to circumvent the associated free energy barriers is to resort to biasing techniques along these degrees of freedom. This is for example the strategy followed in the recently introduced alchemical metadynamics\cite{hsu2023alchemical}. As stated in the introduction, one can also introduce a bias on a ``generic'' orthogonal degree of freedom such as the force acting on $\lambda$ as is done in the OSRW family of methods\cite{zheng2008random,zheng2012practically}.

Note that one can also rely on a combination of an Adaptive Biasing Force and metadynamics\cite{fu2018zooming,fu2019taming,chen2021overcoming}, which has been shown to enhance orthogonal space exploration while keeping the good properties of ABF. Still, this would require some special treatment of the metadynamics bias in our context because of the reflective boundary conditions imposed on $\lambda$.

In the context of alchemical free energy simulations, another approach consists of performing changes in $\lambda$ for a given configuration with the hope that the orthogonal free energy barriers are lower for certain values of $\lambda$, and therefore that these $\lambda$ changes accelerate barrier crossing. Hamiltonian replica exchange methods, in which these changes are made through Monte-Carlo moves in an extended phase space, fall under this category. By making the alchemical variable dynamical, $\lambda-$dynamics-based methods also fall naturally under this category. This is extensively discussed in reference\cite{hahn2020overcoming}. A key difference between these two families of methods lies in the intrinsic dependence of the former to a predetermined $\lambda$-space discretization and to a trial exchange rate when no such choice has to be made in the latter. Nevertheless, this is replaced by the need to choose the parameters driving the dynamics of $\lambda$ in the latter, as discussed above. Furthermore, $\lambda$-dynamics-based methods allow for local relaxation of the alchemical parameter following the local gradient of the potential with respect to $\lambda$ without the condition of accepting a Metropolis criterion, which we expect to facilitate sampling.

In recent years, the emergence of highly parallel computing platforms such as supercomputers, cloud-based resources, or in-house multiple-GPU setups has propelled the development of replica-based simulation techniques that leverage these frameworks \cite{raiteri2006efficient,larson2009folding}. The objective of these multiple-walker methods is to reduce the time-to-solution by capitalizing on the memory-less exponential distribution of the first exit time from metastable states under appropriate assumptions \cite{hedin2019gen}. In practice, the expectation is that running $n$ multiple walkers in parallel diminishes the time required to obtain the first exit event from a metastable state by a factor of $n$.
In the context of lambda-ABF, several approaches can be employed to design such methods. In this study, we adopt one of the simplest strategies, wherein $n$ walkers independently perform lambda-ABF simulations and periodically share their current estimation of the mean force of $\lambda$, updating their local bias accordingly.\cite{minoukadeh2010potential} Although more complex multiple-walker strategies involving selection mechanisms have been proposed \cite{minoukadeh2010potential,comer2014multiple}, they offer only modest efficiency gains at the expense of increased computational overhead and implementation complexity.
Some computational overhead arises from the necessary synchronization between walkers at fixed intervals. Here, we have selected a synchronization interval of 1 picosecond, rendering this overhead negligible.

In summary, we propose lambda-ABF, a technique rooted in Langevin $\lambda$-dynamics, making use of multiple-walker Adaptive Biasing Force simulations using $\lambda$ as a collective variable. The adaptive bias progressively flattens free energy barriers along the alchemical variable, while the crossing and recrossing of orthogonal barriers are favored by the dynamical evolution of $\lambda$ and by the multiple-walker scheme. The free energy profile is then recovered through the Thermodynamic Integration estimator.

\subsection{Standard free energies of binding}

\subsubsection{Thermodynamic cycle}

To obtain standard, absolute free energies of binding, we resort to a now classic thermodynamic cycle\cite{gilson1997statistical} where the desired free energy is computed as the difference between the free energy of decoupling both electrostatic (including polarization) and van der Waals interactions between the ligand and the rest of the system when it is in complex with the host (we will call it  ``complexation'' phase in what follows) and the free energy of decoupling the same non-bonded interactions when the ligand is alone in solution (we will call it ``solvation'' phase in what follows).

Here we opt to follow the respective conventions of the Tinker and CHARMM/NAMD communities as to the definition of the $\lambda$ pathway and the sign of alchemical free energies.
Using the customary Tinker-HP notation, this reads:
\begin{equation}
    \Delta G^\circ_\text{bind} =  -\Delta G_\text{solv} -\Delta G^\circ_\text{V} +\Delta G_\text{restr} + \Delta G_\text{complex},
\end{equation}
and using the SAFEP notation:
\begin{equation}
    \Delta G^\circ_\text{bind} =  \Delta G^*_\text{bulk} - \Delta G^\circ_\text{V} + \Delta G_\text{restr} -\Delta G_\text{site}^*,
\end{equation}
where $\Delta G_{\text{complex}}$ and $\Delta G_\text{solv}$ are the opposite of $\Delta G_\text{site}^*$ and $\Delta G^*_\text{bulk}$, respectively.
The former denote creation free energies, whereas the latter are decoupling free energies.
$\Delta G^\circ_\text{V}$ is the free energy difference of releasing an auxiliary spherical confinement restraint into the standard state volume.\cite{santiago2023computing}
$\Delta G_{\text{restr}}$ is the free energy change of releasing the ligand restraints that must be imposed during the complexation phase in order for these simulations to converge and to properly define binding modes when necessary\cite{hermans1986free,gilson1997statistical,boresch2003absolute,mobley2006use}. We describe our strategy regarding restraints in the following subsection.

\subsubsection{Binding restraints}

The topic of restraints in the context of absolute free energies of binding computed through alchemical methods has been the focus of active research in recent years with three main classes of methods: (flat-bottom) harmonic restraints on the distance between the centers of mass of the ligand and of the binding site\cite{laury2018absolute}, the ``Boresch'' restraints\cite{boresch2003absolute} that limit positional and orientational movement of the ligand with respect to the binding site by imposing 6 harmonic restraints between the ligand and the host (one distance, two angles, and three torsions), and the restraint scheme proposed by Roux, Chipot, and coworkers \cite{wang2006absolute,gumbart2013standard,fu2017new,fu2022accurate}, which also includes a conformational restraint on the ligand.

In the first two cases, the free energy of releasing these restraints in order to complete the thermodynamic cycles can be computed analytically. These methods suffer from distinct limitations. In the common case where only center of mass distances are restrained, the rotational and internal degrees of freedom of the ligand are left fully unrestrained; this leaves a large volume of accessible phase space that can be difficult to sample exhaustively when the ligand is progressively decoupled from its environment, especially for larger ligands. In theory, this issue is corrected by the use of the more complex Boresch restraints, but they still suffer from the difficulty of choosing the atoms involved in their definition, as poor choices may lead to numerical instabilities\cite{clark2023comparison}. The current Roux/Chipot restraints implemented using the Colvars library depend less critically on the choice of reference atoms, yet they retain the complexity of using multiple layered restraints.

Recently, a new restraint scheme has been introduced by some of us based on a Distance to Bound Configuration (DBC) collective variable, which is the RMSD of the ligand in the moving reference frame of the binding site\cite{salari2018}. By nature, it acts on the position and orientation of the ligand with respect to the binding site, as well as its conformation, but not on internal degrees of freedom of the binding site. Furthermore, it natively allows for including as many atoms as desired in the definition of the ligand and the binding site, avoiding some potential issues of Boresch restraints. In practice, it can be easily defined using VMD\cite{HUMP96} in combination with the Colvars Dashboard\cite{henin2022human} to produce a Colvars input file that can be used in conjunction with an MD program such as Tinker-HP or NAMD. Typically, one runs a long Molecular Dynamics trajectory of the fully bound ligand in the desired binding mode, one can then post-process the trajectory with the Colvars Dashboard in order to get the maximum value of the associated DBC sampled by the system in this binding mode. One can then simply impose a flat-bottom harmonic restraint on this DBC starting from this value as described in \cite{santiago2023computing}.

The free energy cost associated with the release of this kind of restraint cannot be computed analytically. However, it is easily computed in the gas phase through Thermodynamic Integration by progressively releasing it to a compatible harmonic distance restraint whose cost can in turn be computed analytically. When intramolecular non-bonded interactions of the ligand are removed, conformational barriers of the ligand are lowered, leading to rapid convergence.

In our study, for the NAMD simulations, we applied a scheme that preserved the self-interactions of the ligand throughout.
In the case of phenol as a ligand, this is of little consequence, as there is little flexibility and self-interactions are limited. Thus, the restraint free energy estimation converges rapidly in that case as well.

\section{Implementation}

\subsection{Alchemical pathway}

\subsubsection{Tinker-HP}

A common practice is to distinguish the non-bonded interactions that are scaled by $\lambda$ by considering separately van der Waals and electrostatic interactions, including polarization \cite{shi2021amoeba} in the context of polarizable force fields\cite{shi2015polarizable,reviewPFF}. One introduces two distinct alchemical parameters $\lambda_v$ and $\lambda_e$ which correspond to the scaling of van der Waals and electrostatic interactions respectively. These two variables are a function of the global alchemical parameter $\lambda$, which is the only one exposed to the Colvars library and thus to the biasing and free energy estimation algorithms. There are then several possibilities to design such a function. We implemented a linear dependence conditioned by two parameters $b_v$ and $b_e$ so that $\lambda_v=0$ at $\lambda=0$ and $\lambda_v=1$ at $\lambda \geq b_v$, while $\lambda_e=0$ at $\lambda \leq b_e$ and $\lambda_e=1$ at $\lambda=1$ (Figure~\ref{fig:alch-path}), which is equivalent to the NAMD scheme.
This allows for a wide range of alchemical pathways, all respecting the boundary conditions of no interactions at $\lambda=0$ and full interactions at $\lambda=1$.

\begin{minipage}{\linewidth}
\makebox[\linewidth]{
\includegraphics[scale=0.7]{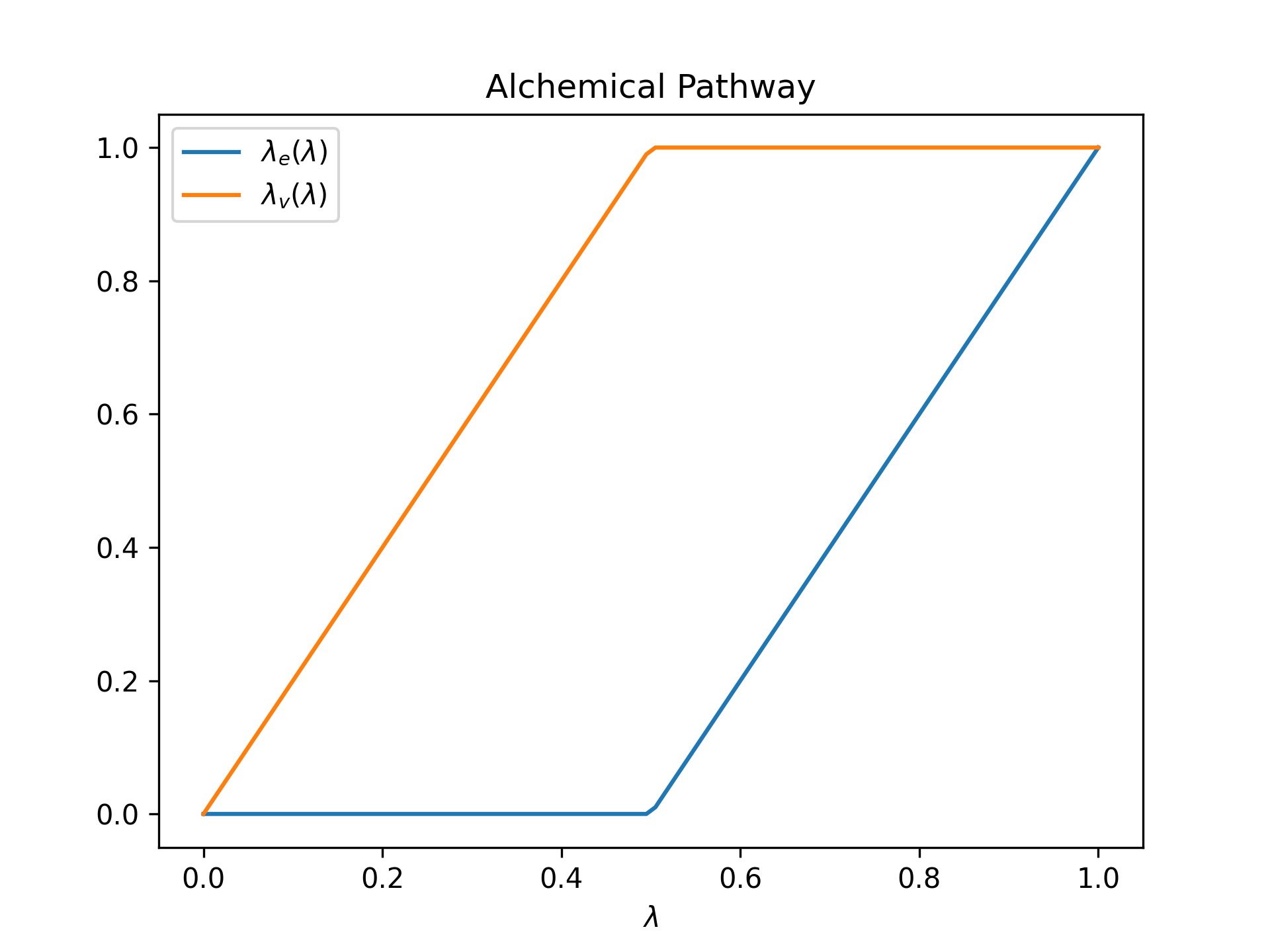}}
\captionof{figure}{Alchemical pathway: evolution of alchemical parameters $\lambda_e$ and $\lambda_v$ as a function of the global  $\lambda$, for the  $b_e=b_v=0.5$.}
\label{fig:alch-path}
\end{minipage}

Furthermore, van der Waals (resp. electrostatics) ``annihilation'' means that no van der Waals (resp. electrostatics) interactions involving the mutated system exist anymore when $\lambda_v=0$ (resp. $\lambda_e$), including the intramolecular ones.

Because lambda-ABF is a TI-based method, it relies on computing averages of the force acting on $\lambda$, which requires special treatment of the many-body interactions, i.e. reciprocal space electrostatics (for simulations using Particle Mesh Ewald\cite{darden1993particle,essmann1995smooth} and polarization with polarizable force fields.\cite{Ewaldpol}). Similar to what is presented in \cite{schnieders2012structure}, we simply interpolate both terms between extreme values of $\lambda_e$ where they are well defined. This of course comes with an additional cost compared to energy-based techniques in which the force on $\lambda$ is not required, such as FEP or BAR. For polarizable FFs, one can limit this overhead by choosing to make the polarization of the mutated system appear only after a fixed value $b_p \geq 0$ of $\lambda_e$\cite{schnieders2012structure}. Because testing the impact of such a parameter is outside the scope of the current study we used $b_p=0$ in everything that follows, even if this possibility is implemented in Tinker-HP.

\subsubsection{NAMD}

For simulations using the CHARMM FF in conjunction with NAMD, we applied linear scaling to Particle Mesh Ewald (PME) electrostatics, and soft-core scaling of Lennard-Jones potentials as described in \cite{zacharias1994separation}. To avoid the instability associated with naked charges, van der Waals and electrostatic interactions were scaled together but in a staggered fashion, corresponding to $b_e=0.5$ and $b_v=1$, or in terms of NAMD parameters: \texttt{alchElecLambdaStart 0.5} and \texttt{alchVdwLambdaEnd 1.0}.

In the present NAMD applications, we used decoupling rather than annihilation (\texttt{alchDecouple on}), meaning that non-bonded interactions within the perturbed groups are retained throughout the perturbation, and the fully decoupled ligand retains its self-interactions.

For fixed-$\lambda$ simulation, we used Interleaved Double-Wide Sampling (IDWS), which is specific to NAMD and allows for on-the-fly computation of comparison energies in the forward and reverse directions. In contrast, in Tinker-HP, comparison energies are computed in post-processing.

\subsection{Software interface}

The lambda-ABF implementation used in this work relies on the combination of a Molecular Dynamics package and the Colvars library.\cite{Fiorin2013}
Here we present a completely novel interface between Colvars and the simulation package Tinker-HP, as well as an extension of the existing NAMD interface for alchemical free energy simulations\cite{Dixit2001namd_fep, Phillips2005} including its GPU implementation.\cite{Chen2020namd_gpu_fep, phillips2020scalable}

Colvars communicates with each back-end program through a specialized implementation of a ``proxy'' C++ class, as documented in \cite{Fiorin2013}.
In contrast with existing interfaces of Colvars with other programs, the new Tinker-HP interface includes a C layer that allows for linking the C++ library and the Fortran code of Tinker-HP.
In practice, some parameters are first initialized, such as the target temperature, the nature of the reaction coordinate of interest with the atoms involved as well as the bias to apply on it (constant or adaptive).
Then, each time the gradients routine of the MD code is called, the current value of the Cartesian coordinates involved in the collective variable of interest is retrieved (as well as the forces acting on them if required) by the Colvars library, enabling the computation of a bias whose forces are sent back to the MD engine.
In the context of multiple time step integrators\cite{tuckerman1992reversible,lagardere2019pushing} within Tinker-HP, this exchange only happens at the outer (largest) timesteps, which is correct as long as the collective variables vary as slowly as the slow force field terms. This is different from the Colvars interfaces to NAMD and other simulation software, which call Colvars every inner timestep.

\begin{figure}
\centering
\includegraphics[width=12cm]{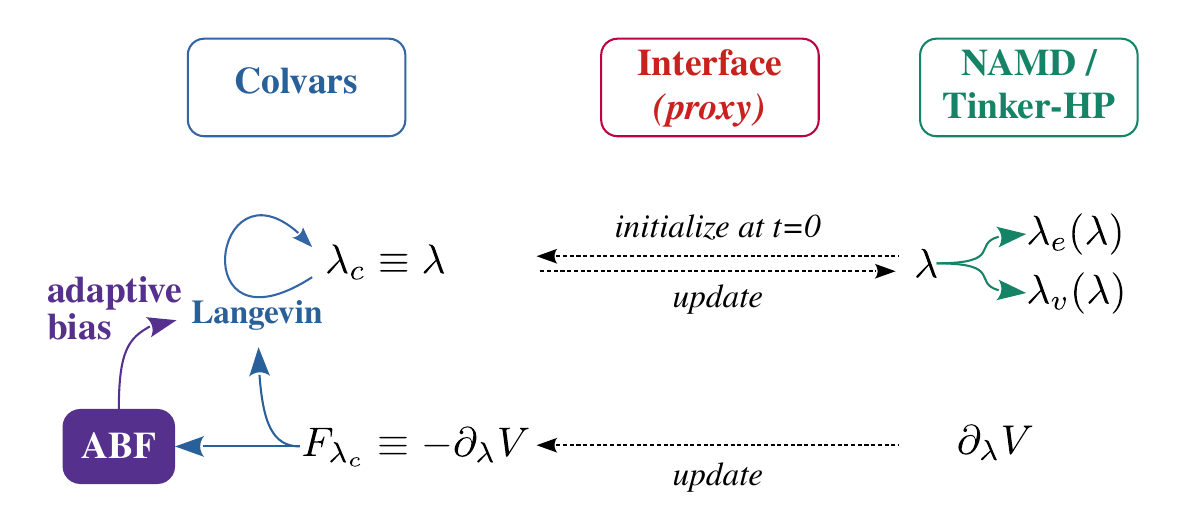}
\caption{Diagram of the communication between Colvars and Tinker-HP (or NAMD) when running a $\lambda$-dynamics simulation and ABF.}\label{fig:Tinker-Colvars-interface}
\end{figure}

The workflow of a $\lambda$-dynamics simulation performed by the Tinker-HP--Colvars and NAMD--Colvars interfaces is depicted in Figure~\ref{fig:Tinker-Colvars-interface}.
At a given time $t$, the MD engine holds a value for $\lambda$ (\texttt{lambda}), and computes the corresponding $\lambda_{e}(\lambda)$ and $\lambda_{v}(\lambda)$, which define the current alchemically perturbed Hamiltonian.
Internally, Colvars keeps a collective variable $\lambda_{c}$ and propagates it dynamically according to Langevin dynamics \eqref{eq:lambda-abf} over one timestep, using a BAOA integrator\cite{kieninger2022gromacs}.
The interface functions are used by Colvars to set the new value of $\lambda$ to  $\lambda_{c}$ after every update and to obtain the updated value of $\partial_{\lambda} V$ computed by the MD engine.
At the end of each update step, the MD engine updates the Hamiltonian of the simulation according to the new $\lambda$.

To enable Tcl scripting features of Colvars, Tinker-HP can now be linked with the Tcl library\cite{ousterhout1989tcl} and a new Colvars command, \texttt{sourceTclFile} has been added to load a script at the beginning of a simulation. This makes Colvars-Tcl scripts portable between Tinker-HP, NAMD, and potentially any other supported MD engine after linking the lightweight and portable Tcl library.

\subsection{ABF parameters}

Aside from the choice of the discretization of the collective variable, the only choice left to the practitioner of ABF simulations lies in the way the bias is progressively turned on after the beginning of the simulation.
In Colvars, this is controlled by the \texttt{fullSamples} parameter (default value 200) which is the number of collected samples in each bin after which the opposite of the estimated mean force is fully applied to the dynamics (this is introduced continuously by a linear ramp). Relying on too small a value of this parameter can lead to the application of a bias that is far from the converged mean force and thus out of equilibrium effects and in turn slow convergence\cite{miao2021avoiding}. This is all the more true in the case of orthogonal degrees of freedom with long decorrelation times that may appear especially during absolute free energies of binding simulations. We recommend the use of the conservative value \texttt{fullSamples}$=5000$ when running lambda-ABF simulations, which proved to be large enough in the most complex systems we studied and which thus constitutes a safe general choice.

\subsection{Availability}

Lambda-ABF is available in the public release of Tinker-HP 1.2 (CPU and GPU versions) and the latest public version of Colvars, both of which are open source and available on their respective GitHub repositories. Similarly, lambda-ABF has been implemented in both the CPU and GPU versions of NAMD 3, and will be integrated into its next public release. These two software interfaces broaden the range of free energy techniques accessible to practitioners by combining them with the many interaction models and sampling techniques already implemented in these software packages, as demonstrated below.

\section{Numerical results}

In this section, we present numerical results obtained on absolute solvation free energies and absolute free energies of binding using our newly developed lambda-ABF method. We start by demonstrating the absence of bias on the free energy estimates obtained with the method by computing the hydration free energies of monoatomic ions and water with the AMOEBA FF. We then further test the efficiency of the method on more complex absolute free energies of binding, from a relatively small host-guest system from the SAMPL6 challenge\cite{laury2018absolute} with the same FF up to more realistic protein-ligand systems: first, the phenol-lysozyme complex using the CHARMM FF\cite{santiago2023computing} and the cyclophilin-D protein (CypD) in complex with a ligand taken from the literature\cite{ahmed2016fragment,gradler2019discovery}, using the AMOEBA FF. Simulations using the AMOEBA model were performed with Tinker-HP and those using CHARMM were run with NAMD.

For all these systems, the goal of the present study is first to assess the efficiency and accuracy of the method given a model and a fixed set of parameters and then to compare the obtained results with experimental ones. To do so, using the same AMOEBA polarizable model, we compare the convergence of lambda-ABF to that of a well-established method: simulation of independent ``windows'' with fixed $\lambda$ values with adjacent free energy differences obtained with the BAR method. In the rest of the paper, this will be referenced as the ``fixed-$\lambda$'' method. No finite size correction is added to the obtained results. Results and statistical uncertainty are assessed by running 3 repeats of each simulation for simulations using the AMOEBA FF, and 5 repeats when using the CHARMM FF\cite{bhati2022large,bhati2023long}.

Because we chose to run simulations using what can be considered now the standard absolute free energy protocol for use with the AMOEBA FF\cite{shi2021amoeba}, we rely on independent simulations at fixed $\lambda$ values that each require some equilibration time. Here, we chose to discard the first 400 picoseconds of simulation for the most simple systems and 1 nanosecond for the more complex ones. One of the merits of $\lambda$-dynamics-based methods in general, and of lambda-ABF in particular is to bypass this requirement.
Detailed simulation parameters as well as histograms of sampled $\lambda$ values can be found in Supporting Information.

\subsection{Hydration free energies}

Hydration free energies of cations such as Na+ and K+ and of small molecules like water are among the simplest case studies for free energy methods as one expects no real challenge in terms of orthogonal space sampling for these systems. This is why we don't rely on the multiple walker strategy in this section. Still, for the case of the cations, these simulations require sampling the annihilation of the strong electrostatic interactions between the ion and its coordinated water molecules in the electrostatic leg and the (small) cavitation due to the annihilation of the associated van der Waals interactions. Conversely, sampling the electrostatic annihilation is expected to be easier for the water molecule but the annihilation of the van der Waals interactions is expected to be more challenging.
In practice, we compare the results obtained with fixed (equispaced) $\lambda$ windows (10 for the van der Waals annihilation and 10 for the electrostatics and polarization annihilation) of 4~ns with a single lambda-ABF trajectory of 10~ns. The goal here is not to assess the rate of convergence of the methods but to validate that lambda-ABF is unbiased with respect to a reference.

 \begin{table}[!ht]
 \center
 \begin{tabular}{|c|c|c|c|}
 \hline
  & \chemform{Na^+}& \chemform{K^+} &Water \\
 \hline
  lambda-ABF& -89.62 ($\pm 10^{-3}$)&-72.21 ($\pm  10^{-3}$) &-5.63 ($\pm 2\times 10^{-5}$)  \\
  \hline
 fixed-$\lambda$& -89.75 ($\pm 5\times 10^{-4}$)  &-72.33 ($\pm 8\times 10^{-2}$)   & -5.67 ($\pm  10^{-2}$)   \\
 \hline
 Experiment\cite{schmid2000new,kelly2005sm6}&-86.8   &-69.3   & -6.31   \\
 \hline
 \end{tabular}
 \caption{ Hydration free energies (in kcal/mol) for the cations \chemform{Na^+},\chemform{K^+} and a water molecule, obtained from lambda-ABF, the fixed-$\lambda$ method and experiment.}
  \label{tab:HFE}
\end{table}
To be able to compare the ion solvation free energies to experimental results that typically choose to use standard states of 1~Atm for the gas and a 1 molar solution a correction of $\beta^{-1}\text{ln}(\frac{V_{1\text{Atm}}}{V_{1M}})=1.9$~kcal/mol was added to the hydration free energies of \chemform{Na^+} and \chemform{K^+}, as described in ref \cite{grossfield2003ion}.

Results shown in Table \ref{tab:HFE} clearly demonstrate that the lambda-ABF approach yields unbiased hydration free energies.

\subsection{Cucurbit[8]uril host-guest complexation}

Before turning to more complex protein-ligand complexes, it is worth benchmarking lambda-ABF on a simpler host-guest system as the ones typically involved in the blind SAMPL challenges\cite{rizzi2018overview}. Here, we focus on the ligand 9 in complex with Cucurbit[8]uril as defined in the 6th iteration of the SAMPL challenges as it has already been extensively studied by some of us\cite{inizan2023scalable}. As described in \cite{laury2018absolute} it involves a hydrophobic central region and a polar exterior.
\\
\medskip

\noindent\textbf{Dealing with metastabilites related to counterions}

\medskip

\begin{figure}[!htb]
    \centering
    \begin{minipage}{.5\textwidth}
        \centering
        \includegraphics[scale=0.5]{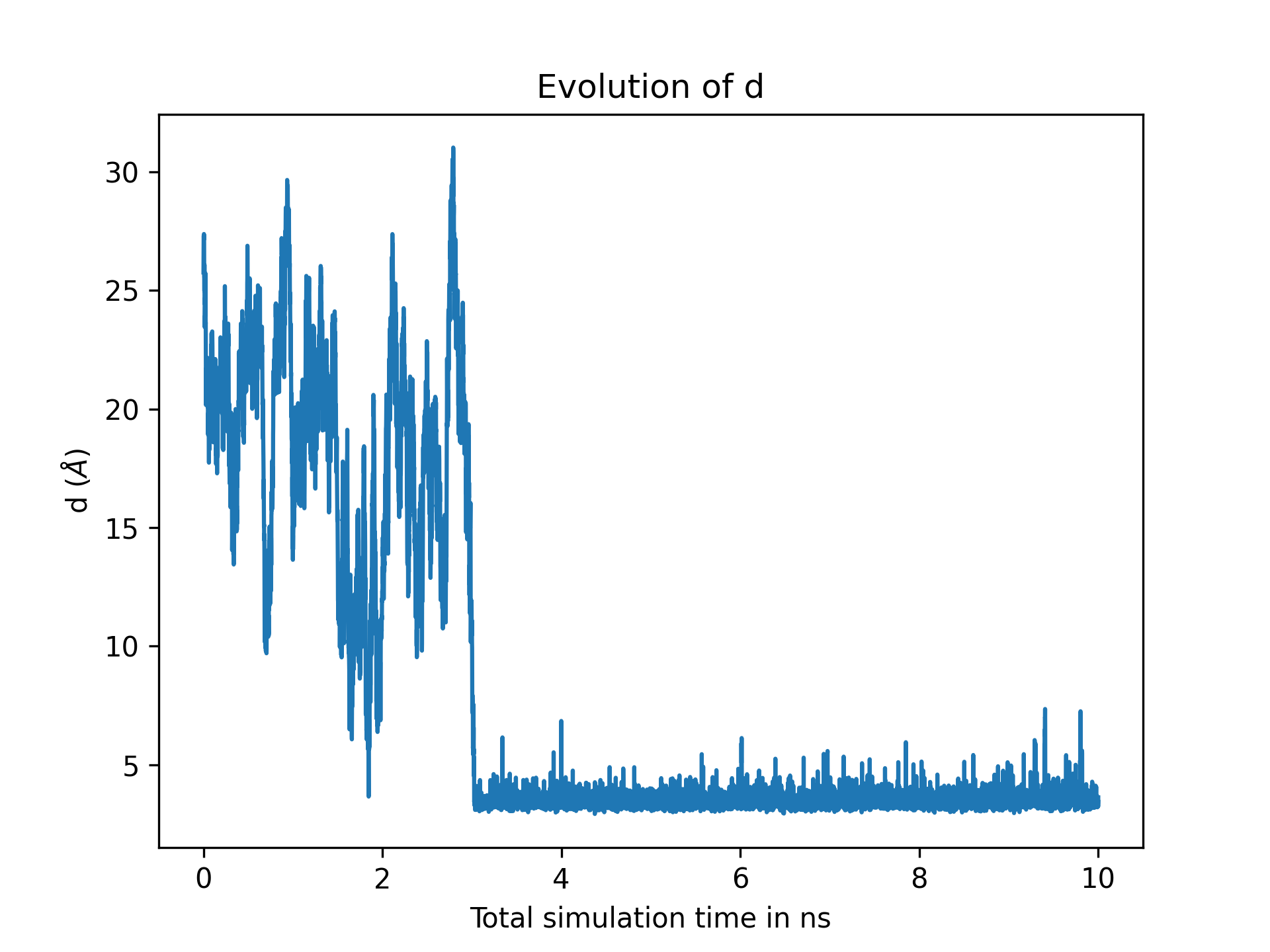}
    \end{minipage}%
    \begin{minipage}{0.5\textwidth}
        \centering
        \includegraphics[scale=0.5]{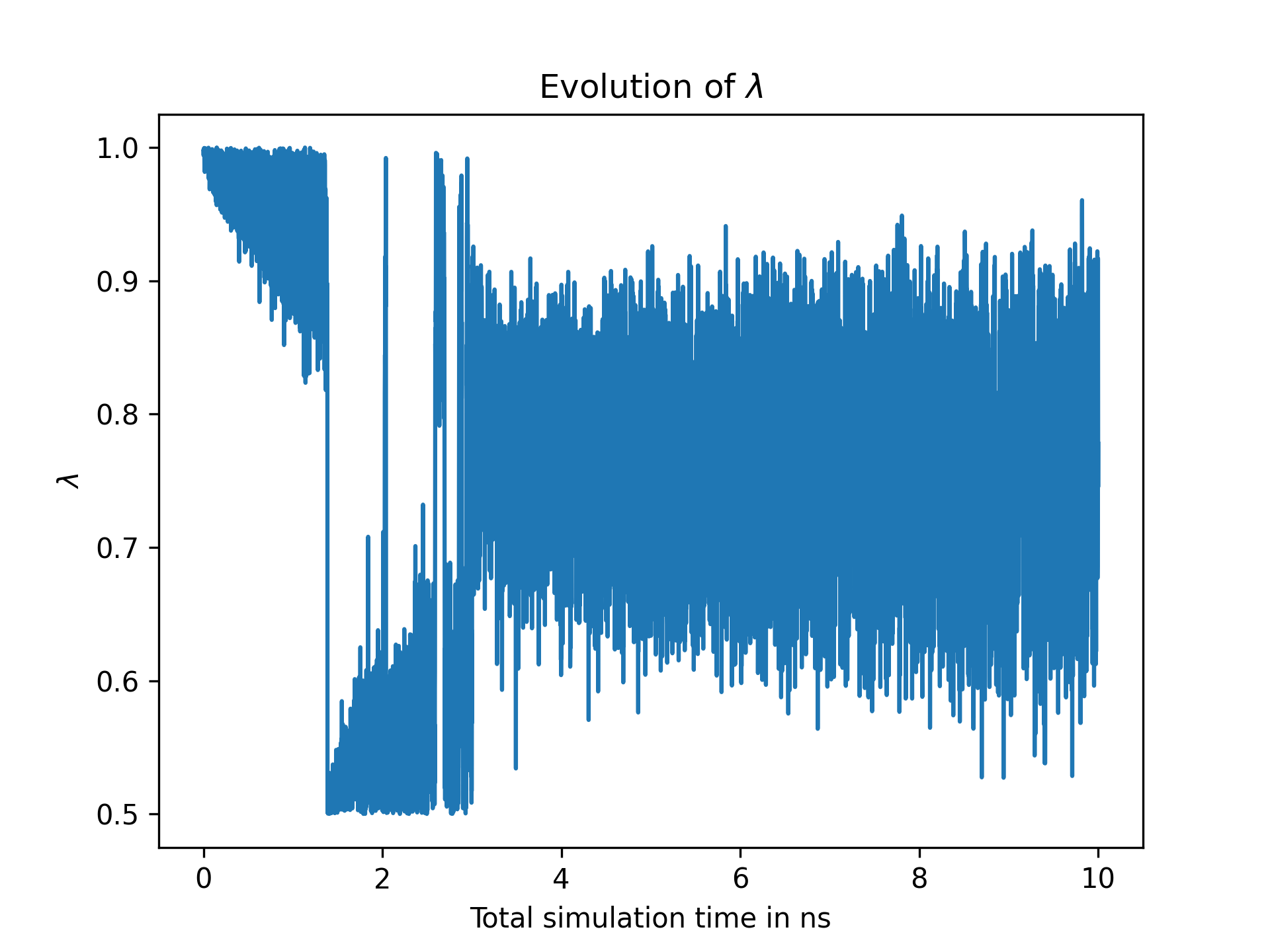}
    \end{minipage}
\caption{Evolution of d (left) and $\lambda$ (right) as a function of the simulation time}
    \label{fig:sampl6-9-d-lambda}
\end{figure}

\begin{figure}[!htb]
\centering
\begin{minipage}{1.0\textwidth}
\includegraphics[width=\textwidth]{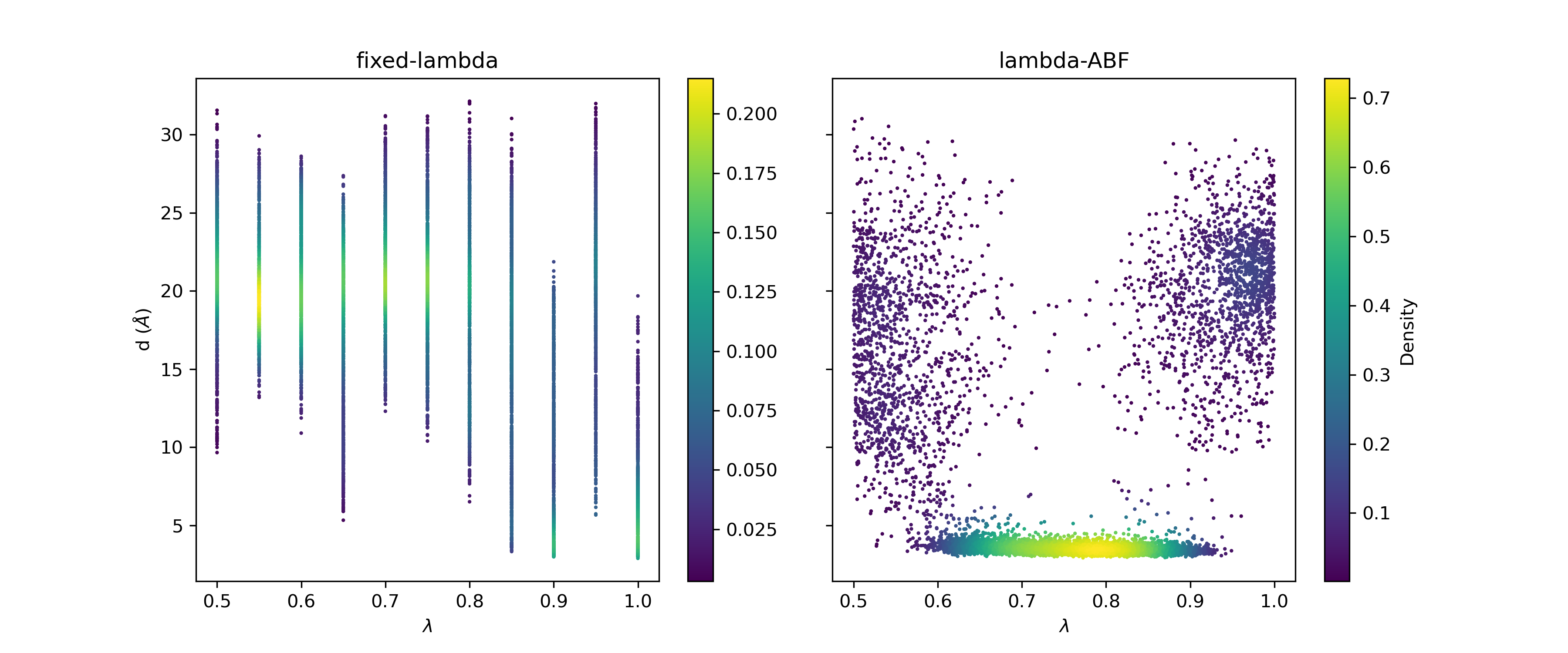}
\end{minipage}\hfill
\vskip\floatsep
\includegraphics[width=0.5\textwidth]{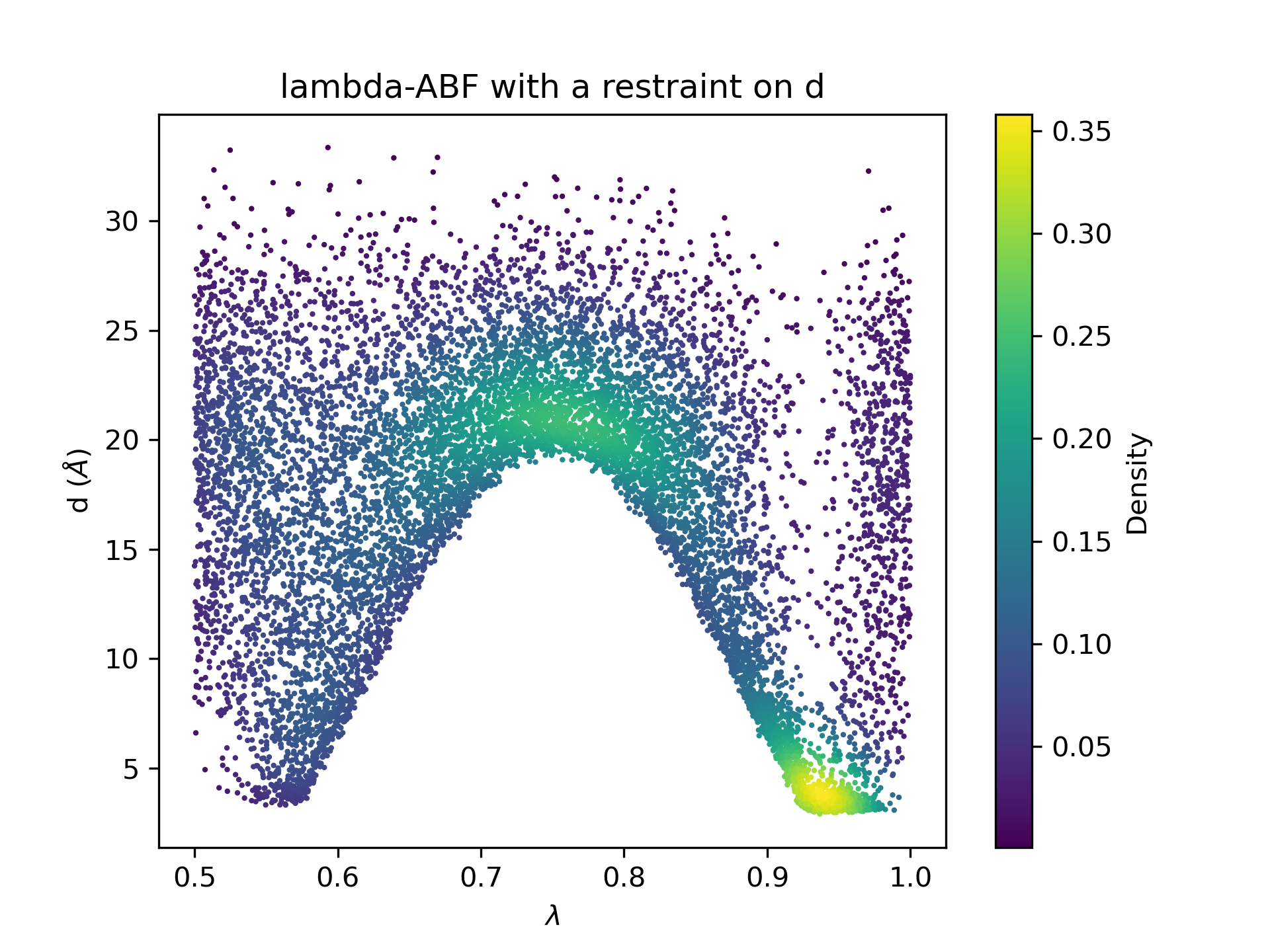}
\caption{($\lambda$,d) distribution with the fixed-$\lambda$ method and the lambda-ABF method, without a restraint on d (up) and with lambda-ABF with a $\lambda$-dependant restraint on d (down)}\label{fig:lambda-d}
\end{figure}

As described in \cite{laury2018absolute}, we included a Cl $^-$ counterion to neutralize the simulation box of both the solvation and the complexation leg of the thermodynamic cycle. Initial lambda-ABF simulations revealed the existence of a deep metastable basin involving this ion at the stage where the ligand is partially discharged. As can be seen in Figure \ref{fig:sampl6-9-d-lambda} (where d is the distance (in \AA\/) between the ion and the nitrogen of the ligand), after a few nanoseconds of simulations, the counterion binds the partially discharged nitrogen of the ligand. As shown in Figure \ref{fig:lambda-d}, this state is not visited by the fixed-$\lambda$ simulations, illustrating the sampling power of the lambda-ABF approach.
However, this process constitutes an orthogonal barrier, leading to a long residence time and limiting sampling of the complete $\lambda$-space. Thus, no meaningful results can be obtained on this system without additional bias in a production setup.

Because this state is associated with intermediate values of $\lambda$ and because we are only interested in getting the free energy, which is a state function, between extreme values of $\lambda$, we can devise a $\lambda$-dependent restraint on the distance d limiting the accessible values in these intermediate stages.
Let us define a threshold distance $d_\text{min}(\lambda)$ as:
\[d_\text{min}(\lambda)=\frac{\delta}{2}(1-\textrm{cos}(2\pi \lambda))\]
Then, $d_\text{min}(0)=d_\text{min}(1)=0$ and $d_\text{min}(0.5)=\delta$, so that a restraint enforcing $d > d_\text{min}$ only acts when $0<\lambda<1$ and not when $\lambda=0$ or $\lambda=1$. In practice, using the Tcl scripting capabilities of the Colvars library, we define the additional energy term:
\[V_\text{rest}(\lambda,d)=\left\{
\begin{array}{ll}
     \frac{k}{2}(d-d_\text{min}(\lambda))^2& \textrm{if } d<d_\text{min}\\
     0& \textrm{if } d \geq d_\text{min}
\end{array}\right.
\]
We use $\delta=20$ \AA{} and a spring constant $k = 5$~kcal/mol.\AA$^{-2}$. By doing so, we limit the distances to be sampled at intermediate values of $\lambda$ without modifying phase space at $\lambda=0$ and $\lambda=1$, as illustrated in Figure \ref{fig:lambda-d}. The use of lambda-ABF in combination with a $\lambda$-dependant restraint illustrates the power and flexibility of the approach as implemented in the Colvars library. Indeed, one can devise a wide range of additional biasing terms on top of a lambda-ABF simulation, in the spirit of the alchemical metadynamics methods\cite{hsu2023alchemical}. This setup will be the focus of future work.

Final values were obtained using 10-nanosecond simulations for each $\lambda$ window for the fixed-$\lambda$ simulations,
using a $\lambda$ schedule described in Supporting Information,
amounting to a total of 130 nanoseconds for the van der Waals legs, and 110 nanoseconds for the electrostatic ones. Lambda-ABF simulations used 4 walkers. 30-nanosecond simulations were run for the electrostatic part of the solvation leg (120 nanoseconds in total), 20 nanoseconds for the corresponding van der Waals part (80 nanoseconds total), 36 nanoseconds for the electrostatic part of the complexation phase (108 nanoseconds total) and 20 nanoseconds for the corresponding van der Waals leg.

Predicted standard binding free energies are shown in Table \ref{tab:ABFEsampl} for both methods, and both match the experimental value published in \cite{laury2018absolute}.

 \begin{table}[h!]
 \center
 \begin{tabular}{|c|c|}
 \hline
  & SAMPL6 ligand 9 \\
 \hline
  lambda-ABF& -8.93 ($\pm 0.2$)  \\
  \hline
 fixed-$\lambda$& -8.87 ($\pm 0.3$)     \\
 \hline
 Exp.&  -8.68  \\
 \hline
 \end{tabular}
 \caption{ Standard free energies of binding (in kcal/mol) for the ligand 9 of the SAMPL6 challenge, obtained from lambda-ABF (with restraint on d) and the fixed-$\lambda$ method.}
 \label{tab:ABFEsampl}
\end{table}

Note that although these two approaches yield the same final estimate,  they exhibit different convergence behaviors. This is illustrated in Figure S3, which shows the running estimates of the various free energy components as a function of the increasing total simulation time. The estimates given by the two methods show comparable variance,  except in the case of the van der Waals part of the solvation leg where it is drastically smaller for lambda-ABF. As will be shown for the more complex case of a protein-ligand system, this reduction in variance in the case of similar converged values is systematic and can also be related to the sampling power of the Multiple Walker lambda-ABF approach. Indeed, a reduced variance in combination with a comparable speed of convergence can be interpreted as an overall comparable sampling when all the replicas are included (comparable average) but with a better sampling per replica (reduced variance). Let us also point out that an intrinsic appreciable feature of lambda-ABF is the ease of getting the running estimates of the free energies involved as they are regularly printed out by the Colvars library and the associated ease of monitoring their convergence. This is not the case for the fixed-$\lambda$ methods that require a post-processing stage requiring access to the previously produced (potentially distributed) data, even if this additional step can be automated.

As described in \cite{salari2018,santiago2023computing}, the free energy cost of releasing the flat bottom harmonic restraint on the DBC collective variable has to be computed numerically, but this drawback is overcome by the fact that this can be done efficiently in the gas phase. Here, we resort to a Thermodynamic Integration approach in which this restraint is combined, with a compatible one (same spring constant and distances) between the center of mass of the ligand and the geometric point corresponding to this center of mass in the bound conformation. At the start of the thermodynamic cycle, the second restraint is not seen by the system (as it is implicitly present in the restraint on the DBC collective variable), but as the restraint on the DBC is progressively removed, the second restraint comes into play. Finally, the free energy cost of releasing the harmonic restraint on the distance is computed analytically.
In principle, estimating the configurational entropy of the ligand with this approach could be challenging for large ligands, but here the associated free energy barriers are strongly reduced because both intramolecular electrostatic and van der Waals are set to zero at this stage.

In practice, this is done through a single gas phase trajectory during which the Colvars library handles the progressive removal of the harmonic restraint on the DBC collective variable (using 20 windows of 1 nanosecond in all the cases presented here) while outputting the mean force required to compute the associated free energy, as shown in Figure S5.

\subsection{Lysozyme-phenol binding}

We validated lambda-ABF simulations of fixed-charge CHARMM models in the NAMD package~\cite{phillips2020scalable}.
The chosen test case is phenol binding to the engineered binding site in the L99A/M102H mutant T4 lysozyme.
The simulation system was identical to that of previous works~\cite{ebrahimi2022symmetry, santiago2023computing}.
 See Supporting information for details.

\begin{figure}[!htb]
    \centering
    \includegraphics[width=0.45\textwidth]{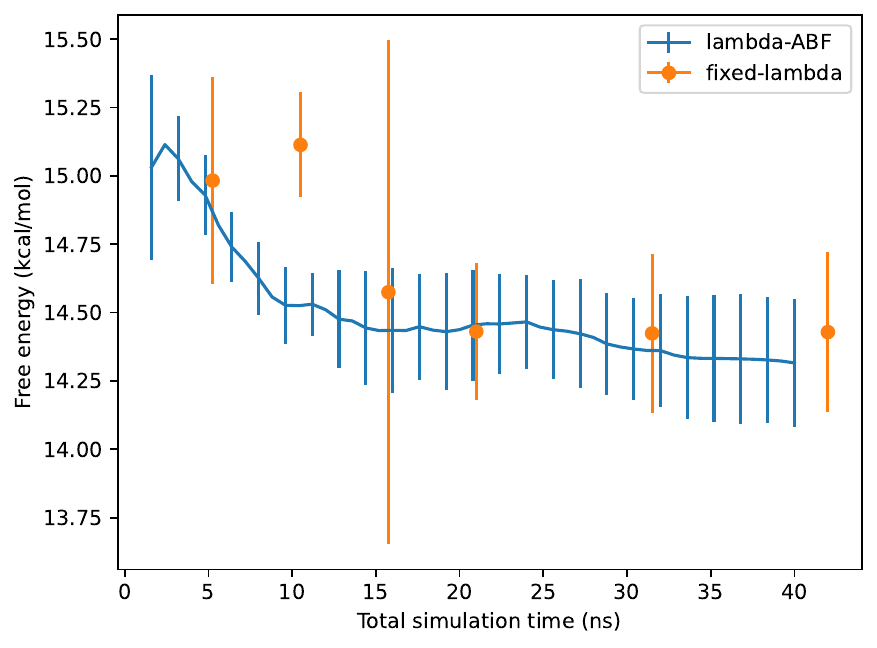}
    \caption{Convergence of free energy estimates for decoupling phenol from the engaverageineered binding site of lysozyme for a fixed-charge force field simulated in NAMD. The estimated decoupling free energy is shown as average and standard deviation over 5 independent repeats, for a lambda-ABF and a fixed-$\lambda$ simulation with interleaved double-wide sampling and the BAR estimator.}
    \label{fig:lysozyme_FE}
\end{figure}

Convergence of the lysozyme/phenol decoupling free energy is illustrated in Figure~\ref{fig:lysozyme_FE}.
The final fixed-$\lambda$/BAR estimate is
$\Delta G^*_\text{site} = 14.4 \pm 0.3$~kcal/mol, whereas the lambda-ABF estimate is
$\Delta G^*_\text{site} = 14.3 \pm 0.2$~kcal/mol.

Lambda-ABF shows faster convergence at short times, reaching an average error of 0.3~kcal/mol after 10~ns of total simulation time when the fixed-$\lambda$ simulations still show an average error of 0.7~kcal/mol.
The dispersion among replicas of both methods is similar, indicating that the source of error is relaxation from the initial condition.

\begin{figure}
    \centering
    \includegraphics[width=0.8\textwidth]{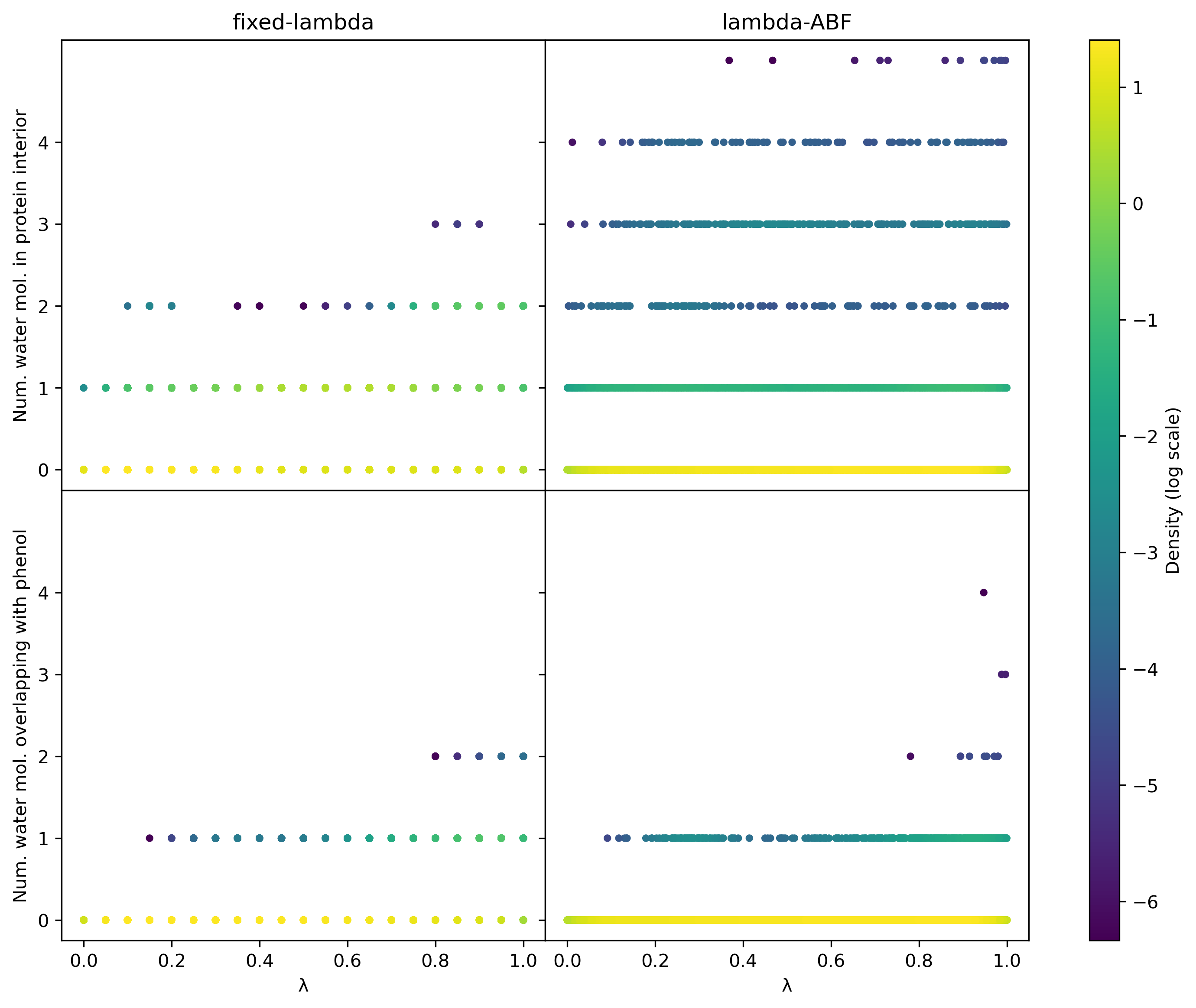}
    \caption{ Hydration of lysozyme interior in phenol decoupling simulations depending on the alchemical sampling scheme.
    Upper panels: hydration of lysozyme interior including the engineered binding site.
    Lower panels: water molecules specifically overlapping with the phenol ligand.
    Left panels show data for fixed-$\lambda$ sampling, whereas right panels contain data from continuous lambda-ABF sampling.
    Point color reflects a kernel estimate of the density, on a logarithmic scale going from purple to yellow.}
    \label{fig:site_water}
\end{figure}

Among the slow relaxing processes is hydration of the predominantly hydrophobic protein interior, and binding site in particular.
In decoupling simulations, the interior of lysozyme is sporadically visited by water molecules. These rare events are expected to contribute to the decoupling free energy.
Occupancy of the site by water molecules is illustrated in Figure~\ref{fig:site_water}, with two metrics: occupancy of the overall protein interior, captured as a sphere around the binding site, and the presence of water molecules overlapping with phenol, which occurs at $\lambda$ values where the soft-core interactions of phenol are sufficiently reduced.
From those metrics, it is clear that for the same, limited simulation time of 200~ns, lambda-ABF enables more extensive sampling of protein hydration.
In contrast, the overlap of a water molecule -- or, for $\lambda > 0.8$, two -- with the partially decoupled ligand occurs with similar probability for both sampling methods. Configurations with 3 or 4 overlapping water molecules are observed in the lambda-ABF simulations, however these are a handful of samples appearing in a single trajectory out of 20 (5 replicas of 4 separate walkers) and therefore are not statistically significant.

The decoupling free energy of phenol in bulk water is found to be $\Delta G^*_\text{bulk} = 4.3 \pm 0.1$~kcal/mol.
This leg of the transformation converges equally rapidly in both the fixed-$\lambda$/BAR and lambda-ABF schemes.
The DBC restraint free energy is estimated using Colvars/TI to be $\Delta G_\text{DBC} = 2.5 \pm 0.1$~kcal/mol, and the combined center-of-mass restraint and standard state correction is found to be $\Delta G_\text{V} = 1.9$~kcal/mol, using the data and analysis methods reported previously.\cite{santiago2023computing}

The predicted standard free energy of binding is therefore $\Delta G^\circ_\text{bind} = -5.6 \pm 0.2$ kcal/mol, which compares favorably with the experimentally reported value of -5.44~kcal/mol.\cite{Merski2013}
This system constitutes an easy case, where fixed-$\lambda$ sampling of the fixed-charge CHARMM force field comes very close to the experimental value, however, lambda-ABF reaches this accurate result faster than fixed-$\lambda$ sampling with BAR free energy estimation.

\subsection{Ligand binding to Cyclophilin D}

Both the host-guest and the lysozyme-phenol complexes already constitute a good illustration of the efficiency of the lambda-ABF method to get absolute free energies of binding. Still, the main domain of interest for these techniques is drug design in which they are applied to real-life protein-ligand complexes. This is why we further test the method on a more realistic (and challenging) system: the cypD protein in complex with a ligand of the literature with a co-crystal structure and experimental affinity values\cite{guichou2016Ncom}. Different cyclophilins are key in biology and we focus ourselves on the cyclophilin-D protein (2472 atoms) in complex with one of the most potent inhibitors of reference \cite{guichou2016Ncom}, ligand 27, which consists of 54 atoms. These numbers make this system a perfect balance between a real-life application in the context of drug design, while still enabling a relatively long sampling time, even for polarizable FFs. Notably, we expect challenges in sampling the orthogonal space formed by the conformations of both the ligand and the protein, especially without being able to identify precisely the associated degrees of freedom beforehand.

All these simulations used the AMOEBA18 FF for cypD and poltype 2\cite{walker2022automation} to get parameters for the ligand.

\medskip

\noindent\textbf{Alchemical decoupling of the ligand in bulk solvent}

\medskip

\begin{figure}[!htb]
    \centering
    \begin{minipage}{.5\textwidth}
        \centering
        \includegraphics[scale=0.5]{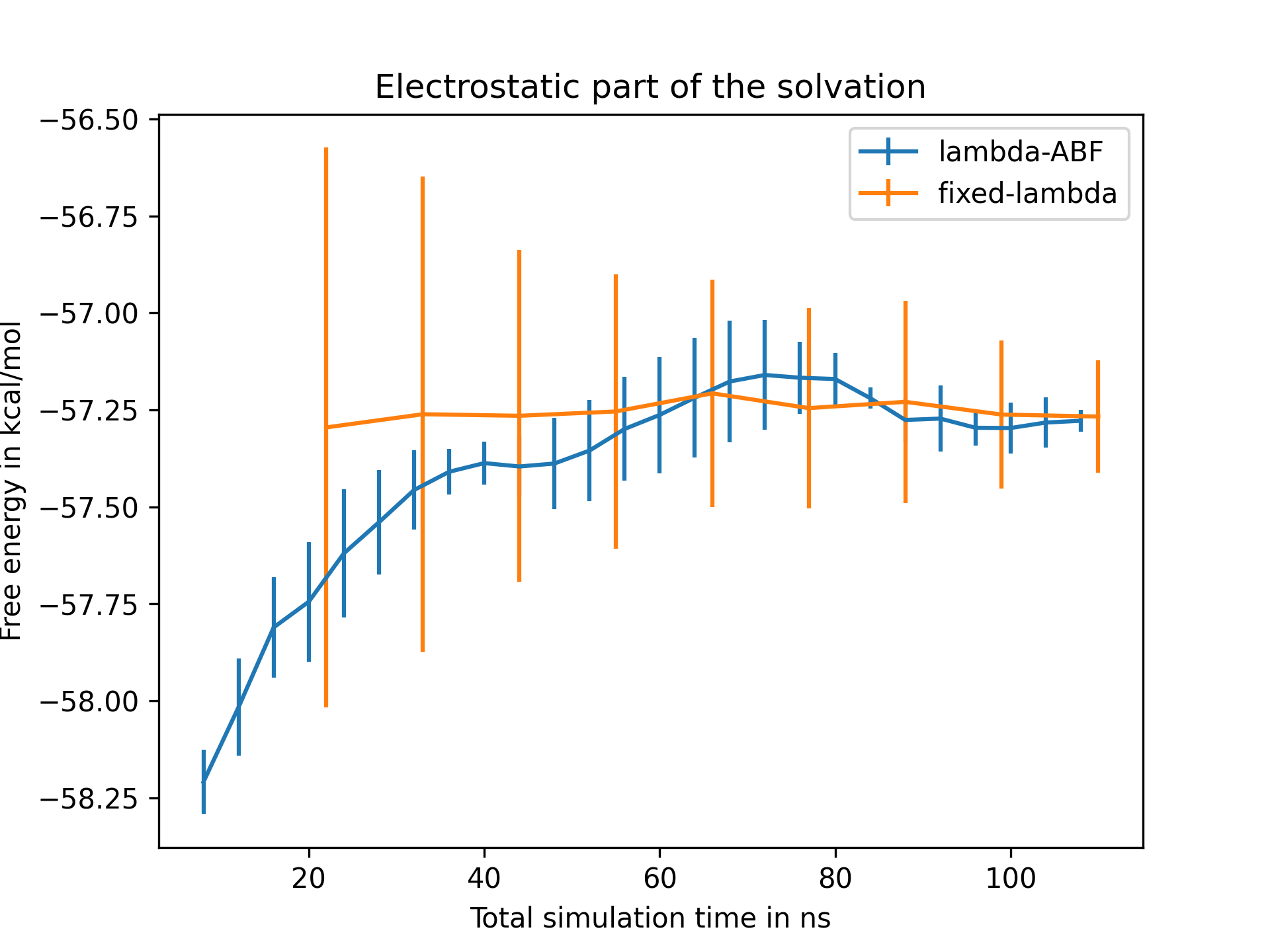}
    \end{minipage}%
    \begin{minipage}{0.5\textwidth}
        \centering
        \includegraphics[scale=0.5]{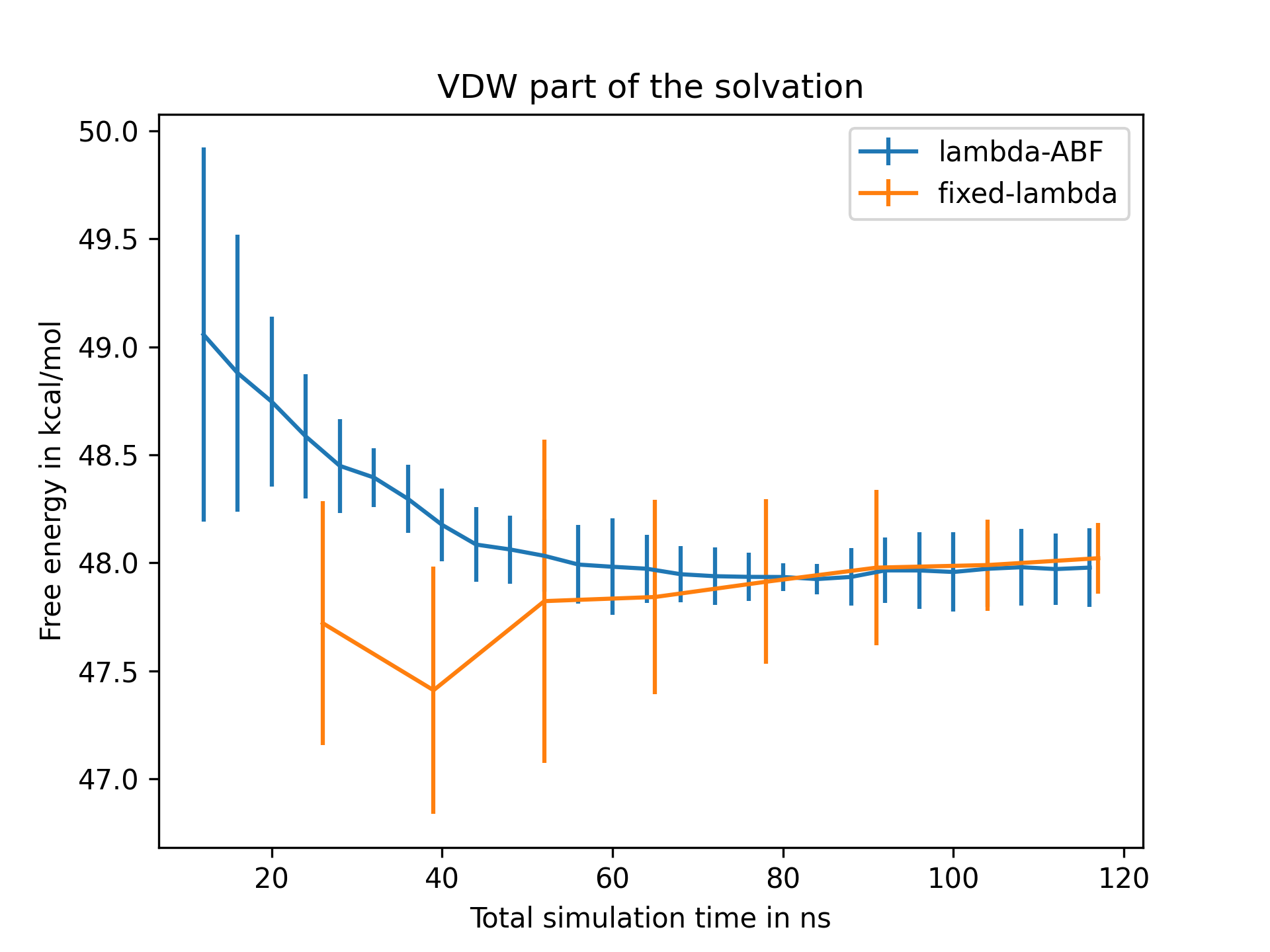}
    \end{minipage}
\caption{Free energy convergence of the electrostatic part (left) and the van der Waals part (right) of the solvation leg of the cypD ligand}
    \label{fig:cyclo-D-solv}
\end{figure}

In addition to the complexation leg of the thermodynamic cycle that relies on the definition of a binding mode, one needs to compute the free energy difference associated with the annihilation of the non-bonded interactions between the ligand and water when the ligand is in bulk solvent.
The convergence graphs of both free energy simulations are shown in Figure \ref{fig:cyclo-D-solv}. We observe a similar rate of convergence when considering the data of the 3 replicas for each method (fixed-$\lambda$ on the one hand and lambda-ABF on the other hand) but with an important and systematic reduction of variance in the case of lambda-ABF. Similar to the situation described in the previous section, this could reflect a similar sampling when one takes into account all 3 repeats of the simulations, but a strong sampling improvement within each replica for lambda-ABF.

\medskip

\noindent\textbf{Simulations predict two metastable binding modes}

\medskip

\begin{figure}[!htb]
    \centering
    \includegraphics[scale=0.45]{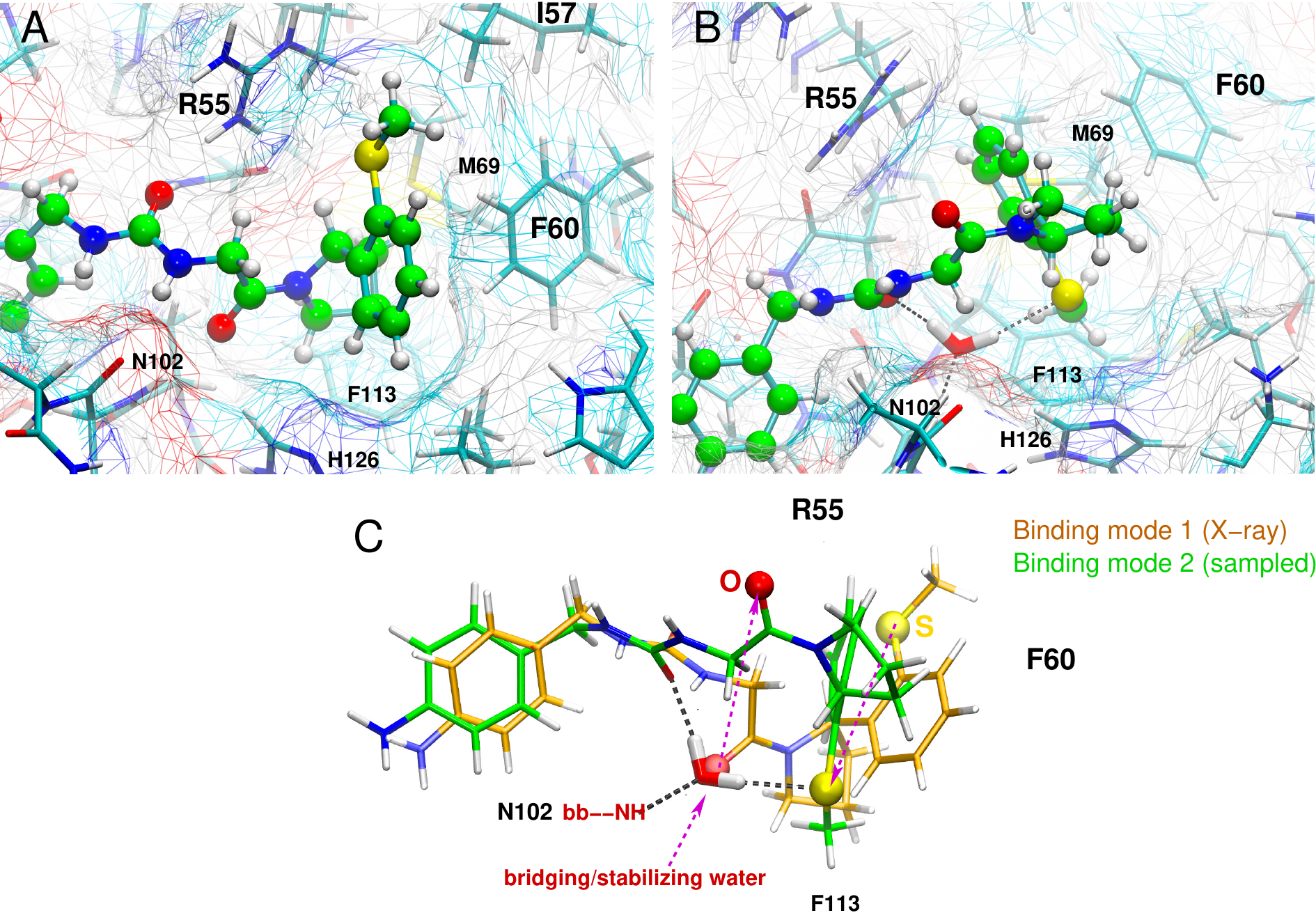}
    \caption{Molecular interactions behind the two binding modes of cypD and its ligand. (A) Binding mode 1 as identified by the co-crystal structure, showing the binding of thiomethyl to R55. (B) The second binding mode as sampled by MD simulation, showing a flip of the thiomethyl group towards F113 and the presence of a bridging water molecule stabilizing the ligand in the new conformation and anchoring it to the NH backbone (bb) of N102. (C) The superposition of the 2 binding modes illustrates how the oxygen atom of the water molecule replaces the carbonyl oxygen of the ligand, and how it coordinates and stabilizes the new binding mode. The adopted binding modes are colored in dark orange and in green for the X-ray and sampled structures, respectively. The black dashed lines depict the 3 H-bonds made by the water molecule. Magenta dashed arrows indicate the displacement of the carbonyl oxygen and thiomethyl sulfur atoms (CPK). }
    \label{fig:interactions-bd}
\end{figure}

Starting from the crystal structure of reference \cite{guichou2016Ncom} (PDB ID 4J5B, X-ray diffraction at 2.01\AA\/), we first resorted to a standard heating and equilibration procedure by first restraining the protein and the complex to their crystallographic positions, then progressively releasing the positions of the lateral chains of the protein while keeping restraints on the positions of the backbone of the protein and on the ligand and finally by releasing all restraints. Interestingly, after 1~ns of plain MD simulation, the ligand adopted a slightly different but stable binding mode where the methylthio-benzene is flipped by 180$^\circ$ as shown in Figure \ref{fig:interactions-bd}.
Hence, some direct interactions shown in the co-crystal structure, such as the hydrogen bond between the carbonyl oxygen of ligand 27 and the backbone (bb) NH of N102, were lost (Figure \ref{fig:interactions-bd}A and B). However, a stabilizing water molecule enters the binding site after the ring flip and its oxygen atom occupies the initial position of the carbonyl oxygen of the ligand, anchoring the ligand again to the backbone NH of N102 with an H-bond and stabilizing the ligand in its new binding mode by engaging in 2 additional H-bonds with the adjacent carbonyl and the sulfur atom of the thiomethyl group (Figure \ref{fig:interactions-bd}B). Figure \ref{fig:interactions-bd}C shows a superposition of the 2 binding modes (X-ray vs sampled) and depicts the ring rotation and the water molecule overlapping the initial position of the carbonyl oxygen.

Apart from the reported co-crystal structure, the binding mode to CypD was investigated and confirmed by means of NMR experiments\cite{guichou2016Ncom}. The CypD $^{15}$N-heteronuclear single quantum coherence (HSQC) spectrum revealed significant chemical shift perturbations on ligand binding only for residues located at or near the catalytic site and gatekeeper pocket, respectively (Supplementary Fig. 4 of the referenced paper \cite{guichou2016Ncom}). The NMR experiments also show that the thiomethyl group is involved in transient contact with R55 N$\epsilon$. This is compatible with either binding mode; in the first one, where the sulfur is oriented towards the guanidinium ring of R55 (X-ray structure, Figure \ref{fig:interactions-bd}A) and after the ring flip (MD sampled structure) the benzyl ring engages the guanidinium ring of R55 T-shaped cation-$\pi$ interaction (Figure \ref{fig:interactions-bd}B).
This identification of an alternate binding mode emphasizes the power of the AMOEBA FF to describe with high accuracy complex interactions, accounting for anisotropy and many-body polarization, especially for water molecules, whose dipole moments change with the environment.

To equilibrate the system in a stable conformation compatible with the crystal structure we used harmonic restraint on a DBC variable which involves all the heavy atoms of the protein (even the lateral chains) that are within a radius of 6\AA~of the ligand, that we released progressively. By doing so, we implicitly force the system to keep all the relative positions and orientations of the ligand with respect to the protein, without having to manually make an inventory of all the protein-ligand interactions keys to the definition of the binding mode. We think that this particular strength of the DBC collective variables in conjunction with the possibility to act on it with a harmonic restraint is of great interest for the free energy practitioners that encounter this situation (relaxing a system in a predefined binding pose).

To extensively benchmark the method we chose to run two separate binding free energy simulations, one for each binding mode, each associated with a compatible DBC variable. Note that we could have run a single simulation with a wider bound state definition (through the DBC variable) encompassing both binding modes.

\medskip

\medskip

\noindent\textbf{Crystallographic binding mode}

\medskip

\begin{figure}[!htb]
    \centering
    \begin{minipage}{.5\textwidth}
        \centering
        \includegraphics[scale=0.5]{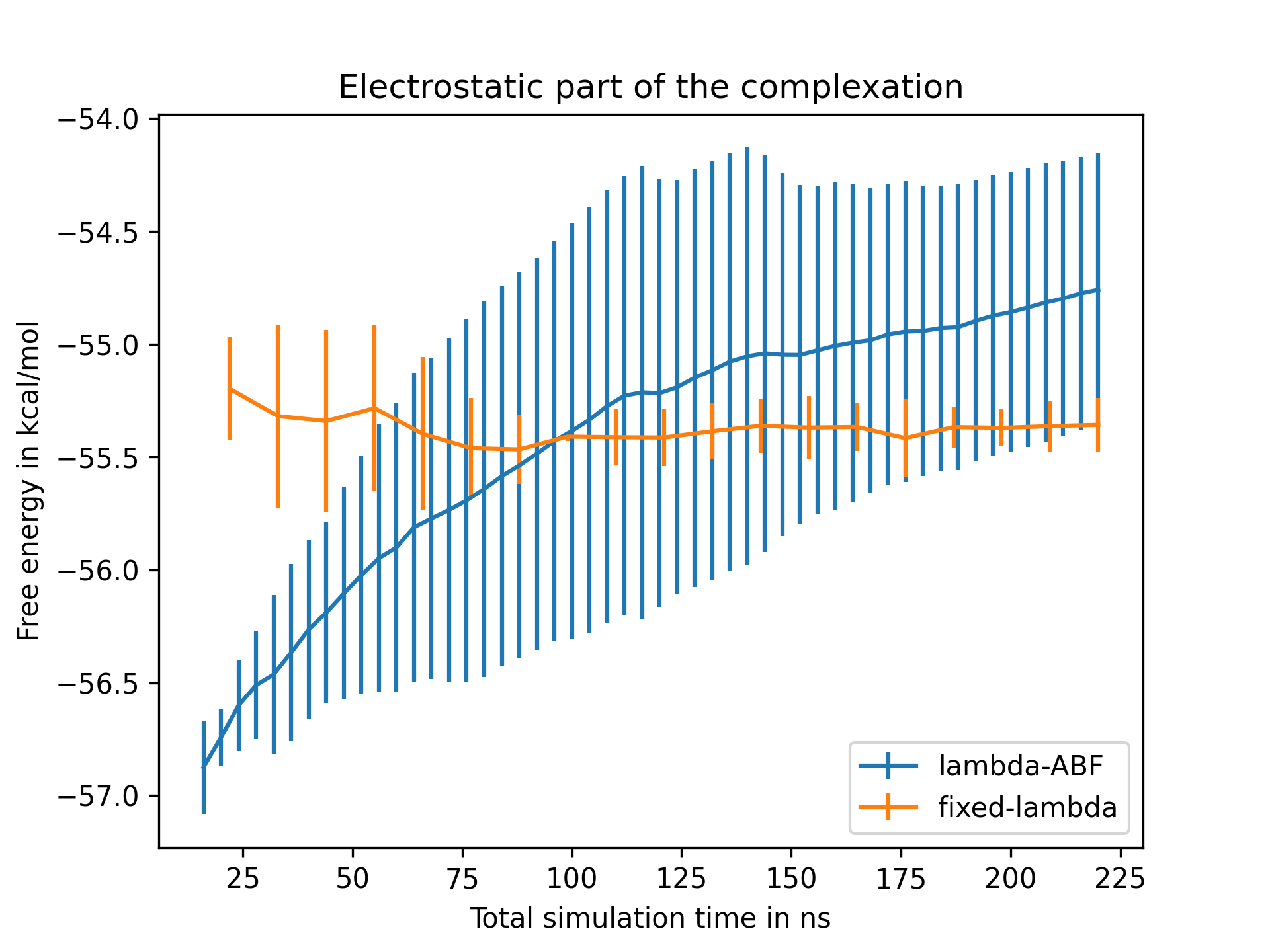}
    \end{minipage}%
    \begin{minipage}{0.5\textwidth}
        \centering
        \includegraphics[scale=0.5]{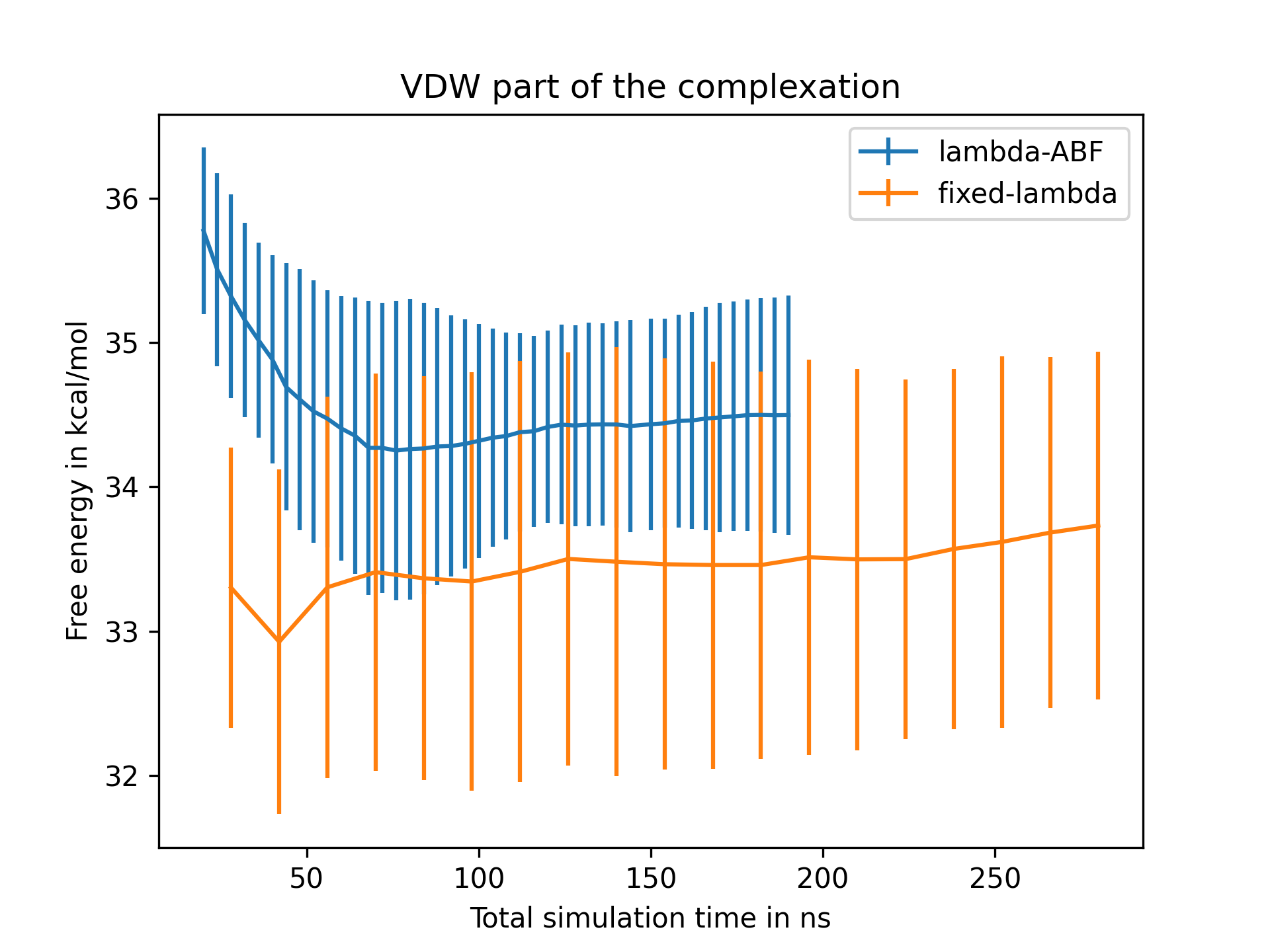}
    \end{minipage}
\caption{Free energy convergence of the electrostatic part (left) and the van der Waals part (right) of the complexation leg of the cypD ligand, first binding mode. The apparently rapid convergence of fixed-lambda sampling is only local (See Figure \ref{fig:lambda-rmsdbs-1}).}
    \label{fig:cyclo-D-complex-1}
\end{figure}

\begin{figure}[!htb]
    \centering
    \begin{minipage}{.5\textwidth}
        \centering
        \includegraphics[scale=0.5]{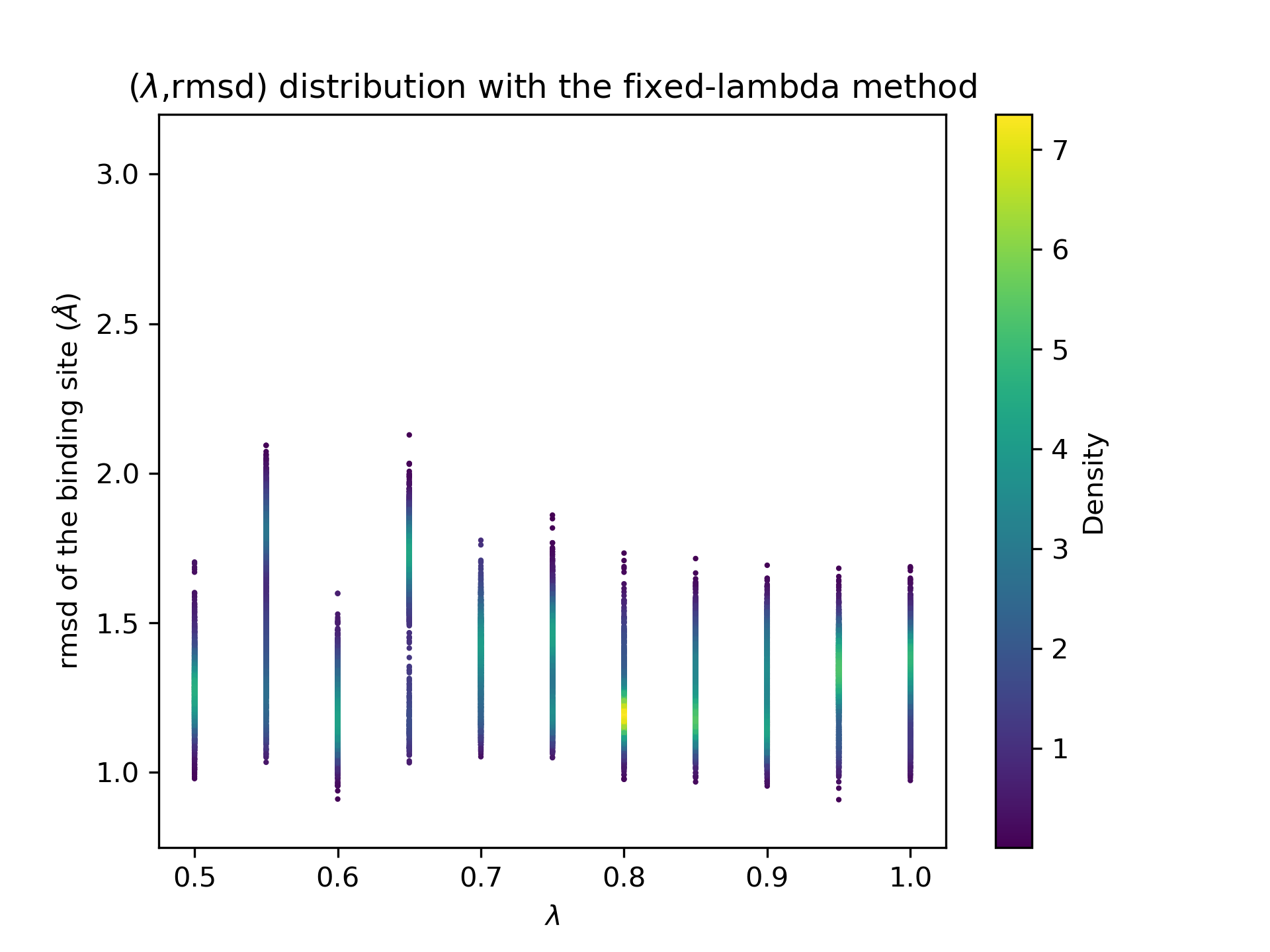}
    \end{minipage}%
    \begin{minipage}{0.5\textwidth}
        \centering
        \includegraphics[scale=0.5]{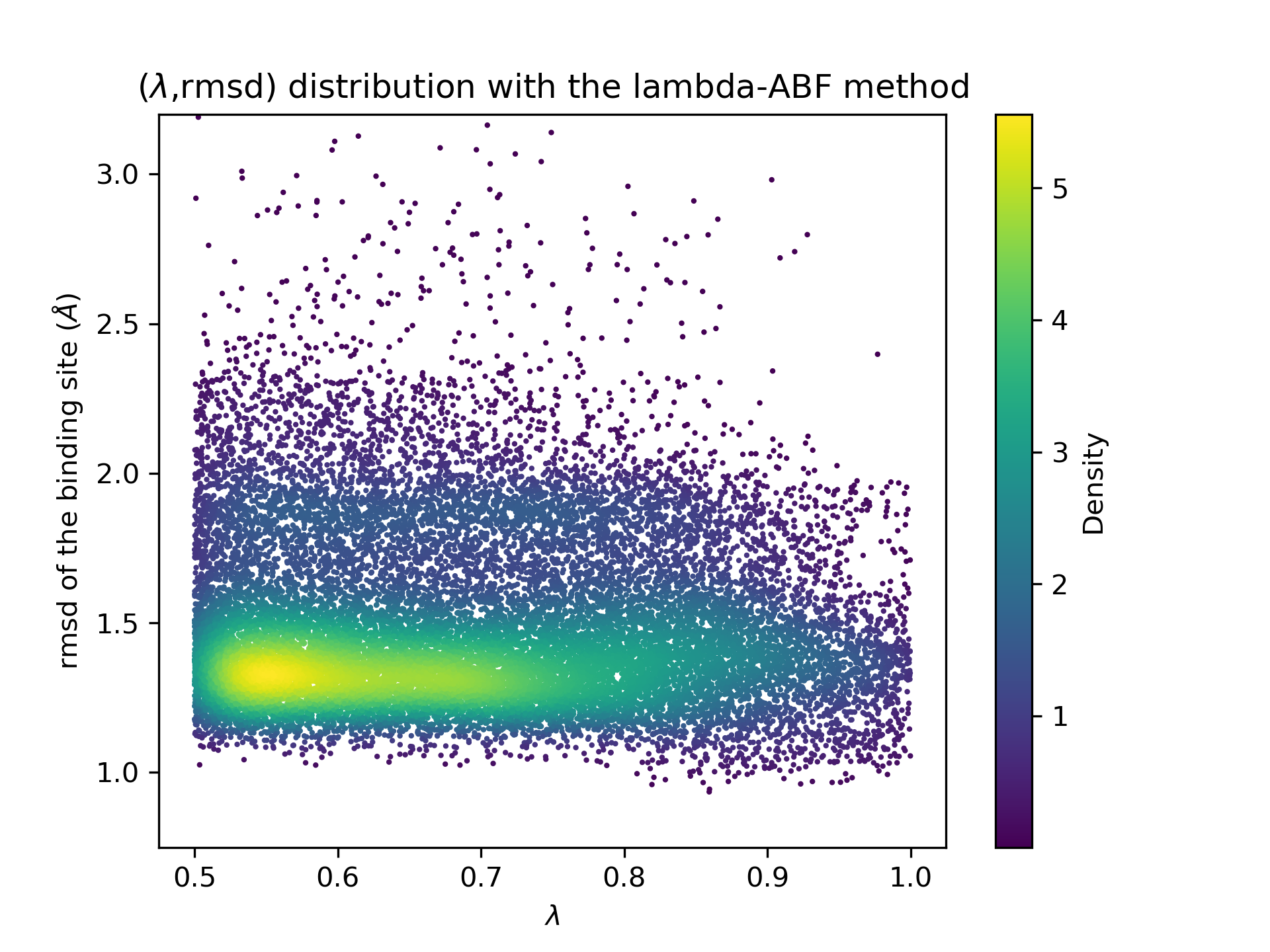}
    \end{minipage}
\\
    \begin{minipage}{.5\textwidth}
        \centering
        \includegraphics[scale=0.5]{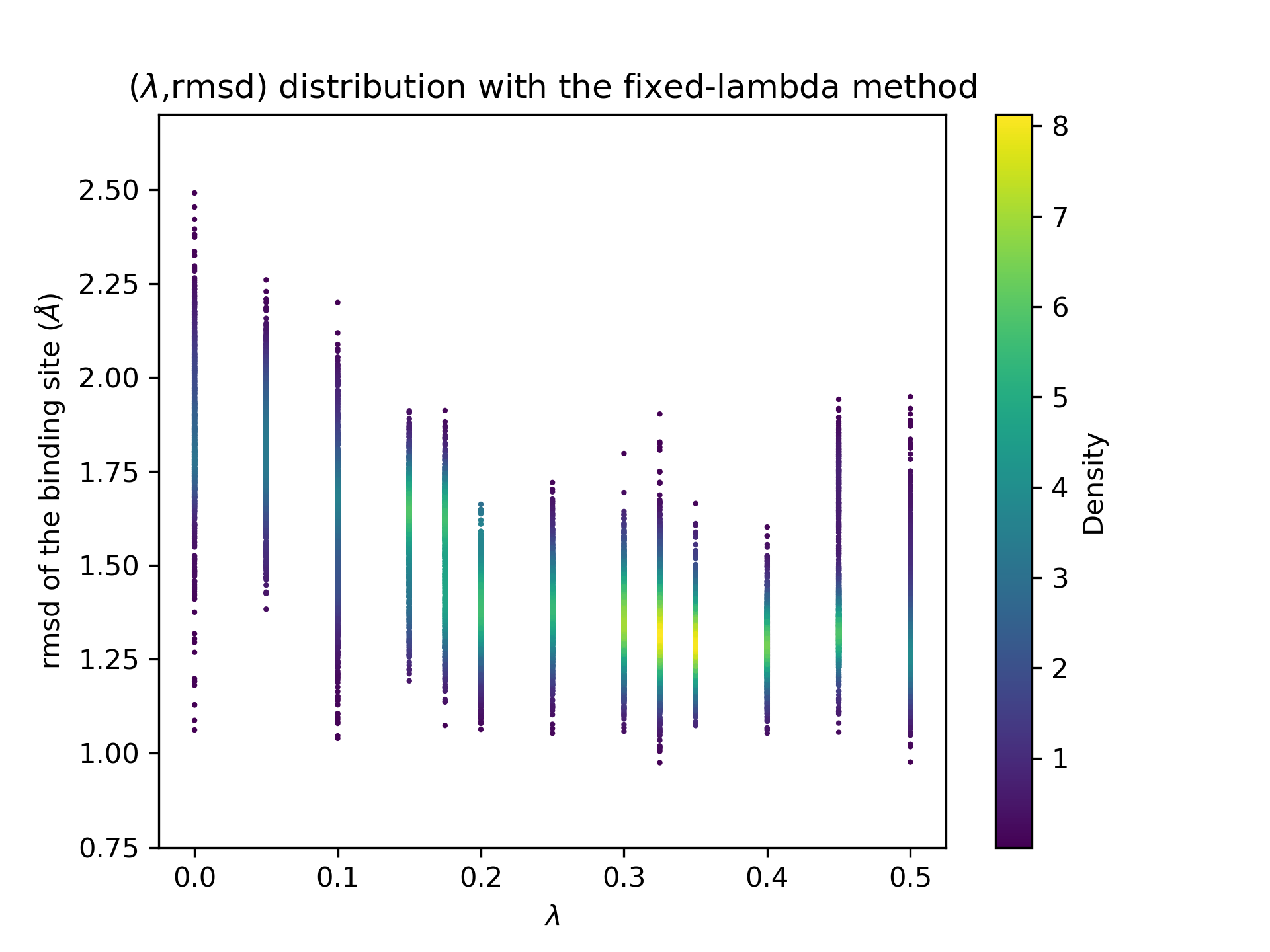}
    \end{minipage}%
    \begin{minipage}{0.5\textwidth}
        \centering
        \includegraphics[scale=0.5]{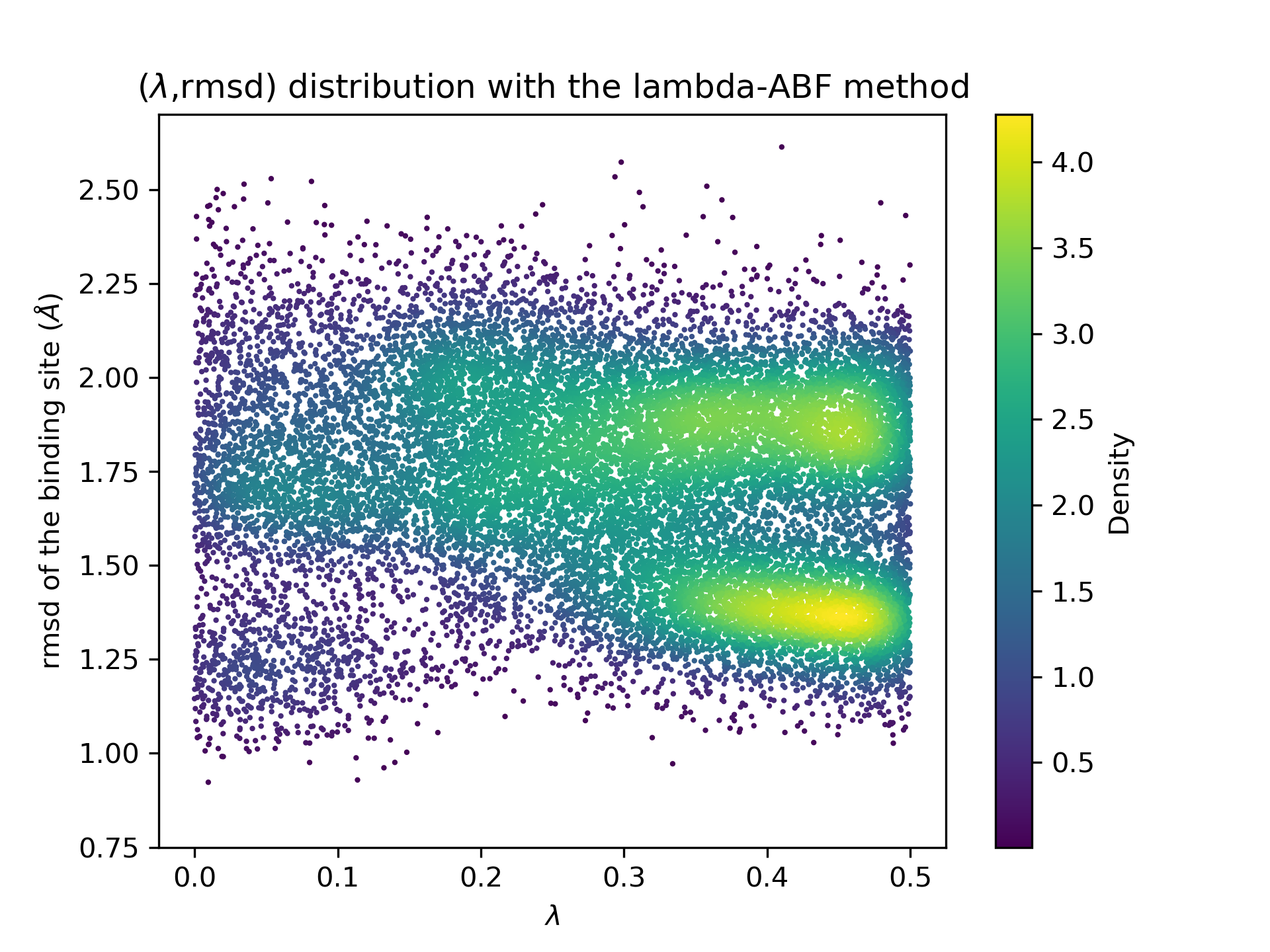}
    \end{minipage}
    \caption{Distribution of ($\lambda$, rmsd of the cyclophilin binding site) sampled with the fixed-$\lambda$ and lambda-ABF methods, for the electrostatic part (up) and the van der Waals part (down), first binding mode.}
    \label{fig:lambda-rmsdbs-1}
\end{figure}

We start by assessing the convergence of the complexation leg of the thermodynamic cycle for the first binding mode. Final values can be found in Table \ref{tab:ABFEcypD1} with their various components shown in Supporting Information. These consolidated results are within statistical error of one another but result from different convergence behaviors, as shown in Figure \ref{fig:cyclo-D-complex-1}. Interestingly, the fixed-$\lambda$ simulations seem to lead to faster convergence for the electrostatic part but it is an indirect consequence of only a partial sampling of the accessible phase space as is illustrated in Figure \ref{fig:lambda-rmsdbs-1} that shows the joint distribution (sampled with the methods) of $\lambda$  and the root mean square deviation (RMSD) of the binding site. This illustrates the biased nature of the free energy estimated by the fixed-$\lambda$ method in this case.

Regarding the convergence of the van der Waals part, lambda-ABF is associated with a smaller variance and a difference of almost 1~kcal/mol in terms of final value. The same reasoning regarding the amount of sampling as for the electrostatic part can be made as shown by Figure \ref{fig:lambda-rmsdbs-1}.

\begin{table}[h!]
 \center
 \begin{tabular}{|c|c|}
 \hline
  & ligand 27-cypD complex \\
 \hline
  lambda-ABF& -5.02 ($\pm 1.03$)  \\
  \hline
 fixed-$\lambda$& -6.32 ($\pm 1.21$)  \\
 \hline
  Exp.& -9.06\\
 \hline
 \end{tabular}
 \caption{ Standard free energies of binding (in kcal/mol) for the ligand 27-cypD complex, obtained from lambda-ABF and the fixed-$\lambda$ method, first binding mode.}
 \label{tab:ABFEcypD1}
\end{table}

\medskip
\medskip
\medskip

\noindent\textbf{Simulation-predicted binding mode}

\begin{figure}[!htb]
    \centering
    \begin{minipage}{.5\textwidth}
        \centering
        \includegraphics[scale=0.5]{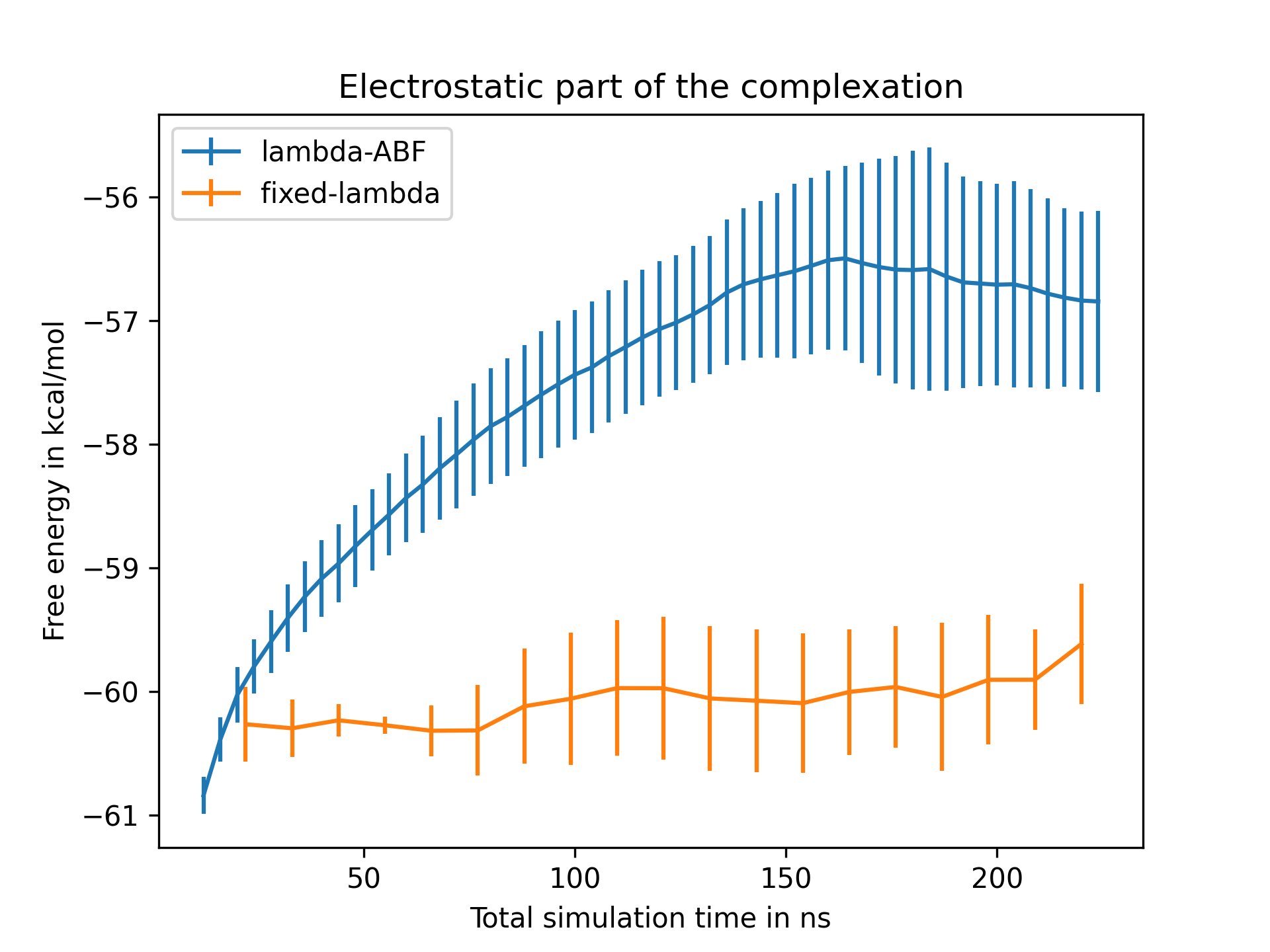}
    \end{minipage}%
    \begin{minipage}{0.5\textwidth}
        \centering
        \includegraphics[scale=0.5]{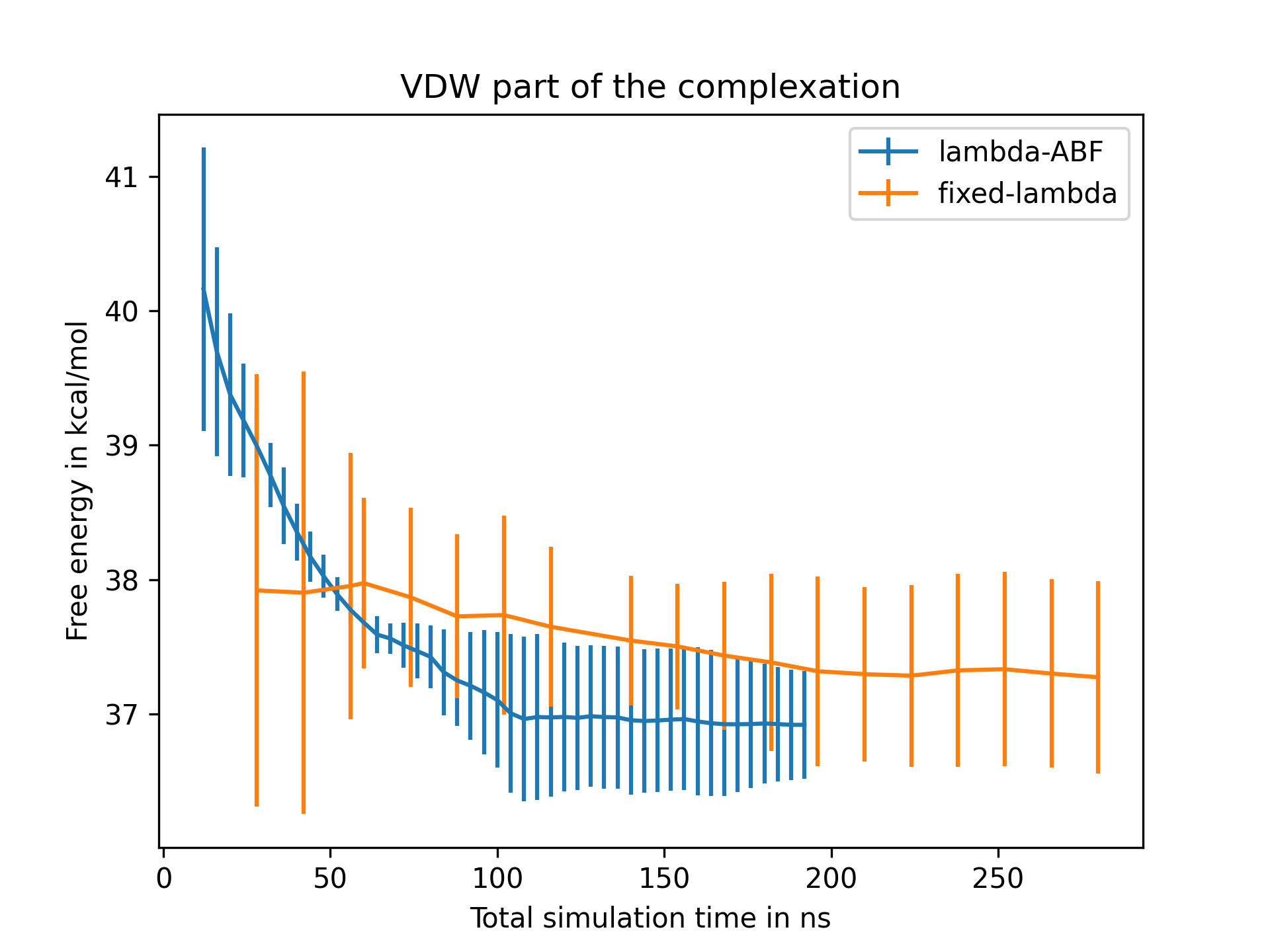}
    \end{minipage}
\caption{Free energy convergence of the electrostatic part (left) and the van der Waals part (right) of the complexation leg of the cypD ligand, simulation-predicted binding mode.}
    \label{fig:cyclo-D-complex-2}
\end{figure}

\begin{figure}[!htb]
    \centering
    \begin{minipage}{.5\textwidth}
        \centering
        \includegraphics[scale=0.5]{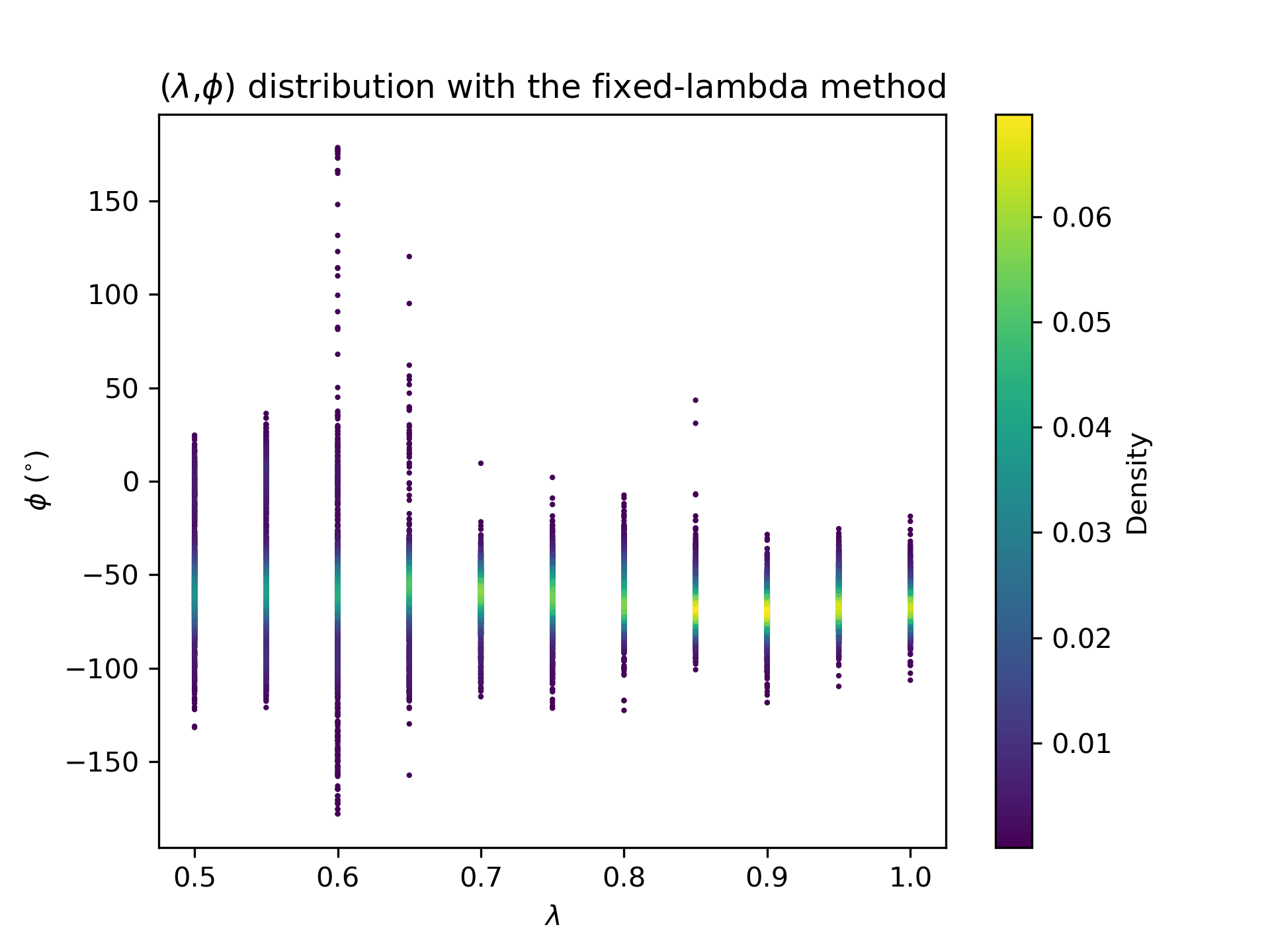}
    \end{minipage}%
    \begin{minipage}{0.5\textwidth}
        \centering
        \includegraphics[scale=0.5]{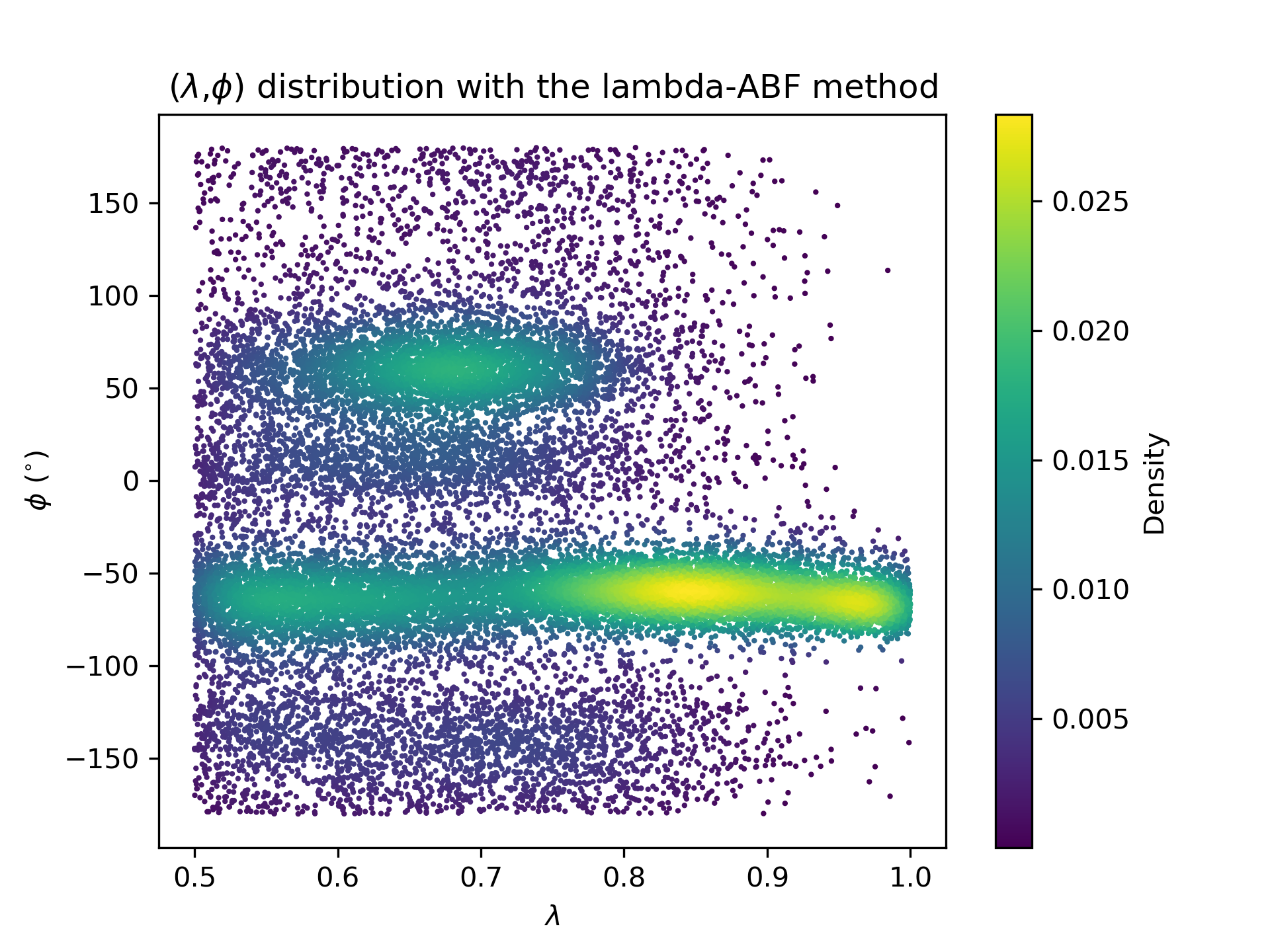}
    \end{minipage}
    \caption{($\lambda$,$\phi$)  distribution with the fixed-$\lambda$ and the lambda-ABF (electrostatic leg of the complexation, alternate binding mode).}
    \label{fig:lambda-phi-2}
\end{figure}

\medskip
Results for the second binding mode can be found in Table \ref{tab:ABFEcypD2} with their various components shown in Supporting Information. For the electrostatic part, similar to the first binding mode, fixed-$\lambda$ simulations could appear to converge faster than with the other methods at first sight (Figure \ref{fig:cyclo-D-complex-2}). But one can see an increasing trend in convergence appearing after 200ns of total sampling. Interestingly, the lambda-ABF results seem converged after 150~ns of sampling to a value that differs by more than 3 kcal/mol. As for the other situations described above, these differences can be linked to a richer sampling power of the lambda-ABF methods as is shown by Figure \ref{fig:lambda-phi-2} that shows the joint distribution of $\lambda$ and a significant dihedral of the ligand
(whose involved atoms are described in Supporting Information)
, $\phi$. This observation is reinforced by Figure S7 showing the distribution of $\phi$ sampled with both methods.

For this stronger binding mode, the van der Waals part of the thermodynamic cycle is associated with similar converged values, with smaller variance for lambda-ABF as shown in Figure \ref{fig:cyclo-D-complex-2}.

 \begin{table}[h!]
 \center
 \begin{tabular}{|c|c|}
 \hline
  & ligand 27-cypD complex \\
 \hline
  lambda-ABF& -8.13 ($\pm 0.83$)  \\
  \hline
 fixed-$\lambda$& -10.77 ($\pm 0.83$)     \\
 \hline
  Exp.& -9.06\\
 \hline
 \end{tabular}
 \caption{ Standard free energies of binding (in kcal/mol) for the ligand 27-cypD complex, obtained from lambda-ABF and the fixed-$\lambda$ method, simulation-predicted binding mode.}
 \label{tab:ABFEcypD2}
\end{table}

The contribution of the two binding modes can be combined to get a global absolute free energy of binding that takes both of them into account by using the formula:
\begin{equation}
    \Delta G_\text{tot}=-\beta^{-1}\ln(e^{-\beta \Delta G_1} + e^{-\beta \Delta G_2})
    \label{bindcumul}
\end{equation}
which leads to the final estimates that are indistinguishable from the ones of the second binding modes given the free energy difference compared to the first one.

Finally, the absolute free energy of binding obtained with lambda-ABF is within 1 kcal/mol of the experimental value, whereas that obtained with fixed-$\lambda$ exhibits an error of almost 2 kcal/mol, illustrating the predictive power of the combination of the polarizable AMOEBA FF and the lambda-ABF sampling approach for this system.

\section{Conclusion}\label{sec:conclusion}

We introduce a combined sampling and free energy estimation approach for alchemical transformations, lambda-ABF, and its implementation in the Colvars open source library~\cite{Fiorin2013}.
This alchemical feature is made available in the novel interface of Colvars with Tinker-HP and as an extension to the existing NAMD/Colvars interface.
This constitutes the first portable implementation --- and unified user interface --- for alchemical simulations.
This will be ported to other molecular dynamics packages in the future.

By design, the lambda-ABF method is conceptually and practically simple and requires fewer user parameters than standard fixed-$\lambda$ approaches. Its implementation is readily usable by practitioners of molecular simulations.
The method relies on state-of-the-art $\lambda$-dynamics combined with multiple-walker Adaptive Biasing Force on the alchemical variable, and Thermodynamic Integration to recover free energy differences.

Thanks to Colvars' modular design and its extensibility through scripts, users may readily tune the alchemical pathway to avoid intermediate metastable basins (such as transient ion pairs), without biasing the target free energy differences. We provide one such example that proved critical for convergence and can be adapted to other similar situations.

For binding applications, a distance-to-bound-configuration (DBC) restraint is used within the alchemical step, as proposed previously in the SAFEP framework ~\cite{salari2018, santiago2023computing}.
It is easy to set up and has the benefit of focusing sampling onto a narrower, relevant region of configuration space.
In addition, we introduce the use of DBC restraints for equilibrating receptor-ligand complexes while preserving a known binding pose.
The Colvars Dashboard extension of VMD~\cite{henin2022human} can be used to easily set up the DBC variable and to analyze thoroughly the exploration of relevant degrees of freedom through the alchemical simulations. To that effect,  we provide a script for loading the $\lambda$ trajectory within VMD to inspect it visually and correlate it with other observables.\cite{Henin2023load_cv_traj}

The sampling power of the method is demonstrated in real-life applications, from hydration free energies of simple molecules, up to absolute protein-ligand binding free energies.
We find that in addition to leveling of the free energy surface by ABF in the $\lambda$ direction, this scheme results in improved orthogonal relaxation thanks to both the dynamical character of $\lambda$ and orthogonal space coverage by multiple walkers.

Comparing lambda-ABF with fixed-$\lambda$ sampling, we observe two regimes: either similar converged estimates with a strongly reduced error or variance, or significantly different predictions associated with more extensive sampling by lambda-ABF.
Because it is generic and its implementation is portable, the method has a broad spectrum of potential applications, from physical chemistry to drug design.

\section*{Supporting Information}

Additional figures S1-S3 showing detailed numerical results; table with results of parameter sensitivity analysis; scripts with simulation parameters.

\section*{Data availability}

Simulation scripts used for this study are provided as Supplementary information.
Raw data generated during the study is available upon request from the authors\\
(E-mail:
louis.lagardere@sorbonne-universite.fr).

\section*{Code availability}

The code used during the study is available on GitHub:

\url{https://github.com/TinkerTools/tinker-hp}, \\
\url{https://github.com/Colvars/colvars}

\section*{Competing interests}
L.L. and J-P.P. are co-founders and shareholders of Qubit Pharmaceuticals. All other authors declare no competing interests.

\section*{Acknowledgements}

This work has been funded in part by the European Research Council (ERC) under the European Union’s Horizon 2020 research and innovation program (grant No 810367), project EMC2 (J-P.P.).
Computations have been performed at GENCI
(IDRIS, Orsay, France) on grant no A0130712052.

\section*{Author contributions statement}

L.L., L.M, O.A, J.H developed the code;
L.L, L.M., K.E.H., J.H. performed research;
L. L., L.M, P.M, JP.-P, J.H. contributed new theoretical tools;
all authors analyzed the results;
L. L., P.M, JP.-P.,K.E.H, J.H. wrote the paper with input from all authors;
L.L., P.M, J-P.P. and J.H. conceived and supervised the project.

\bibliographystyle{unsrt}
\bibliography{main}

\begin{thebibliography}{100}

\bibitem{jorgensen1985monte}
William~L Jorgensen and C~Ravimohan.
\newblock Monte carlo simulation of differences in free energies of hydration.
\newblock {\em The Journal of Chemical Physics}, 83(6):3050--3054, 1985.

\bibitem{chipot2007free}
Christophe Chipot and Andrew Pohorille.
\newblock {\em Free energy calculations}, volume~86.
\newblock Springer, 2007.

\bibitem{wang2015accurate}
Lingle Wang, Yujie Wu, Yuqing Deng, Byungchan Kim, Levi Pierce, Goran Krilov,
  Dmitry Lupyan, Shaughnessy Robinson, Markus~K Dahlgren, Jeremy Greenwood,
  et~al.
\newblock Accurate and reliable prediction of relative ligand binding potency
  in prospective drug discovery by way of a modern free-energy calculation
  protocol and force field.
\newblock {\em Journal of the American Chemical Society}, 137(7):2695--2703,
  2015.

\bibitem{woo2005calculation}
Hyung-June Woo and Beno{\^\i}t Roux.
\newblock Calculation of absolute protein--ligand binding free energy from
  computer simulations.
\newblock {\em Proceedings of the National Academy of Sciences},
  102(19):6825--6830, 2005.

\bibitem{cournia2017relative}
Zoe Cournia, Bryce Allen, and Woody Sherman.
\newblock Relative binding free energy calculations in drug discovery: recent
  advances and practical considerations.
\newblock {\em Journal of Chemical Information and Modeling},
  57(12):2911--2937, 2017.

\bibitem{kong1996lambda}
Xianjun Kong and Charles~L Brooks~III.
\newblock $\lambda$-dynamics: A new approach to free energy calculations.
\newblock {\em The Journal of Chemical Physics}, 105(6):2414--2423, 1996.

\bibitem{zheng2008random}
Lianqing Zheng, Mengen Chen, and Wei Yang.
\newblock Random walk in orthogonal space to achieve efficient free-energy
  simulation of complex systems.
\newblock {\em Proceedings of the National Academy of Sciences},
  105(51):20227--20232, 2008.

\bibitem{gapsys2021accurate}
Vytautas Gapsys, Ahmet Yildirim, Matteo Aldeghi, Yuriy Khalak, David Van~der
  Spoel, and Bert~L de~Groot.
\newblock Accurate absolute free energies for ligand--protein binding based on
  non-equilibrium approaches.
\newblock {\em Communications Chemistry}, 4(1):61, 2021.

\bibitem{cruz2020combining}
Jeffrey Cruz, Lauren Wickstrom, Danzhou Yang, Emilio Gallicchio, and Nanjie
  Deng.
\newblock Combining alchemical transformation with a physical pathway to
  accelerate absolute binding free energy calculations of charged ligands to
  enclosed binding sites.
\newblock {\em Journal of Chemical Theory and Computation}, 16(4):2803--2813,
  2020.

\bibitem{zheng2012practically}
Lianqing Zheng and Wei Yang.
\newblock Practically efficient and robust free energy calculations:
  double-integration orthogonal space tempering.
\newblock {\em Journal of Chemical Theory and Computation}, 8(3):810--823,
  2012.

\bibitem{knight2011multisite}
Jennifer~L Knight and Charles~L Brooks~III.
\newblock Multisite $\lambda$ dynamics for simulated structure--activity
  relationship studies.
\newblock {\em Journal of Chemical Theory and Computation}, 7(9):2728--2739,
  2011.

\bibitem{hayes2017adaptive}
Ryan~L Hayes, Kira~A Armacost, Jonah~Z Vilseck, and Charles~L Brooks~III.
\newblock Adaptive landscape flattening accelerates sampling of alchemical
  space in multisite $\lambda$ dynamics.
\newblock {\em The Journal of Physical Chemistry B}, 121(15):3626--3635, 2017.

\bibitem{chodera2011alchemical}
John~D Chodera, David~L Mobley, Michael~R Shirts, Richard~W Dixon, Kim Branson,
  and Vijay~S Pande.
\newblock Alchemical free energy methods for drug discovery: progress and
  challenges.
\newblock {\em Current Opinion in Structural Biology}, 21(2):150--160, 2011.

\bibitem{song2020evolution}
Lin~Frank Song and Kenneth~M Merz~Jr.
\newblock Evolution of alchemical free energy methods in drug discovery.
\newblock {\em Journal of Chemical Information and Modeling},
  60(11):5308--5318, 2020.

\bibitem{zwanzig1954high}
Robert~W Zwanzig.
\newblock High-temperature equation of state by a perturbation method. i.
  nonpolar gases.
\newblock {\em The Journal of Chemical Physics}, 22(8):1420--1426, 1954.

\bibitem{bennett1976efficient}
Charles~H Bennett.
\newblock Efficient estimation of free energy differences from monte carlo
  data.
\newblock {\em Journal of Computational Physics}, 22(2):245--268, 1976.

\bibitem{shirts2008statistically}
Michael~R Shirts and John~D Chodera.
\newblock Statistically optimal analysis of samples from multiple equilibrium
  states.
\newblock {\em The Journal of Chemical Physics}, 129(12):124105, 2008.

\bibitem{straatsma1991multiconfiguration}
TP~Straatsma and JA~McCammon.
\newblock Multiconfiguration thermodynamic integration.
\newblock {\em The Journal of Chemical Physics}, 95(2):1175--1188, 1991.

\bibitem{cuendet2014free}
Michel~A Cuendet and Mark~E Tuckerman.
\newblock Free energy reconstruction from metadynamics or adiabatic free energy
  dynamics simulations.
\newblock {\em Journal of Chemical Theory and Computation}, 10(8):2975--2986,
  2014.

\bibitem{gapsys2012new}
Vytautas Gapsys, Daniel Seeliger, and Bert~L de~Groot.
\newblock New soft-core potential function for molecular dynamics based
  alchemical free energy calculations.
\newblock {\em Journal of Chemical Theory and Computation}, 8(7):2373--2382,
  2012.

\bibitem{procacci2014fast}
Piero Procacci and Chiara Cardelli.
\newblock Fast switching alchemical transformations in molecular dynamics
  simulations.
\newblock {\em Journal of Chemical Theory and Computation}, 10(7):2813--2823,
  2014.

\bibitem{jarzynski1997nonequilibrium}
Christopher Jarzynski.
\newblock Nonequilibrium equality for free energy differences.
\newblock {\em Physical Review Letters}, 78(14):2690, 1997.

\bibitem{crooks1999entropy}
Gavin~E Crooks.
\newblock Entropy production fluctuation theorem and the nonequilibrium work
  relation for free energy differences.
\newblock {\em Physical Review E}, 60(3):2721, 1999.

\bibitem{park2004calculating}
Sanghyun Park and Klaus Schulten.
\newblock Calculating potentials of mean force from steered molecular dynamics
  simulations.
\newblock {\em The Journal of Chemical Physics}, 120(13):5946--5961, 2004.

\bibitem{deRuiter2013}
Anita de~Ruiter, Stefan Boresch, and Chris Oostenbrink.
\newblock Comparison of thermodynamic integration and bennett acceptance ratio
  for calculating relative protein‐ligand binding free energies.
\newblock {\em Journal of Computational Chemistry}, 34(12):1024–1034, January
  2013.

\bibitem{Zeng2023}
Jin Zeng and Yue Qian.
\newblock Adaptive lambda schemes for efficient relative binding free energy
  calculation.
\newblock {\em Journal of Computational Chemistry}, December 2023.

\bibitem{fukunishi2002hamiltonian}
Hiroaki Fukunishi, Osamu Watanabe, and Shoji Takada.
\newblock On the hamiltonian replica exchange method for efficient sampling of
  biomolecular systems: Application to protein structure prediction.
\newblock {\em The Journal of Chemical Physics}, 116(20):9058--9067, 2002.

\bibitem{jang2003replica}
Soonmin Jang, Seokmin Shin, and Youngshang Pak.
\newblock Replica-exchange method using the generalized effective potential.
\newblock {\em Physical Review Letters}, 91(5):058305, 2003.

\bibitem{jiang2010free}
Wei Jiang and Beno{\^\i}t Roux.
\newblock Free energy perturbation hamiltonian replica-exchange molecular
  dynamics (fep/h-remd) for absolute ligand binding free energy calculations.
\newblock {\em Journal of Chemical Theory and Computation}, 6(9):2559--2565,
  2010.

\bibitem{rizzi2018overview}
Andrea Rizzi, Steven Murkli, John~N McNeill, Wei Yao, Matthew Sullivan,
  Michael~K Gilson, Michael~W Chiu, Lyle Isaacs, Bruce~C Gibb, David~L Mobley,
  et~al.
\newblock Overview of the sampl6 host--guest binding affinity prediction
  challenge.
\newblock {\em Journal of Computer-aided Molecular Design}, 32(10):937--963,
  2018.

\bibitem{tidor1993simulated}
Bruce Tidor.
\newblock Simulated annealing on free energy surfaces by a combined molecular
  dynamics and monte carlo approach.
\newblock {\em The Journal of Physical Chemistry}, 97(5):1069--1073, 1993.

\bibitem{hahn2020overcoming}
David~F Hahn, Gerhard König, and Philippe~H Hünenberger.
\newblock Overcoming orthogonal barriers in alchemical free energy
  calculations: On the relative merits of $\lambda$-variations,
  $\lambda$-extrapolations, and biasing.
\newblock {\em Journal of Chemical Theory and Computation}, 16(3):1630--1645,
  2020.

\bibitem{wu2011lambda}
Pan Wu, Xiangqian Hu, and Weitao Yang.
\newblock $\lambda$-metadynamics approach to compute absolute solvation free
  energy.
\newblock {\em The journal of Physical Chemistry Letters}, 2(17):2099--2103,
  2011.

\bibitem{abrams2006efficient}
Jerry~B Abrams, Lula Rosso, and Mark~E Tuckerman.
\newblock Efficient and precise solvation free energies via alchemical
  adiabatic molecular dynamics.
\newblock {\em The Journal of Chemical Physics}, 125(7):074115, 2006.

\bibitem{bieler2014local}
Noah~S Bieler, Rico Häuselmann, and Philippe~H Hünenberger.
\newblock Local elevation umbrella sampling applied to the calculation of
  alchemical free-energy changes via $\lambda$-dynamics: The $\lambda$-leus
  scheme.
\newblock {\em Journal of Chemical Theory and Computation}, 10(8):3006--3022,
  2014.

\bibitem{bieler2015orthogonal}
Noah~S Bieler and Philippe~H H{\"u}nenberger.
\newblock Orthogonal sampling in free-energy calculations of residue mutations
  in a tripeptide: Ti versus $\lambda$-leus.
\newblock {\em Journal of Computational Chemistry}, 36(22):1686--1697, 2015.

\bibitem{hahn2019alchemical}
David~F Hahn and Philippe~H Hünenberger.
\newblock Alchemical free-energy calculations by multiple-replica
  $\lambda$-dynamics: The conveyor belt thermodynamic integration scheme.
\newblock {\em Journal of Chemical Theory and Computation}, 15(4):2392--2419,
  2019.

\bibitem{lee2023aces}
Tai-Sung Lee, Hsu-Chun Tsai, Abir Ganguly, and Darrin~M York.
\newblock Aces: Optimized alchemically enhanced sampling.
\newblock {\em Journal of Chemical Theory and Computation}, 2023.

\bibitem{hsu2023alchemical}
Wei-Tse Hsu, Valerio Piomponi, Pascal~T Merz, Giovanni Bussi, and Michael~R
  Shirts.
\newblock Alchemical metadynamics: Adding alchemical variables to metadynamics
  to enhance sampling in free energy calculations.
\newblock {\em Journal of Chemical Theory and Computation}, 19(6):1805--1817,
  2023.

\bibitem{barducci2011metadynamics}
Alessandro Barducci, Massimiliano Bonomi, and Michele Parrinello.
\newblock Metadynamics.
\newblock {\em Wiley Interdisciplinary Reviews: Computational Molecular
  Science}, 1(5):826--843, 2011.

\bibitem{Darve2001}
E.~Darve and A.~Pohorille.
\newblock Calculating free energies using average force.
\newblock {\em J. Chem. Phys.}, 115:9169--9183, 2001.

\bibitem{darve2008adaptive}
Eric Darve, David Rodr{\'i}guez-G{\'o}mez, and Andrew Pohorille.
\newblock Adaptive biasing force method for scalar and vector free energy
  calculations.
\newblock {\em The Journal of Chemical Physics}, 128(14):144120, 2008.

\bibitem{comer2015adaptive}
Jeffrey Comer, James~C Gumbart, J{\'e}r{\^o}me H{\'e}nin, Tony Leli{\`e}vre,
  Andrew Pohorille, and Christophe Chipot.
\newblock The adaptive biasing force method: Everything you always wanted to
  know but were afraid to ask.
\newblock {\em The Journal of Physical Chemistry B}, 119(3):1129--1151, 2015.

\bibitem{Fiorin2013}
Giacomo Fiorin, Michael~L. Klein, and J{\'e}r{\^o}me H{\'e}nin.
\newblock {Using collective variables to drive molecular dynamics simulations}.
\newblock {\em {Mol. Phys.}}, 111(22-23):3345--3362, 2013.

\bibitem{salari2018}
Reza Salari, Thomas Joseph, Ruchi Lohia, Jérôme Hénin, and Grace Brannigan.
\newblock A streamlined, general approach for computing ligand binding free
  energies and its application to gpcr-bound cholesterol.
\newblock {\em Journal of Chemical Theory and Computation}, 14(12):6560--6573,
  2018.
\newblock PMID: 30358394.

\bibitem{phillips2020scalable}
James~C Phillips, David~J Hardy, Julio~DC Maia, John~E Stone, Jo{\~a}o~V
  Ribeiro, Rafael~C Bernardi, Ronak Buch, Giacomo Fiorin, J{\'e}r{\^o}me
  H{\'e}nin, Wei Jiang, et~al.
\newblock Scalable molecular dynamics on cpu and gpu architectures with namd.
\newblock {\em The Journal of Chemical Physics}, 153(4):044130, 2020.

\bibitem{lagardere2018tinker}
Louis Lagard{\`e}re, Luc-Henri Jolly, Filippo Lipparini, F{\'e}lix Aviat,
  Benjamin Stamm, Zhifeng~F Jing, Matthew Harger, Hedieh Torabifard,
  G~Andr{\'e}s Cisneros, Michael~J Schnieders, et~al.
\newblock Tinker-hp: a massively parallel molecular dynamics package for
  multiscale simulations of large complex systems with advanced point dipole
  polarizable force fields.
\newblock {\em Chemical science}, 9(4):956--972, 2018.

\bibitem{adjoua2021tinker}
Olivier Adjoua, Louis Lagard{\`e}re, Luc-Henri Jolly, Arnaud Durocher, Thibaut
  Very, Isabelle Dupays, Zhi Wang, Th{\'e}o~Jaffrelot Inizan, Fr{\'e}d{\'e}ric
  C{\'e}lerse, Pengyu Ren, et~al.
\newblock Tinker-hp: Accelerating molecular dynamics simulations of large
  complex systems with advanced point dipole polarizable force fields using
  gpus and multi-gpu systems.
\newblock {\em Journal of Chemical Theory and Computation}, 17(4):2034--2053,
  2021.

\bibitem{mackerell2002charmm}
Alexander~D MacKerell~Jr, Bernard Brooks, Charles~L Brooks~III, Lennart
  Nilsson, Benoit Roux, Youngdo Won, and Martin Karplus.
\newblock Charmm: the energy function and its parameterization.
\newblock {\em Encyclopedia of computational chemistry}, 1, 2002.

\bibitem{vanommeslaeghe2010charmm}
Kenno Vanommeslaeghe, Elizabeth Hatcher, Chayan Acharya, Sibsankar Kundu,
  Shijun Zhong, Jihyun Shim, Eva Darian, Olgun Guvench, P~Lopes, Igor Vorobyov,
  et~al.
\newblock Charmm general force field: A force field for drug-like molecules
  compatible with the charmm all-atom additive biological force fields.
\newblock {\em Journal of Computational Chemistry}, 31(4):671--690, 2010.

\bibitem{ren2003polarizable}
Pengyu Ren and Jay~W Ponder.
\newblock Polarizable atomic multipole water model for molecular mechanics
  simulation.
\newblock {\em The Journal of Physical Chemistry B}, 107(24):5933--5947, 2003.

\bibitem{ponder2010current}
Jay~W Ponder, Chuanjie Wu, Pengyu Ren, Vijay~S Pande, John~D Chodera, Michael~J
  Schnieders, Imran Haque, David~L Mobley, Daniel~S Lambrecht, Robert~A
  DiStasio~Jr, et~al.
\newblock Current status of the amoeba polarizable force field.
\newblock {\em The Journal of Physical Chemistry B}, 114(8):2549--2564, 2010.

\bibitem{ren2011polarizable}
Pengyu Ren, Chuanjie Wu, and Jay~W Ponder.
\newblock Polarizable atomic multipole-based molecular mechanics for organic
  molecules.
\newblock {\em Journal of Chemical Theory and Computation}, 7(10):3143--3161,
  2011.

\bibitem{shi2013polarizable}
Yue Shi, Zhen Xia, Jiajing Zhang, Robert Best, Chuanjie Wu, Jay~W Ponder, and
  Pengyu Ren.
\newblock Polarizable atomic multipole-based amoeba force field for proteins.
\newblock {\em Journal of Chemical Theory and Computation}, 9(9):4046--4063,
  2013.

\bibitem{AMOEBAnucleic}
Changsheng Zhang, Chao Lu, Zhifeng Jing, Chuanjie Wu, Jean-Philip Piquemal,
  Jay~W. Ponder, and Pengyu Ren.
\newblock Amoeba polarizable atomic multipole force field for nucleic acids.
\newblock {\em Journal of Chemical Theory and Computation}, 14(4):2084--2108,
  2018.
\newblock PMID: 29438622.

\bibitem{knight2011applying}
Jennifer~L Knight and Charles~L Brooks~III.
\newblock Applying efficient implicit nongeometric constraints in alchemical
  free energy simulations.
\newblock {\em Journal of Computational Chemistry}, 32(16):3423--3432, 2011.

\bibitem{lelievre2008long}
Tony Leli{\`e}vre, Mathias Rousset, and Gabriel Stoltz.
\newblock Long-time convergence of an adaptive biasing force method.
\newblock {\em Nonlinearity}, 21(6):1155, 2008.

\bibitem{Henin2021}
J.~H\'enin.
\newblock Fast and accurate multidimensional free energy integration.
\newblock {\em J. Chem. Theory Comput.}, 17(11):6789--6798, 2021.
\newblock PMID: 34665624.

\bibitem{Darve2002}
E.~Darve, M.~Wilson, and A.~Pohorille.
\newblock Calculating free energies using a scaled-force molecular dynamics
  algorithm.
\newblock {\em Mol. Sim.}, 28:113--144, 2002.

\bibitem{Henin2004}
J.~H\'enin and C.~Chipot.
\newblock Overcoming free energy barriers using unconstrained molecular
  dynamics simulations.
\newblock {\em J. Chem. Phys.}, 121:2904--2914, 2004.

\bibitem{Bruckner2010}
Stefan Bruckner and Stefan Boresch.
\newblock Efficiency of alchemical free energy simulations. i. a practical
  comparison of the exponential formula, thermodynamic integration, and
  bennett’s acceptance ratio method.
\newblock {\em Journal of Computational Chemistry}, 32(7):1303–1319, December
  2010.

\bibitem{Bruckner2010a}
Stefan Bruckner and Stefan Boresch.
\newblock Efficiency of alchemical free energy simulations. ii. improvements
  for thermodynamic integration.
\newblock {\em Journal of Computational Chemistry}, 32(7):1320–1333, December
  2010.

\bibitem{fu2018zooming}
Haohao Fu, Hong Zhang, Haochuan Chen, Xueguang Shao, Christophe Chipot, and
  Wensheng Cai.
\newblock Zooming across the free-energy landscape: shaving barriers, and
  flooding valleys.
\newblock {\em The Journal of Physical Chemistry Letters}, 9(16):4738--4745,
  2018.

\bibitem{fu2019taming}
Haohao Fu, Xueguang Shao, Wensheng Cai, and Christophe Chipot.
\newblock Taming rugged free energy landscapes using an average force.
\newblock {\em Accounts of Chemical Research}, 52(11):3254--3264, 2019.

\bibitem{chen2021overcoming}
Haochuan Chen, Haohao Fu, Christophe Chipot, Xueguang Shao, and Wensheng Cai.
\newblock Overcoming free-energy barriers with a seamless combination of a
  biasing force and a collective variable-independent boost potential.
\newblock {\em Journal of Chemical Theory and Computation}, 17(7):3886--3894,
  2021.

\bibitem{raiteri2006efficient}
Paolo Raiteri, Alessandro Laio, Francesco~Luigi Gervasio, Cristian Micheletti,
  and Michele Parrinello.
\newblock Efficient reconstruction of complex free energy landscapes by
  multiple walkers metadynamics.
\newblock {\em The Journal of Physical Chemistry B}, 110(8):3533--3539, 2006.

\bibitem{larson2009folding}
Stefan~M Larson, Christopher~D Snow, Michael Shirts, and Vijay~S Pande.
\newblock Folding@ home and genome@ home: Using distributed computing to tackle
  previously intractable problems in computational biology.
\newblock {\em arXiv preprint arXiv:0901.0866}, 2009.

\bibitem{hedin2019gen}
Florent H{\'e}din and Tony Leli{\`e}vre.
\newblock gen. parrep: a first implementation of the generalized parallel
  replica dynamics for the long time simulation of metastable biochemical
  systems.
\newblock {\em Computer Physics Communications}, 239:311--324, 2019.

\bibitem{minoukadeh2010potential}
Kimiya Minoukadeh, Christophe Chipot, and Tony Leli{\`e}vre.
\newblock Potential of mean force calculations: a multiple-walker adaptive
  biasing force approach.
\newblock {\em Journal of Chemical Theory and Computation}, 6(4):1008--1017,
  2010.

\bibitem{comer2014multiple}
Jeffrey Comer, James~C Phillips, Klaus Schulten, and Christophe Chipot.
\newblock Multiple-replica strategies for free-energy calculations in namd:
  Multiple-walker adaptive biasing force and walker selection rules.
\newblock {\em Journal of Chemical Theory and Computation}, 10(12):5276--5285,
  2014.

\bibitem{gilson1997statistical}
Michael~K Gilson, James~A Given, Bruce~L Bush, and J~Andrew McCammon.
\newblock The statistical-thermodynamic basis for computation of binding
  affinities: a critical review.
\newblock {\em Biophysical journal}, 72(3):1047--1069, 1997.

\bibitem{santiago2023computing}
Ezry Santiago-McRae, Mina Ebrahimi, Jesse~W. Sandberg, Grace Brannigan, and
  Jérôme Hénin.
\newblock Computing absolute binding affinities by streamlined alchemical free
  energy perturbation (safep) [article v1.0].
\newblock {\em Living Journal of Computational Molecular Science}, 5(1):2067,
  Oct. 2023.

\bibitem{hermans1986free}
Jan Hermans and S~Shankar.
\newblock The free energy of xenon binding to myoglobin from molecular dynamics
  simulation.
\newblock {\em Israel Journal of Chemistry}, 27(2):225--227, 1986.

\bibitem{boresch2003absolute}
Stefan Boresch, Franz Tettinger, Martin Leitgeb, and Martin Karplus.
\newblock Absolute binding free energies: a quantitative approach for their
  calculation.
\newblock {\em The Journal of Physical Chemistry B}, 107(35):9535--9551, 2003.

\bibitem{mobley2006use}
David~L Mobley, John~D Chodera, and Ken~A Dill.
\newblock On the use of orientational restraints and symmetry corrections in
  alchemical free energy calculations.
\newblock {\em The Journal of Chemical Physics}, 125(8):084902, 2006.

\bibitem{laury2018absolute}
Marie~L Laury, Zhi Wang, Aaron~S Gordon, and Jay~W Ponder.
\newblock Absolute binding free energies for the sampl6 cucurbit [8] uril
  host--guest challenge via the amoeba polarizable force field.
\newblock {\em Journal of Computer-aided Molecular Design}, 32:1087--1095,
  2018.

\bibitem{wang2006absolute}
Jiyao Wang, Yuqing Deng, and Beno{\^\i}t Roux.
\newblock Absolute binding free energy calculations using molecular dynamics
  simulations with restraining potentials.
\newblock {\em Biophysical journal}, 91(8):2798--2814, 2006.

\bibitem{gumbart2013standard}
James~C Gumbart, Beno{\^\i}t Roux, and Christophe Chipot.
\newblock Standard binding free energies from computer simulations: What is the
  best strategy?
\newblock {\em Journal of Chemical Theory and Computation}, 9(1):794--802,
  2013.

\bibitem{fu2017new}
Haohao Fu, Wensheng Cai, J{\'e}r{\^o}me H{\'e}nin, Beno{\^\i}t Roux, and
  Christophe Chipot.
\newblock New coarse variables for the accurate determination of standard
  binding free energies.
\newblock {\em Journal of Chemical Theory and Computation}, 13(11):5173--5178,
  2017.

\bibitem{fu2022accurate}
Haohao Fu, Haochuan Chen, Marharyta Blazhynska, Emma Goulard Coderc~de Lacam,
  Florence Szczepaniak, Anna Pavlova, Xueguang Shao, James~C Gumbart,
  Fran{\c{c}}ois Dehez, Beno{\^\i}t Roux, et~al.
\newblock Accurate determination of protein: ligand standard binding free
  energies from molecular dynamics simulations.
\newblock {\em Nature Protocols}, 17(4):1114--1141, 2022.

\bibitem{clark2023comparison}
Finlay Clark, Graeme Robb, Daniel~J Cole, and Julien Michel.
\newblock Comparison of receptor--ligand restraint schemes for alchemical
  absolute binding free energy calculations.
\newblock {\em Journal of Chemical Theory and Computation}, 2023.

\bibitem{HUMP96}
William Humphrey, Andrew Dalke, and Klaus Schulten.
\newblock {VMD} -- {V}isual {M}olecular {D}ynamics.
\newblock {\em Journal of Molecular Graphics}, 14:33--38, 1996.

\bibitem{henin2022human}
J{\'e}r{\^o}me H{\'e}nin, Laura~JS Lopes, and Giacomo Fiorin.
\newblock Human learning for molecular simulations: the collective variables
  dashboard in vmd.
\newblock {\em Journal of Chemical Theory and Computation}, 18(3):1945--1956,
  2022.

\bibitem{shi2021amoeba}
Yuanjun Shi, Marie~L Laury, Zhi Wang, and Jay~W Ponder.
\newblock Amoeba binding free energies for the sampl7 trimertrip host--guest
  challenge.
\newblock {\em Journal of Computer-aided Molecular Design}, 35:79--93, 2021.

\bibitem{shi2015polarizable}
Yue Shi, Pengyu Ren, Michael Schnieders, and Jean-Philip Piquemal.
\newblock Polarizable force fields for biomolecular modeling.
\newblock {\em Reviews in Computational Chemistry Volume 28}, pages 51--86,
  2015.

\bibitem{reviewPFF}
Josef Melcr and Jean-Philip Piquemal.
\newblock Accurate biomolecular simulations account for electronic
  polarization.
\newblock {\em Frontiers in Molecular Biosciences}, 6:143, 2019.

\bibitem{darden1993particle}
Tom Darden, Darrin York, and Lee Pedersen.
\newblock Particle mesh ewald: An n log (n) method for ewald sums in large
  systems.
\newblock {\em The Journal of Chemical Physics}, 98(12):10089--10092, 1993.

\bibitem{essmann1995smooth}
Ulrich Essmann, Lalith Perera, Max~L Berkowitz, Tom Darden, Hsing Lee, and
  Lee~G Pedersen.
\newblock A smooth particle mesh ewald method.
\newblock {\em The Journal of Chemical Physics}, 103(19):8577--8593, 1995.

\bibitem{Ewaldpol}
Louis Lagardère, Filippo Lipparini, Etienne Polack, Benjamin Stamm, Eric
  Cancés, Michael Schnieders, Pengyu Ren, Yvon Maday, and Jean-Philip
  Piquemal.
\newblock Scalable evaluation of polarization energy and associated forces in
  polarizable molecular dynamics: Ii. toward massively parallel computations
  using smooth particle mesh ewald.
\newblock {\em Journal of Chemical Theory and Computation}, 11(6):2589--2599,
  2015.

\bibitem{schnieders2012structure}
Michael~J Schnieders, Jonas Baltrusaitis, Yue Shi, Gaurav Chattree, Lianqing
  Zheng, Wei Yang, and Pengyu Ren.
\newblock The structure, thermodynamics, and solubility of organic crystals
  from simulation with a polarizable force field.
\newblock {\em Journal of Chemical Theory and Computation}, 8(5):1721--1736,
  2012.

\bibitem{zacharias1994separation}
M~Zacharias, TP~Straatsma, and JA~McCammon.
\newblock Separation-shifted scaling, a new scaling method for lennard-jones
  interactions in thermodynamic integration.
\newblock {\em The Journal of Chemical Physics}, 100(12):9025--9031, 1994.

\bibitem{Dixit2001namd_fep}
S.~B. Dixit and C.~Chipot.
\newblock Can absolute free energies of association be estimated from molecular
  mechanical simulations~? the biotin-streptavidin system revisited.
\newblock {\em J. Phys. Chem. A}, 105:9795--9799, 2001.

\bibitem{Phillips2005}
J.~C. Phillips, R.~Braun, W.~Wang, J.~Gumbart, E.~Tajkhorshid, E.~Villa,
  C.~Chipot, L.~Skeel, R. D.~Kal\'e, and K.~Schulten.
\newblock Scalable molecular dynamics with {\sc namd}.
\newblock {\em J. Comput. Chem.}, 26:1781--1802, 2005.

\bibitem{Chen2020namd_gpu_fep}
Haochuan Chen, Julio D.~C. Maia, Brian~K. Radak, David~J. Hardy, Wensheng Cai,
  Christophe Chipot, and Emad Tajkhorshid.
\newblock Boosting free-energy perturbation calculations with gpu-accelerated
  namd.
\newblock {\em Journal of Chemical Information and Modeling},
  60(11):5301--5307, 2020.
\newblock PMID: 32805108.

\bibitem{tuckerman1992reversible}
MBBJM Tuckerman, Bruce~J Berne, and Glenn~J Martyna.
\newblock Reversible multiple time scale molecular dynamics.
\newblock {\em The Journal of Chemical Physics}, 97(3):1990--2001, 1992.

\bibitem{lagardere2019pushing}
Louis Lagard{\`e}re, F{\'e}lix Aviat, and Jean-Philip Piquemal.
\newblock Pushing the limits of multiple-time-step strategies for polarizable
  point dipole molecular dynamics.
\newblock {\em The Journal of Physical Chemistry Letters}, 10(10):2593--2599,
  2019.

\bibitem{kieninger2022gromacs}
Stefanie Kieninger and Bettina~G Keller.
\newblock Gromacs stochastic dynamics and baoab are equivalent configurational
  sampling algorithms.
\newblock {\em Journal of Chemical Theory and Computation}, 18(10):5792--5798,
  2022.

\bibitem{ousterhout1989tcl}
John~K. Ousterhout.
\newblock Tcl: An embeddable command language.
\newblock Technical Report UCB/CSD-89-541, EECS Department, University of
  California, Berkeley, Nov 1989.

\bibitem{miao2021avoiding}
Mengyao Miao, Haohao Fu, Hong Zhang, Xueguang Shao, Christophe Chipot, and
  Wensheng Cai.
\newblock Avoiding non-equilibrium effects in adaptive biasing force
  calculations.
\newblock {\em Molecular Simulation}, 47(5):390--394, 2021.

\bibitem{ahmed2016fragment}
Abdelhakim Ahmed-Belkacem, Lionel Colliandre, Nazim Ahnou, Quentin Nevers,
  Muriel Gelin, Yannick Bessin, Rozenn Brillet, Olivier Cala, Dominique
  Douguet, William Bourguet, et~al.
\newblock Fragment-based discovery of a new family of non-peptidic
  small-molecule cyclophilin inhibitors with potent antiviral activities.
\newblock {\em Nature Communications}, 7(1):12777, 2016.

\bibitem{gradler2019discovery}
Ulrich Gr{\"a}dler, Daniel Schwarz, Michael Blaesse, Birgitta Leuthner,
  Theresa~L Johnson, Frederic Bernard, Xuliang Jiang, Andreas Marx, Marine
  Gilardone, Hugues Lemoine, et~al.
\newblock Discovery of novel cyclophilin d inhibitors starting from three
  dimensional fragments with millimolar potencies.
\newblock {\em Bioorganic \& Medicinal Chemistry Letters}, 29(23):126717, 2019.

\bibitem{bhati2022large}
Agastya~P Bhati and Peter~V Coveney.
\newblock Large scale study of ligand--protein relative binding free energy
  calculations: Actionable predictions from statistically robust protocols.
\newblock {\em Journal of Chemical Theory and Computation}, 18(4):2687--2702,
  2022.

\bibitem{bhati2023long}
Agastya~P Bhati, Art Hoti, Andrew Potterton, Mateusz~K Bieniek, and Peter~V
  Coveney.
\newblock Long time scale ensemble methods in molecular dynamics:
  Ligand--protein interactions and allostery in sars-cov-2 targets.
\newblock {\em Journal of Chemical Theory and Computation}, 2023.

\bibitem{schmid2000new}
Roland Schmid, Arzu~M Miah, and Valentin~N Sapunov.
\newblock A new table of the thermodynamic quantities of ionic hydration:
  values and some applications (enthalpy--entropy compensation and born radii).
\newblock {\em Physical Chemistry Chemical Physics}, 2(1):97--102, 2000.

\bibitem{kelly2005sm6}
Casey~P Kelly, Christopher~J Cramer, and Donald~G Truhlar.
\newblock Sm6: A density functional theory continuum solvation model for
  calculating aqueous solvation free energies of neutrals, ions, and solute-
  water clusters.
\newblock {\em Journal of chemical theory and computation}, 1(6):1133--1152,
  2005.

\bibitem{grossfield2003ion}
Alan Grossfield, Pengyu Ren, and Jay~W Ponder.
\newblock Ion solvation thermodynamics from simulation with a polarizable force
  field.
\newblock {\em Journal of the American Chemical Society}, 125(50):15671--15682,
  2003.

\bibitem{inizan2023scalable}
Th{\'e}o~Jaffrelot Inizan, Thomas Pl{\'e}, Olivier Adjoua, Pengyu Ren, Hatice
  G{\"o}kcan, Olexandr Isayev, Louis Lagard{\`e}re, and Jean-Philip Piquemal.
\newblock Scalable hybrid deep neural networks/polarizable potentials
  biomolecular simulations including long-range effects.
\newblock {\em Chemical Science}, 14(20):5438--5452, 2023.

\bibitem{ebrahimi2022symmetry}
Mina Ebrahimi and J{\'{e}}r{\^{o}}me H{\'{e}}nin.
\newblock Symmetry-adapted restraints for binding free energy calculations.
\newblock {\em Journal of Chemical Theory and Computation}, 18(4):2494--2502,
  March 2022.

\bibitem{Merski2013}
Matthew Merski and Brian~K Shoichet.
\newblock {The impact of introducing a histidine into an apolar cavity site on
  docking and ligand recognition}.
\newblock {\em Journal of Medicinal Chemistry}, 56(7):2874--2884, 2013.

\bibitem{guichou2016Ncom}
Abdelhakim Ahmed-Belkacem, Lionel Colliandre, Nazim Ahnou, Quentin Nevers,
  Muriel Gelin, Yannick Bessin, Rozenn Brillet, Olivier Cala, Dominique
  Douguet, William Bourguet, Isabelle Krimm, Jean-Michel Pawlotsky, and
  Jean-François Guichou.
\newblock Fragment-based discovery of a new family of non-peptidic
  small-molecule cyclophilin inhibitors with potent antiviral activities.
\newblock {\em Nature Communications}, 7(1):12777, 2016.

\bibitem{walker2022automation}
Brandon Walker, Chengwen Liu, Elizabeth Wait, and Pengyu Ren.
\newblock Automation of amoeba polarizable force field for small molecules:
  Poltype 2.
\newblock {\em Journal of Computational Chemistry}, 43(23):1530--1542, 2022.

\bibitem{Henin2023load_cv_traj}
J{\'e}r{\^o}me H{\'e}nin.
\newblock Script for loading a colvars trajectory into vmd, 2023.

\end{thebibliography}


\begin{thebibliography}{10}

\bibitem{leimkuhler2013rational}
Benedict Leimkuhler and Charles Matthews.
\newblock Rational construction of stochastic numerical methods for molecular
  sampling.
\newblock {\em Applied Mathematics Research eXpress}, 2013(1):34--56, 2013.

\bibitem{feller1995constant}
Scott~E Feller, Yuhong Zhang, Richard~W Pastor, and Bernard~R Brooks.
\newblock Constant pressure molecular dynamics simulation: The langevin piston
  method.
\newblock {\em The Journal of Chemical Physics}, 103(11):4613--4621, 1995.

\bibitem{chung2023accurate}
Moses~KJ Chung, Ryan~J Miller, Borna Novak, Zhi Wang, and Jay~W Ponder.
\newblock Accurate host--guest binding free energies using the amoeba
  polarizable force field.
\newblock {\em Journal of Chemical Information and Modeling}, 63(9):2769--2782,
  2023.

\bibitem{essmann1995smooth}
Ulrich Essmann, Lalith Perera, Max~L Berkowitz, Tom Darden, Hsing Lee, and
  Lee~G Pedersen.
\newblock A smooth particle mesh ewald method.
\newblock {\em The Journal of Chemical Physics}, 103(19):8577--8593, 1995.

\bibitem{Jo2008}
Sunhwan Jo, Taehoon Kim, Vidyashankara~G Iyer, and Wonpil Im.
\newblock Charmm-gui: a web-based graphical user interface for charmm.
\newblock {\em Journal of computational chemistry}, 29(11):1859--1865, 2008.

\bibitem{Lee2016}
Jumin Lee, Xi~Cheng, Sunhwan Jo, Alexander~D MacKerell, Jeffery~B Klauda, and
  Wonpil Im.
\newblock Charmm-gui input generator for namd, gromacs, amber, openmm, and
  charmm/openmm simulations using the charmm36 additive force field.
\newblock {\em Biophysical journal}, 110(3):641a, 2016.

\bibitem{Jorgensen1983}
W.~L. Jorgensen, J.~Chandrasekhar, J.~D. Madura, R.~W. Impey, and M.~L. Klein.
\newblock Comparison of simple potential functions for simulating liquid water.
\newblock {\em J. Chem. Phys.}, 79:926--935, 1983.

\bibitem{phillips2020scalable}
James~C Phillips, David~J Hardy, Julio~DC Maia, John~E Stone, Jo{\~a}o~V
  Ribeiro, Rafael~C Bernardi, Ronak Buch, Giacomo Fiorin, J{\'e}r{\^o}me
  H{\'e}nin, Wei Jiang, et~al.
\newblock Scalable molecular dynamics on cpu and gpu architectures with namd.
\newblock {\em The Journal of Chemical Physics}, 153(4):044130, 2020.

\bibitem{Phillips2005}
J.~C. Phillips, R.~Braun, W.~Wang, J.~Gumbart, E.~Tajkhorshid, E.~Villa,
  C.~Chipot, L.~Skeel, R. D.~Kal\'e, and K.~Schulten.
\newblock Scalable molecular dynamics with {\sc namd}.
\newblock {\em J. Comput. Chem.}, 26:1781--1802, 2005.

\bibitem{darden1993particle}
Tom Darden, Darrin York, and Lee Pedersen.
\newblock Particle mesh ewald: An n log (n) method for ewald sums in large
  systems.
\newblock {\em The Journal of Chemical Physics}, 98(12):10089--10092, 1993.

\bibitem{Huang2016}
Jing Huang, Sarah Rauscher, Grzegorz Nawrocki, Ting Ran, Michael Feig, Bert~L
  de~Groot, Helmut Grubm\"{u}ller, and Alexander~D MacKerell.
\newblock {CHARMM}36m: an improved force field for folded and intrinsically
  disordered proteins.
\newblock {\em Nature Methods}, 14(1):71--73, November 2016.

\bibitem{Fiorin2013}
Giacomo Fiorin, Michael~L. Klein, and J{\'e}r{\^o}me H{\'e}nin.
\newblock {Using collective variables to drive molecular dynamics simulations}.
\newblock {\em {Mol. Phys.}}, 111(22-23):3345--3362, 2013.

\bibitem{ebrahimi2022symmetry}
Mina Ebrahimi and J{\'{e}}r{\^{o}}me H{\'{e}}nin.
\newblock Symmetry-adapted restraints for binding free energy calculations.
\newblock {\em Journal of Chemical Theory and Computation}, 18(4):2494--2502,
  March 2022.

\bibitem{zacharias1994separation}
M~Zacharias, TP~Straatsma, and JA~McCammon.
\newblock Separation-shifted scaling, a new scaling method for lennard-jones
  interactions in thermodynamic integration.
\newblock {\em The Journal of Chemical Physics}, 100(12):9025--9031, 1994.

\bibitem{Bernardi2020}
R~Bernardi, M~Bhandarkar, A~Bhatele, E~Bohm, R~Brunner, R~Buch, F~Buelens,
  H~Chen, C~Chipot, A~Dalke, S~Dixit, G~Fiorin, P~Freddolino, H~Fu, P~Grayson,
  J~Gullingsrud, A~Gursoy, D~Hardy, C~Harrison, J~Hénin, W~Humphrey,
  D~Hurwitz, A~Hynninen, N~Jain, W~Jiang, N~Krawetz, S~Kumar, D~Kunzman, J~Lai,
  C~Lee, J~Maia, R~McGreevy, C~Mei, M~Melo, M~Nelson, J~Phillips, B~Radak,
  J~Ribeiro, T~Rudack, O~Sarood, A~Shinozaki, D~Tanner, P~Wang, D~Wells,
  G~Zheng, and F~Zhu.
\newblock {NAMD User's Guide version 2.14}, 2020.
\newblock Accessed Aug 8, 2022s.

\bibitem{santiago2023computing}
Ezry Santiago-McRae, Mina Ebrahimi, Jesse~W. Sandberg, Grace Brannigan, and
  Jérôme Hénin.
\newblock Computing absolute binding affinities by streamlined alchemical free
  energy perturbation (safep) [article v1.0].
\newblock {\em Living Journal of Computational Molecular Science}, 5(1):2067,
  Oct. 2023.

\bibitem{bennett1976efficient}
Charles~H Bennett.
\newblock Efficient estimation of free energy differences from monte carlo
  data.
\newblock {\em Journal of Computational Physics}, 22(2):245--268, 1976.

\end{thebibliography}




\end{document}


\maketitle

\section{Setup used for numerical results}

All simulations were made at 298K and 1 Atmosphere. The Tinker-HP simulations used a BAOAB integrator, with a 1fs timestep, to integrate the Langevin equations of motion\cite{leimkuhler2013rational} and a Langevin Piston barostat\cite{feller1995constant}.
In all cases, no additional strategy was used to ensure neutrality of the system during alchemical simulations, as it has been shown to have little effects in recent studies\cite{chung2023accurate}.
\subsection{Hydration Free Energies}
All the simulations yielding the hydration free energies presented in the text used a cubic water box of 18.643\AA{} edge. Electrostatics and polarization interactions were computed using the Smooth Particle Mesh Ewald method\cite{essmann1995smooth} with a real space cutoff of 7 \AA{} and a 24x24x24 grid. Van der Waals interactions were cutoff at 7 \AA.

\begin{figure}[!htb]
    \centering
    \begin{minipage}{.5\textwidth}
        \centering
        \includegraphics[scale=0.5]{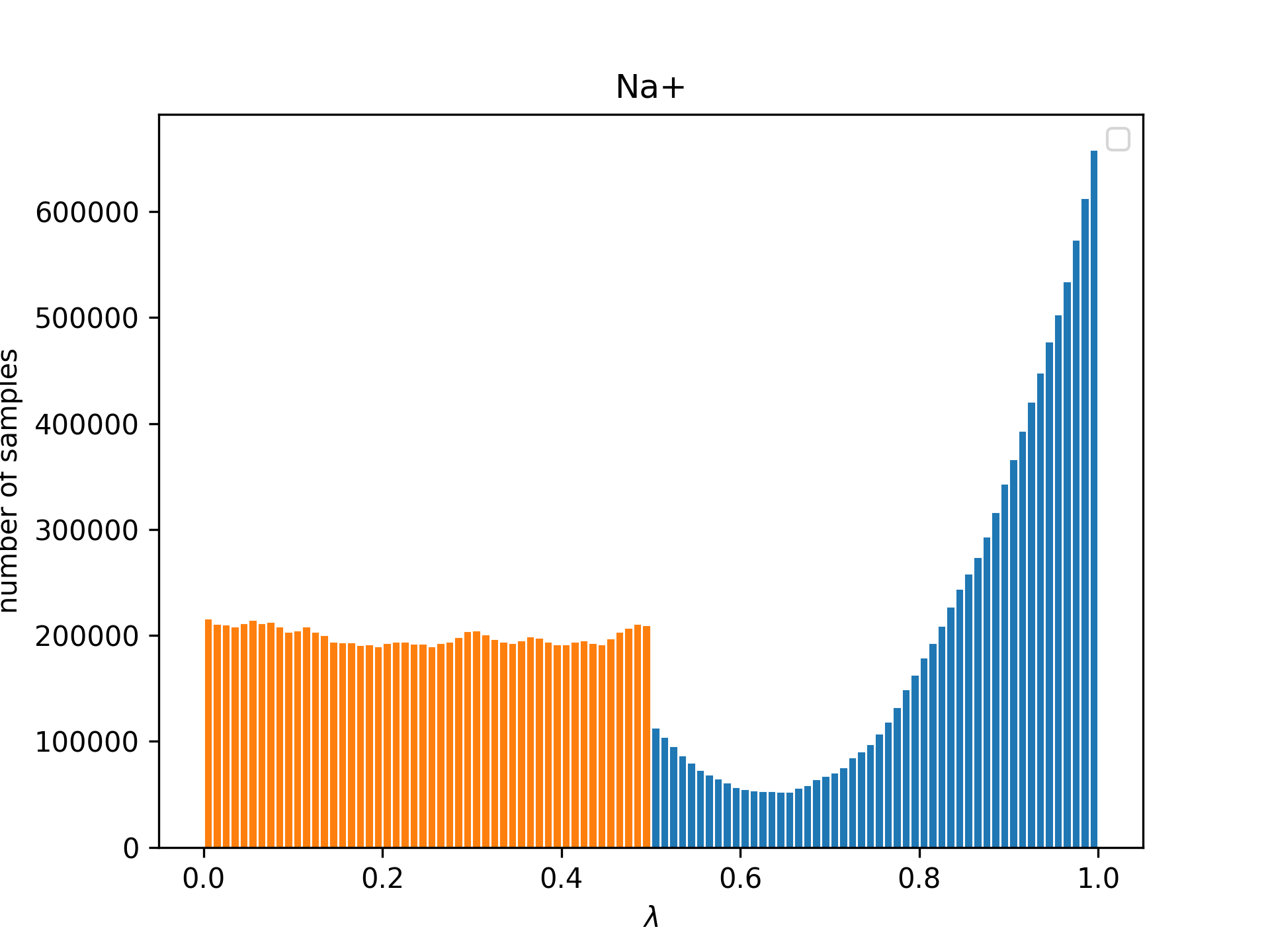}
    \end{minipage}%
    \begin{minipage}{0.5\textwidth}
        \centering
        \includegraphics[scale=0.5]{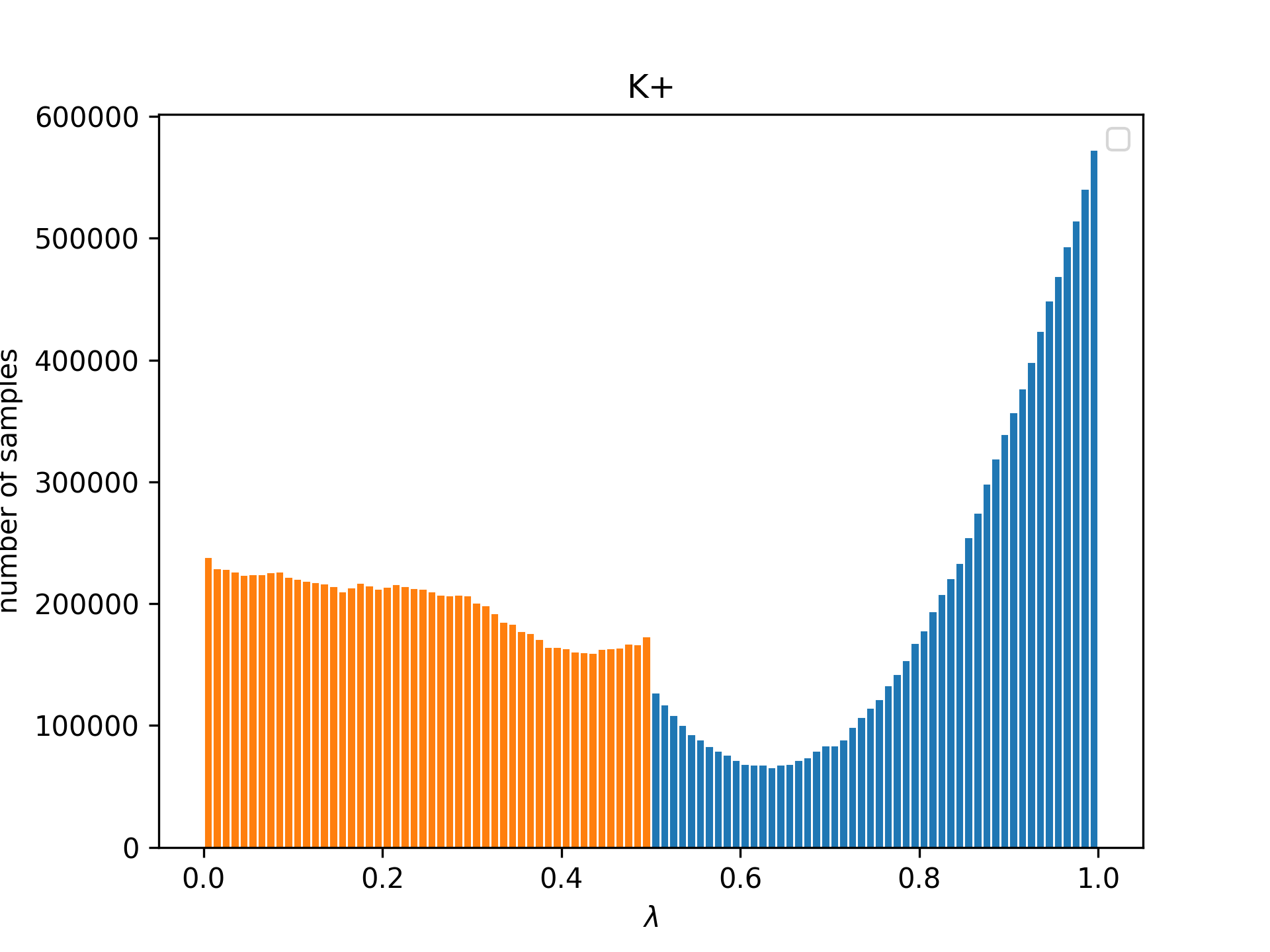}
    \end{minipage}
    \\
    \begin{minipage}{.5\textwidth}
        \centering
        \includegraphics[scale=0.5]{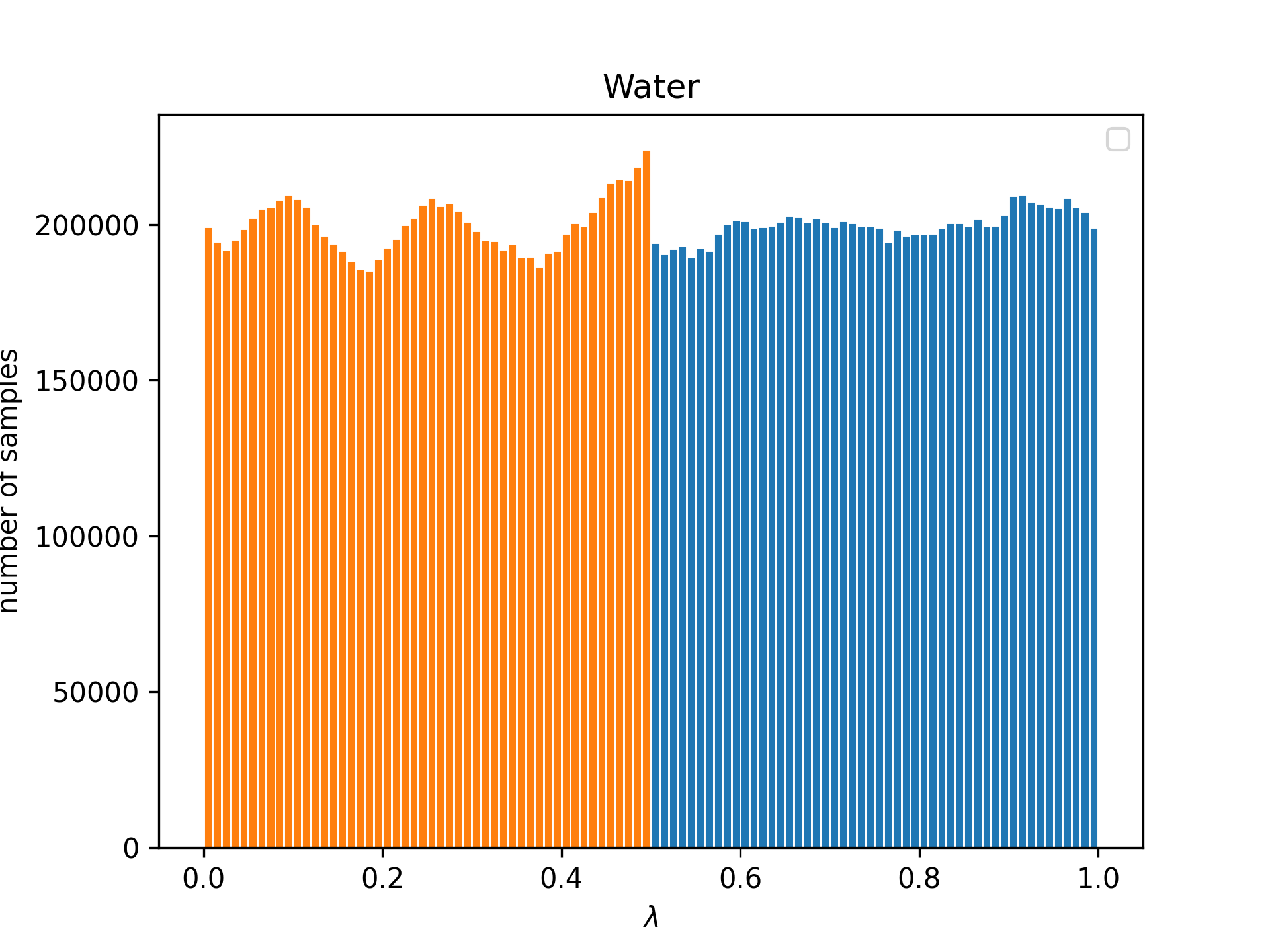}
    \end{minipage}
\caption{Histograms of the different lambda values sampled during the lambda-ABF simulations for the hydration free energies of the sodium ion, the potassium ion and water}
    \label{fig:count-hfe}
\end{figure}

\subsection{Lambda schedule}

\subsubsection{Tinker-HP}

In all Tinker-HP simulations, the lambda schedule used was the following:
In fixed-$\lambda$ simulations, the $\lambda$ schedule for both solvent phase and complex phase was the following:
\begin{table}[h!]
 \center
 \begin{tabular}{|c|c|c|c|c|c|c|c|c|c|c|c|c|c|c|c|c|c|c|c|c|c|c|c|}
 \hline
  $\lambda$ (global)&1.0&0.95&0.9&0.85&0.8&0.75&0.7&0.65&0.6&0.55&0.5 \\
 \hline
  $\lambda_e$ (electrostatics)&1.0&0.9&0.8&0.7&0.6&0.5&0.4&0.3&0.2&0.1&0.0\\
  \hline
  $\lambda_v$ (van der Waals)&1.0&1.0&1.0&1.0&1.0&1.0&1.0&1.0&1.0&1.0&1.0  \\
  \hline
 \end{tabular}
  \begin{tabular}{|c|c|c|c|c|c|c|c|c|c|c|c|c|c|c|c|c|c|c|c|c|c|c|c|}
 \hline
  $\lambda$ (global)&0.5&0.45&0.4&0.35&0.325&0.3&0.25&0.2&0.175&0.15&0.1&0.05&0.0 \\
 \hline
  $\lambda_e$ (electrostatics)&0.0&0.0&0.0&0.0&0.0&0.0&0.0&0.0&0.0&0.0&0.0&0.0&0.0\\
  \hline
  $\lambda_v$ (van der Waals)&1.0&0.9&0.8&0.7&0.65&0.6&0.5&0.4&0.35&0.3&0.2&0.1&0.0  \\
  \hline
 \end{tabular}
 \caption{\label{tab:lambdaschedule} lambda-schedule used for both solvent and complex phase of the free energies of binding simulations using Tinker-HP.}
\end{table}

\begin{minipage}{\linewidth}
\makebox[\linewidth]{
\includegraphics[scale=0.7]{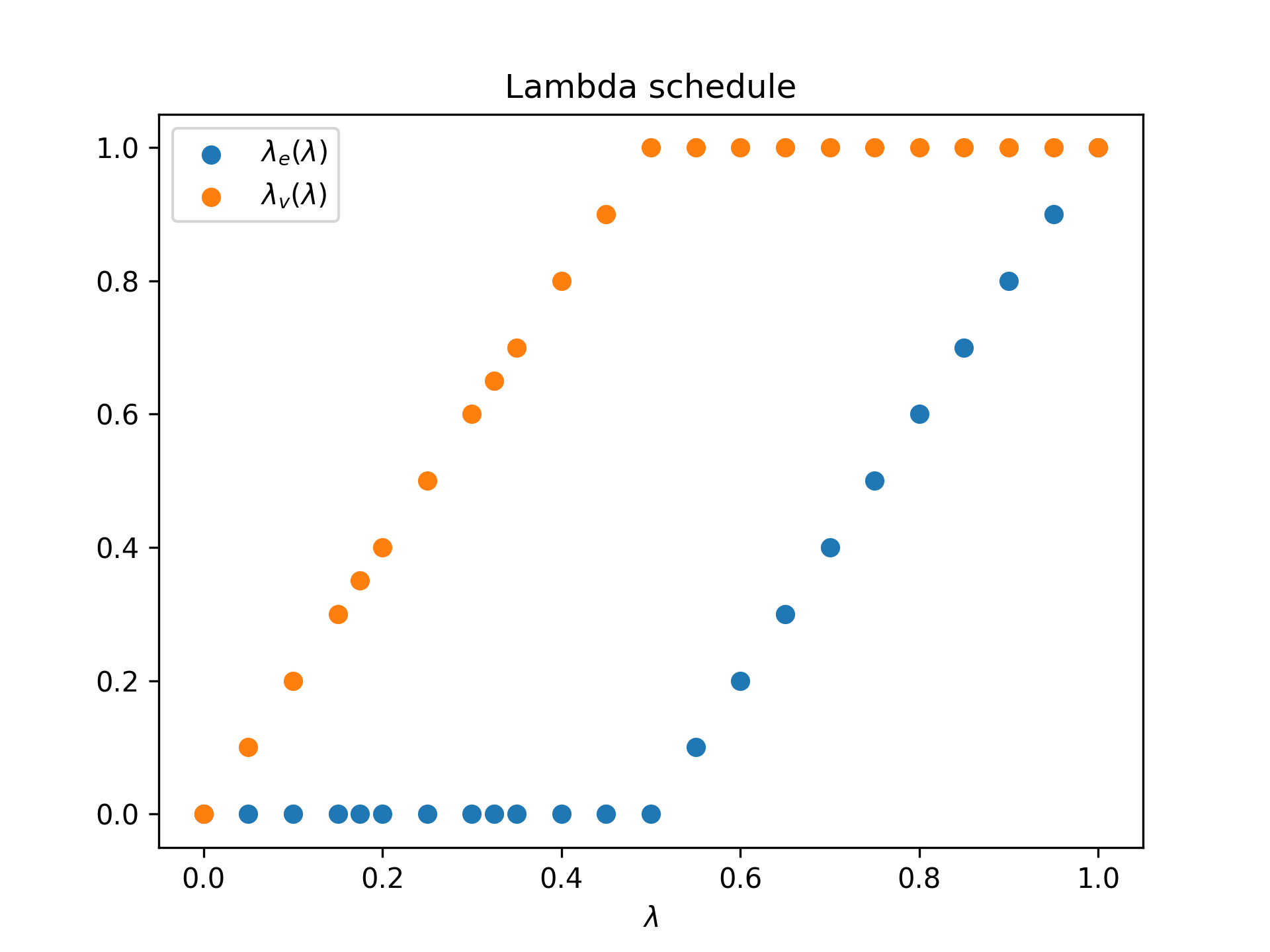}}
\captionof{figure}{lambda-schedule used in Tinker-HP simulations}\label{fig:lambda-schedule-tinker}
\end{minipage}\\

\subsection{Cucurbit[8]uril host-guest complex}
Both solvent and complex phase simulations were made using a cubic simulation box of edge 40 \AA. Electrostatics and polarization interactions were computed using the Smooth Particle Mesh Ewald method\cite{essmann1995smooth} with a real space cutoff of 7 \AA and a 48x48x48 grid.  Van der Waals interactions were cutoff at 12 \AA.

The movements of the ligand was restrained using a DBC variable associated to a harmonic restrained as defined in the following Colvars configuration:
\begin{verbatim}
colvar {
    name DBC

    rmsd {
        # Reference coordinates (for ligand RMSD computation)
        refPositionsFile cb8-cg9-ref.xyz

        atoms {
            # Define ligand atoms used for RMSD calculation
            atomNumbers 145 149 150 151 153 154 156 157 158 160

            # Moving frame of reference is defined below
            centerToReference yes
            rotateToReference yes
            fittingGroup {
            atomNumbers 1 2 3 4 5 6 7 8 9 10 11 12 13 14 15 16 17 20 21 22 27 28 29 30 33 34 36 37 38 39 40
            42 43 44 45 46 48 49 53 54 55 56 57 58 59 61 62 64 65 66 68 69 70 71 72 74 75 76 77 78 80 81 85
            86 87 88 89 90 91 93 94 96 97 98 100 101 102 103 104 106 107 108 109 110 112 113 117 118 119 120
            121 122 123 125 126 128
            }
            # Reference coordinates for binding site atoms
            # (can be the same file as ligand coordinates above)
        refPositionsFile cb8-cg9-ref.xyz

        }
    }
}



harmonicwalls {
    colvars DBC
    upperWalls 3.5
    forceConstant 100.0
}
\end{verbatim}

The distance between the Cl- counterion and the nitrogen atom of the ligand was restrained by using a TCL script which reads:
\begin{verbatim}
proc calc_colvar_forces { ts } {
  #  puts "Running calc_colvar_forces at timestep $ts"
  set PI [expr atan(1.)*4.]
  set k 10
  set delta 10.0
  set nd 1
  set ne 0

  set l [cv colvar l value]
  set l1 [expr 2.0*($l)]
  set dmin [expr $delta/2.0 * (1-cos(2*$PI*$l1))]

  set d [cv colvar d1 value]
  if { $d < $dmin } {
    set Fd [expr -$k*($d - $dmin)]
    cv colvar d1 addforce $Fd
    cv colvar l   addforce [expr - $Fd * $delta * $PI * sin(2.*$PI*$l1)*2.]
    # incompatible with subtractAppliedForce if doing ABF on lambda
    puts "Running calc_colvar_forces at timestep $ts, dmin $dmin, l $l"

    cv addenergy [expr $k / 2.0 * ($d-$dmin) * ($d-$dmin)]
  }
}
\end{verbatim}
associated to the d1 collective variable defined with the following Colvars configuration:
\begin{verbatim}
colvar {
  name d1
  width 0.1
  distance {
    group1 {
      atomnumbers 145
      }
    group2 {
      atomnumbers 6345
      }
  }
}
\end{verbatim}
The dynamical evolution of $\lambda$ was defined by the following Colvars configuration:
\begin{verbatim}
colvar {
  name l
  extendedLagrangian on
  extendedLangevinDamping 1000
  extendedtemp 300
  extendedmass  150000


  lowerBoundary 0.0
  upperBoundary 0.5
  #lowerBoundary 0.5
  #upperBoundary 1.0
  reflectingLowerBoundary
  reflectingUpperBoundary
  width 0.01
  alchLambda {
  }
  outputTotalForce
  outputAppliedforce
}


abf {
  colvars l
  fullSamples 5000
  historyFreq 1000
  shared
  sharedfreq 1000
}
\end{verbatim}

The following lines were included in the Tinker-HP *key input file in order to define the soft-core van der Waals parameters with van der Waals annihilation:
\begin{verbatim}
vdw-annihilate
vdw-sc-exp 2
vdw-sc-alpha 0.3
\end{verbatim}
The following lines were added to the same input files in order to start lambda-dynamics simulations with $b_p\approx 0$ and 4 walkers in the multiple-walker scheme:
\begin{verbatim}
lambdadyn
bound-pol-lambda 0.01
replicas 4
\end{verbatim}

\begin{figure}[!htb]
    \centering
    \begin{minipage}{.5\textwidth}
        \centering
        \includegraphics[scale=0.5]{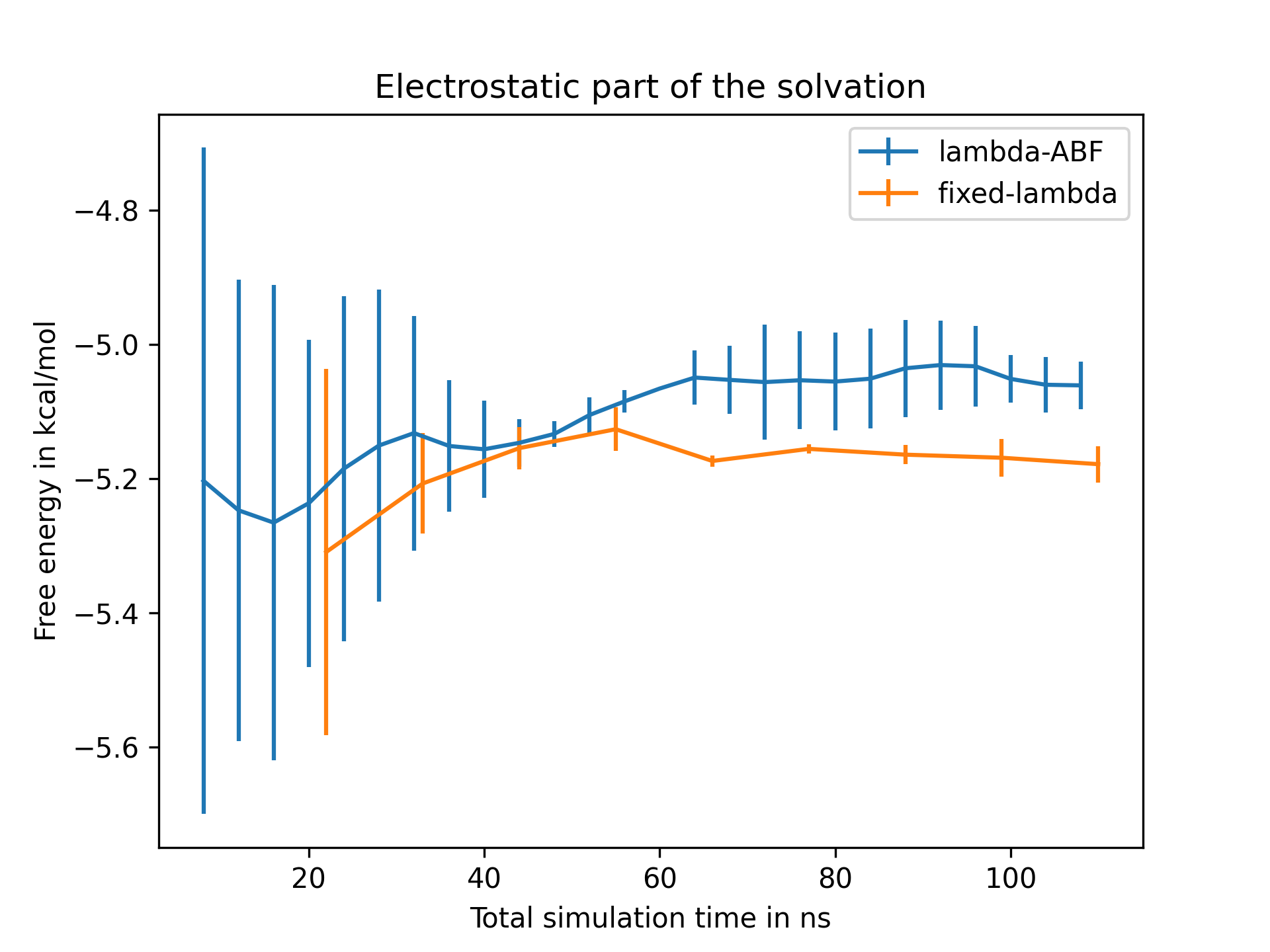}
    \end{minipage}%
    \begin{minipage}{0.5\textwidth}
        \centering
        \includegraphics[scale=0.5]{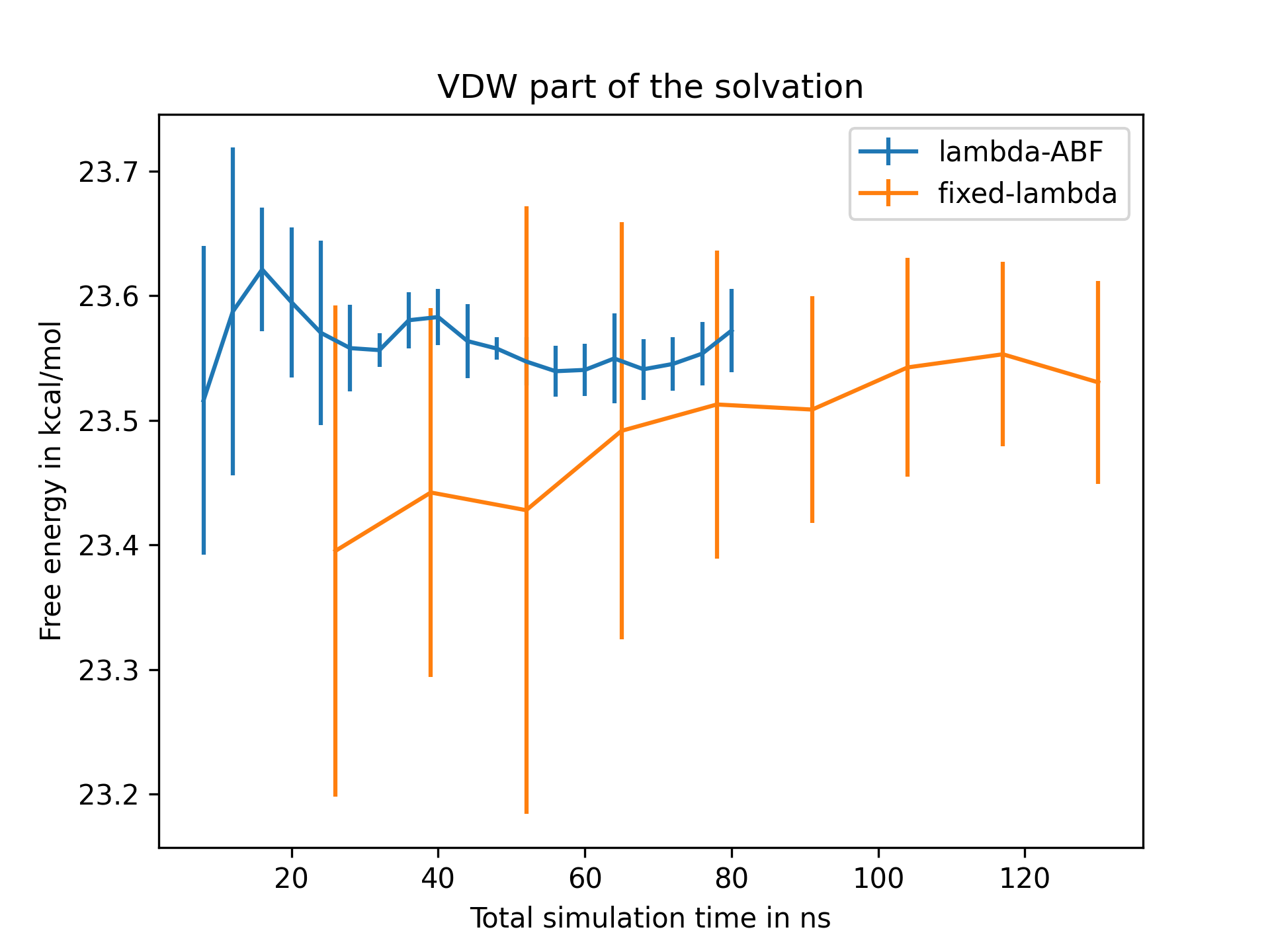}
    \end{minipage}
    \\
    \begin{minipage}{.5\textwidth}
        \centering
        \includegraphics[scale=0.5]{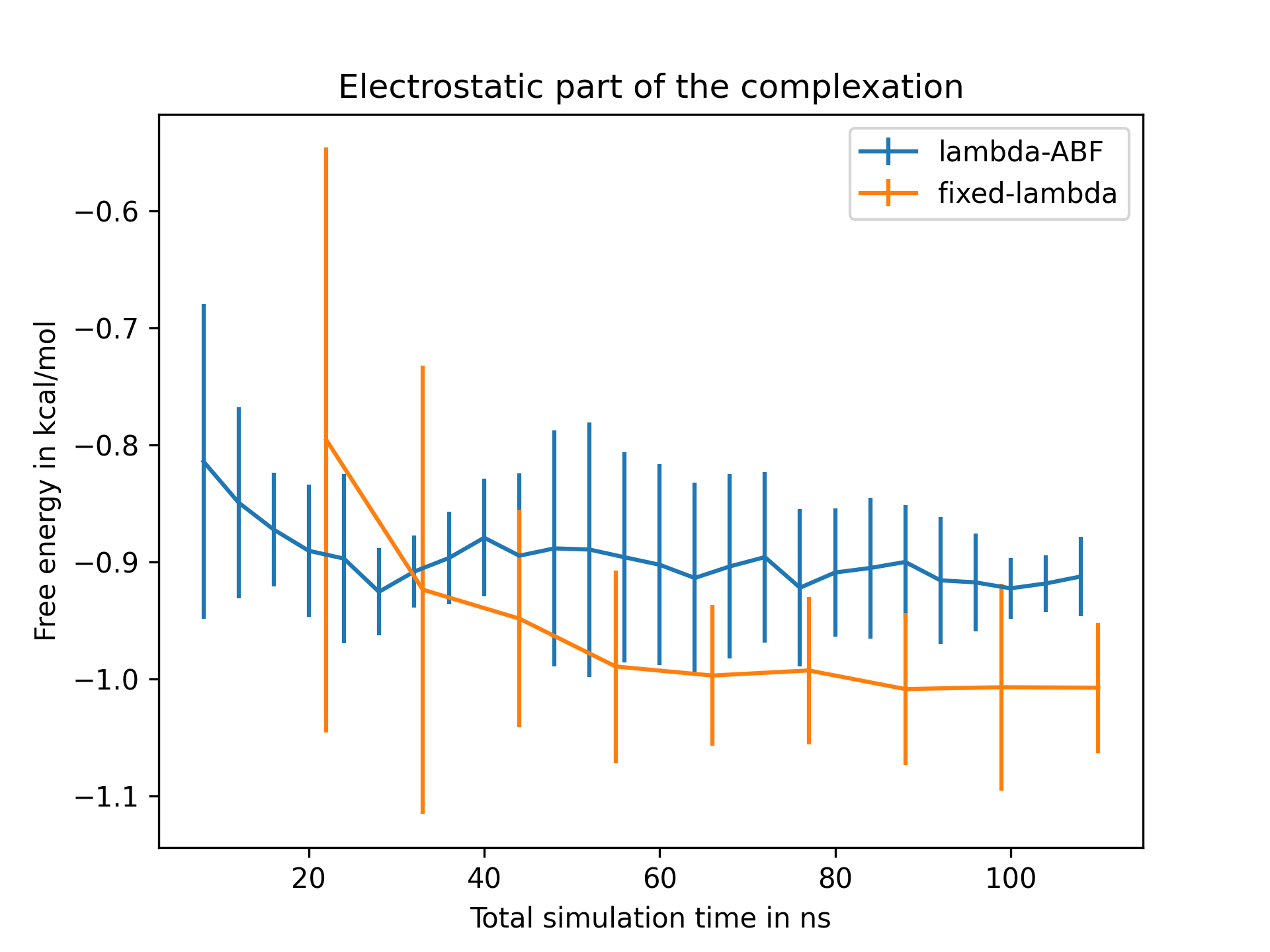}
    \end{minipage}%
    \begin{minipage}{0.5\textwidth}
        \centering
        \includegraphics[scale=0.5]{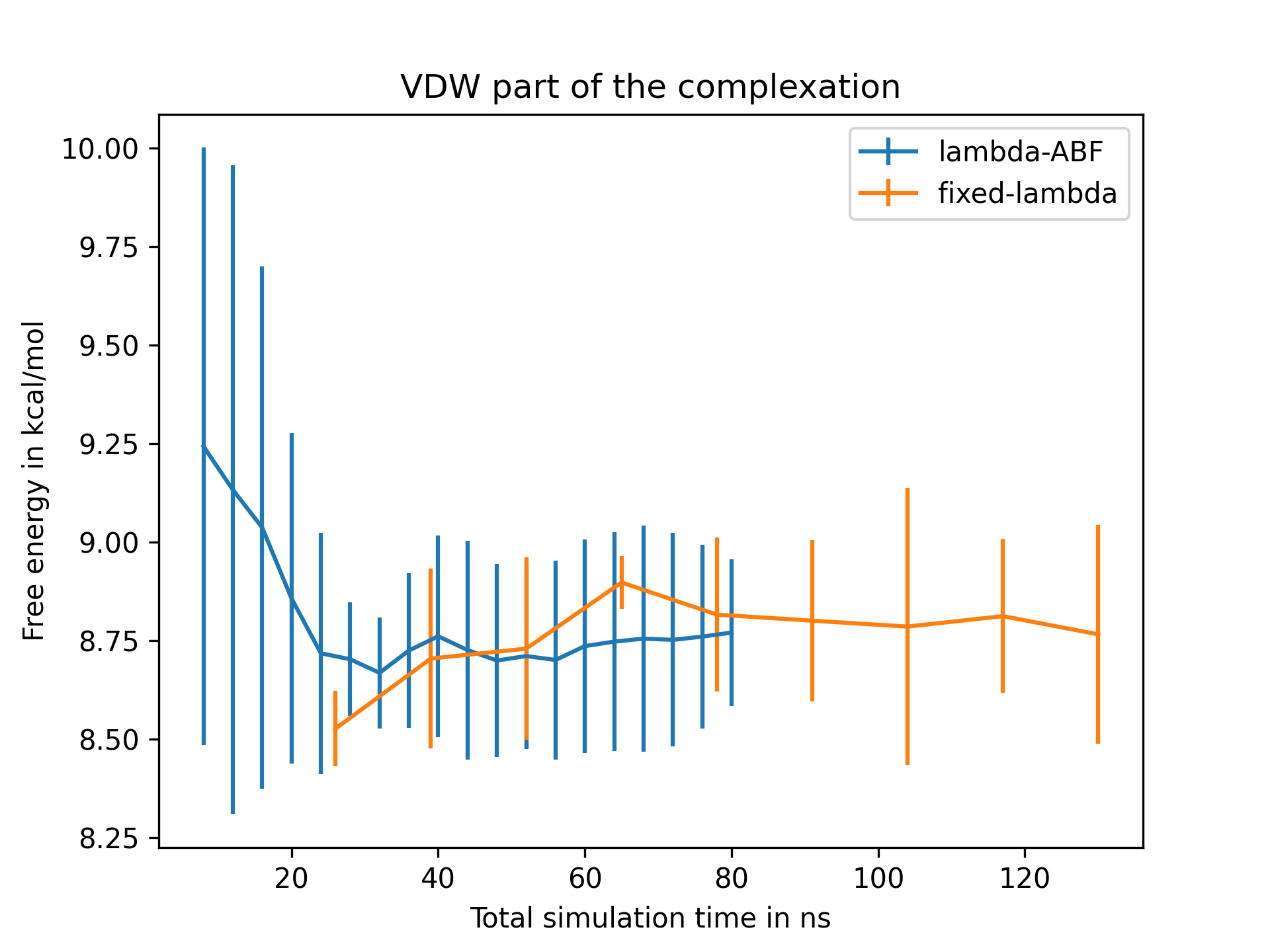}
    \end{minipage}
\caption{Free energy convergence for the SAMPL6 system: electrostatic part of the solvation (upper left), van der Waals part of the solvation (upper right), electrostatic part of the complexation (lower left), van der Waals part of the complexation (lower right)}
    \label{fig:sampl6-9-fe}
\end{figure}

\begin{figure}[!htb]
    \centering
    \begin{minipage}{.5\textwidth}
        \centering
        \includegraphics[scale=0.5]{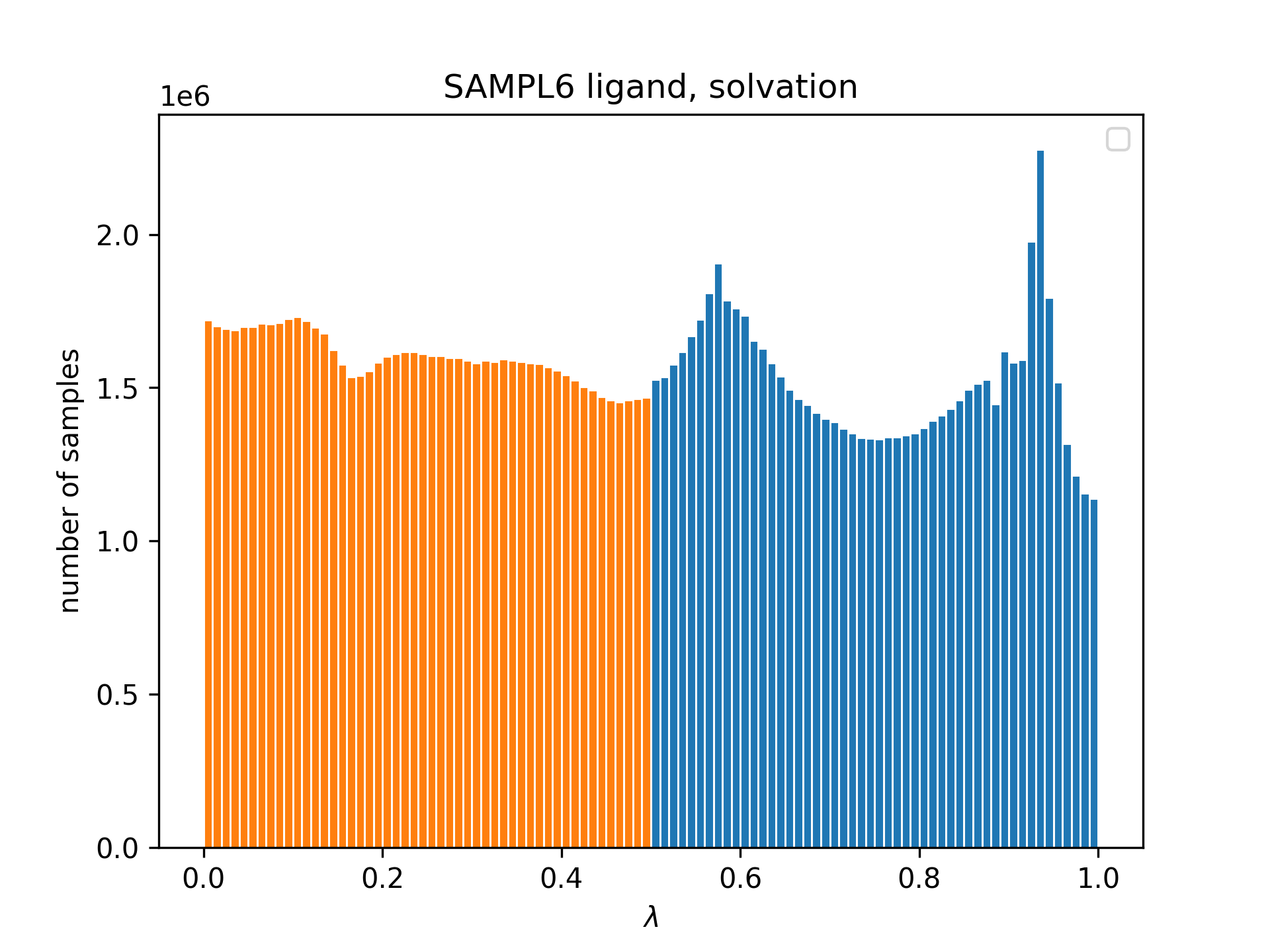}
    \end{minipage}%
    \begin{minipage}{0.5\textwidth}
        \centering
        \includegraphics[scale=0.5]{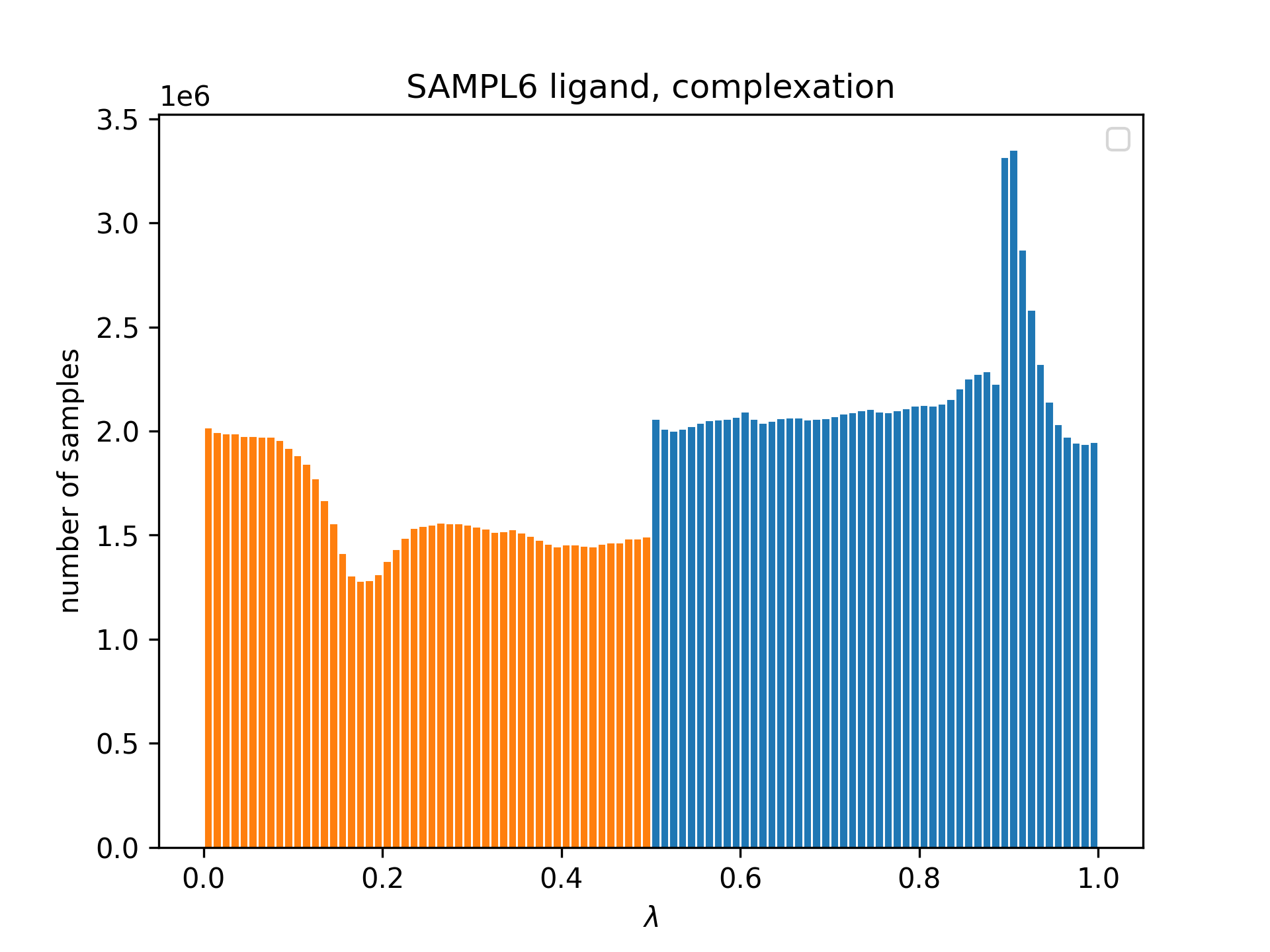}
    \end{minipage}
\caption{Histograms of the different lambda values sampled during the lambda-ABF simulations for the solvation and complexation leg of the SAMPL6 ligand}
    \label{fig:count-sampl6}
\end{figure}

\begin{figure}[!htb]
    \centering
\begin{minipage}{\linewidth}
\makebox[\linewidth]{
\includegraphics[scale=0.7]{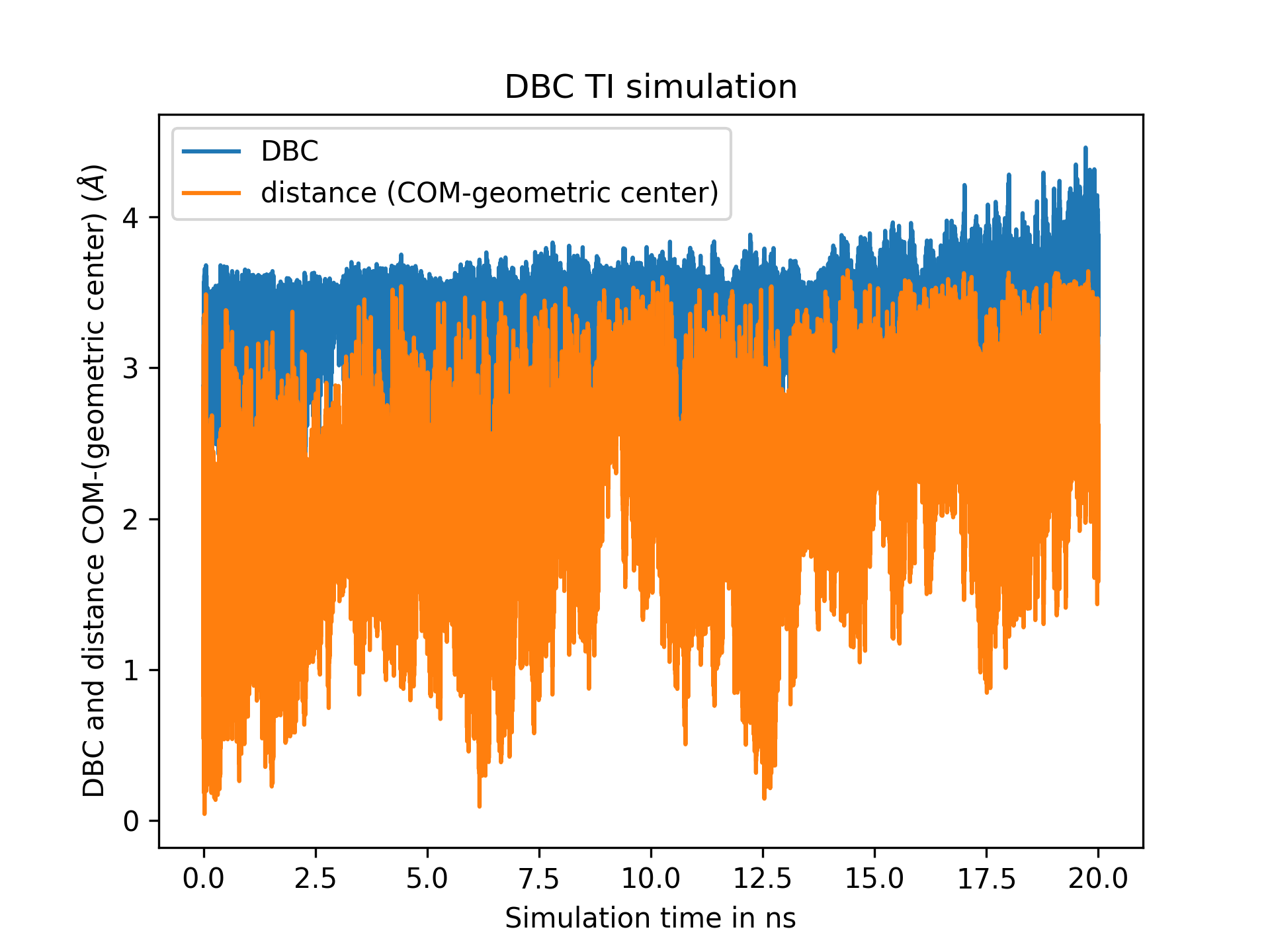}}
\captionof{figure}{DBC and distance COM-(geometric center) evolution during the TI simulation to get the free energy cost of releasing the DBC restraints}\label{fig:sampl6-9-DBC}
\end{minipage}
\end{figure}

\begin{table}[h!]
 \center
 \begin{tabular}{|c|c|c|c|c|c|}
 \hline
  & elec (solvation)& vdW (solvation) & elec (complexation) & vdW (complexation) & restraint \\
 \hline
  lambda-ABF& -5.06 ($\pm$ 0.003)  & 23.57 ($\pm$ 0.001 ) & -0.91 ($\pm$ 0.001) & 8.77 ($\pm$ 0.035) & 1.73 ($\pm 3.10^{-4}$)\\
  \hline
 fixed-lambda&  -5.18 ($\pm$ 0.001) & 23.53 ($\pm$ 0.007) &-1.01 ($\pm$ 0.003) & 8.77 ($\pm$ 0.077)& 1.73  ($\pm 3.10^{-4}$)  \\
 \hline
 \end{tabular}
 \caption{\label{tab:ABFEsampl-detail} terms used to compute the absolute free energies of binding (in kcal/mol) for the ligand 9 of the SAMPL6 challenge, obtained from lambda-ABF and the fixed-lambda method. The restraint free energy includes a standard state correction.}
\end{table}

\subsection{Cyclophilin D}
\subsubsection{General MD parameters}
The solvent phase simulations were made using a cubic simulation box of edge 40 \AA and the complex phase ones using a cubic box of edge 58.34 \AA. In both cases, Electrostatics and polarization interactions were computed using the Smooth Particle Mesh Ewald method\cite{essmann1995smooth} with a real space cutoff of 7 \AA and a 48x48x48 grid.  Van der Waals interactions were cutoff at 12 \AA.
All CypD simulations used a concentration of NaCl of 0.15~M.

The movements of the ligand was restrained using a DBC variable associated to a harmonic restrained as defined in the following Colvars configuration, for the first binding mode:
\begin{verbatim}
colvar {
    name DBC

    rmsd {
        # Reference coordinates (for ligand RMSD computation)
        refPositionsFile cyclo-D-ref-cryst.xyz
        # refPositionsCol O
        # refPositionsColValue 1

        atoms {
            # Define ligand atoms used for RMSD calculation
            atomNumbersRange 2473-2500

            # Moving frame of reference is defined below
            centerToReference yes
            rotateToReference yes
            fittingGroup {
                # Define binding site atoms used for fitting
            atomNumbers 801 802 815 816 817 818 819 820 855 857 898 899 900 901 902 903 904 912 913 914 915
            918 919 920 921 931 940 941 946 947 948 949 950 1065 1066 1067 1068 1072 1073 1074 1075 1078 1080
            1086 1087 1088 1089 1095 1096 1183 1192 1197 1198 1199 1200 1201 1202 1203 1474 1475 1489 1490
            1491 1492 1495 1499 1500 1501 1502 1505 1506 1507 1508 1513 1514 1515 1516 1519 1523 1559 1560
            1561 1564 1565 1572 1573 1574 1575 1578 1586 1587 1588 1589 1593 1594 1595 1604 1605 1606 1607
            1610 1611 1612 1613 1614 1623 1624 1641 1642 1647 1648 1649 1650 1651 1652 1653 1687 1780 1782
            1802 1803 1804 1805 1862 1863 1864 1865 1866 1867 2197 2218 2219
            }
            # Reference coordinates for binding site atoms
            # (can be the same file as ligand coordinates above)
        refPositionsFile cyclo-D-ref-cryst.xyz
        }
    }
}

harmonicwalls {
    colvars DBC
    upperWalls 2.0
    forceConstant 100.0
}
\end{verbatim}

and for the second one:
\begin{verbatim}
colvar {
    name DBC

    rmsd {
        # Reference coordinates (for ligand RMSD computation)
        refPositionsFile cyclo-D-ref-1ns.xyz
        # refPositionsCol O
        # refPositionsColValue 1

        atoms {
            # Define ligand atoms used for RMSD calculation
            atomNumbersRange 2473-2500

            # Moving frame of reference is defined below
            centerToReference yes
            rotateToReference yes
            fittingGroup {
                # Define binding site atoms used for fitting
            atomNumbers 1066 1073 1087 1490 1500 1514 1559 1573 1587 1605 1642

            }
            # Reference coordinates for binding site atoms
            # (can be the same file as ligand coordinates above)
        refPositionsFile cyclo-D-ref-1ns.xyz
        # refPositionsCol O
        # refPositionsColValue 1
        }
    }

}

harmonicwalls {
    colvars DBC
    upperWalls 2.2
    forceConstant 100.0
}
\end{verbatim}

In both cases, the dynamical evolution of the alchemical parameter was controlled similarly as for the previous system.

\begin{table}[h!]
 \center
 \begin{tabular}{|c|c|c|c|c|c|}
 \hline
  & elec (solvation)& vdW (solvation) & elec (complexation) & vdW (complexation) & restraint \\
 \hline
  lambda-ABF& -57.17 ($\pm$ 0.005)  & 47.93 ($\pm$ 0.004 ) & -55.49 ($\pm$ 0.13) & 34.5 ($\pm$ 0.69) & 6.0 ($\pm$ 0.013)\\
  \hline
 fixed-lambda&  -57.27 ($\pm$ 0.02) & 47.96 ($\pm$ 0.001) &-55.36 ($\pm$ 0.014) & 33.73 ($\pm$ 1.45)& 6.0  ($\pm$ 0.013)  \\
 \hline
 \end{tabular}
 \caption{\label{tab:ABFEcypD-detail-1} terms used to compute  the standard free energies of binding (in kcal/mol) for the ligand 27-cypD complex, obtained from lambda-ABF and the fixed-lambda method, first binding mode. The restraint free energy includes a standard state correction.}
\end{table}

\begin{table}[h!]
 \center
 \begin{tabular}{|c|c|c|c|c|c|}
 \hline
  & elec (solvation)& vdW (solvation) & elec (complexation) & vdW (complexation) & restraint \\
 \hline
  lambda-ABF& -57.17 ($\pm$ 0.005)  & 47.93 ($\pm$ 0.004 ) & -56.84 ($\pm$ 0.53) & 36.92 ($\pm$ 0.16) & 2.56 ($\pm$ 0.0)\\
  \hline
 fixed-lambda&  -57.27 ($\pm$ 0.02) & 47.96 ($\pm$ 0.001) &-59.9 ($\pm$ 0.16) & 37.27 ($\pm$ 0.51)& 2.56  ($\pm$ 0.0)  \\
 \hline
 \end{tabular}
 \caption{\label{tab:ABFEcypD-detail-2} Terms used to compute the absolute free energies of binding (in kcal/mol) for the ligand 27-cypD complex, obtained from lambda-ABF and the fixed-lambda method, second binding mode.}
\end{table}

\subsection{Diedral angle of the second binding mode}
\begin{figure}[!htb]
\begin{tabular}{ccc}
\includegraphics[scale=0.25]{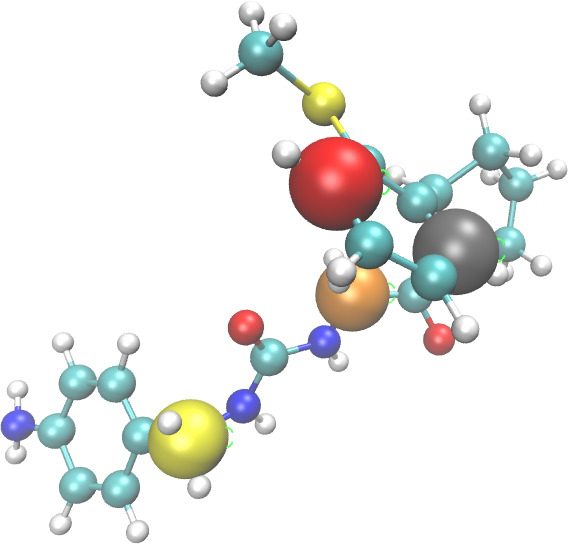} &
\includegraphics[scale=0.25]{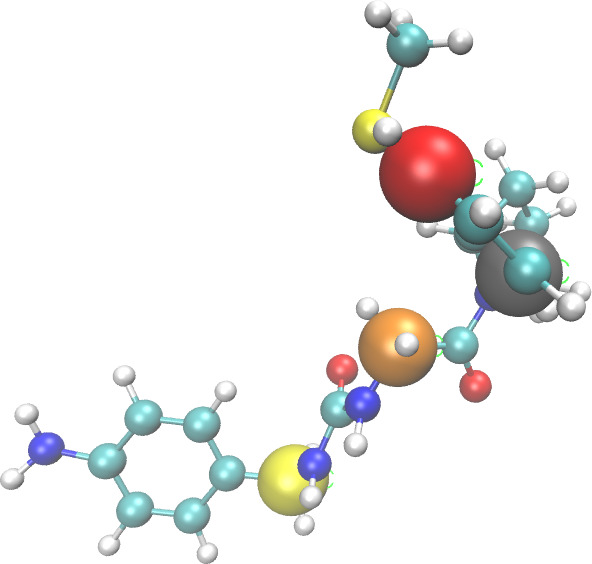} &
\includegraphics[scale=0.25]{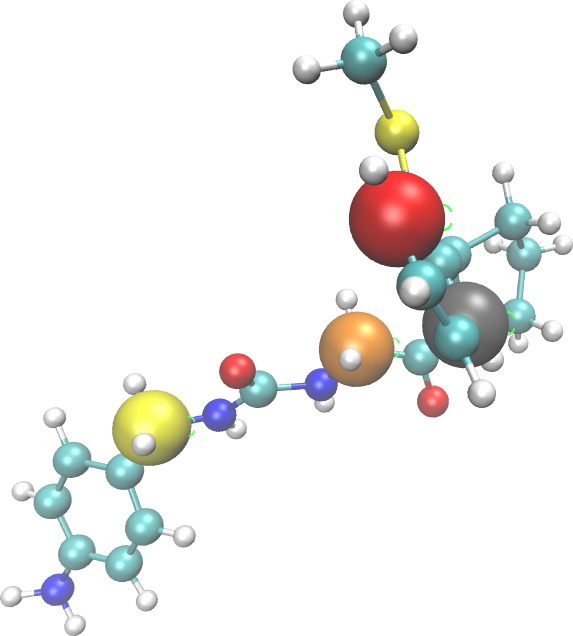} \\
\includegraphics[scale=0.25]{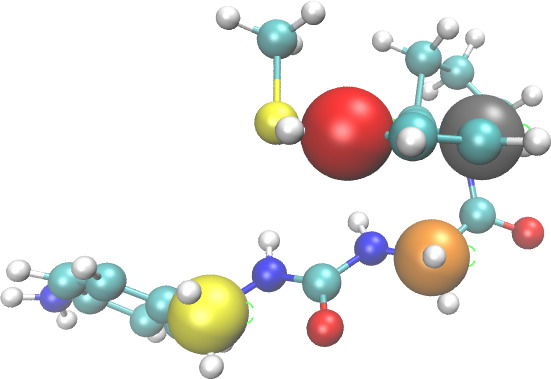} &
\includegraphics[scale=0.25]{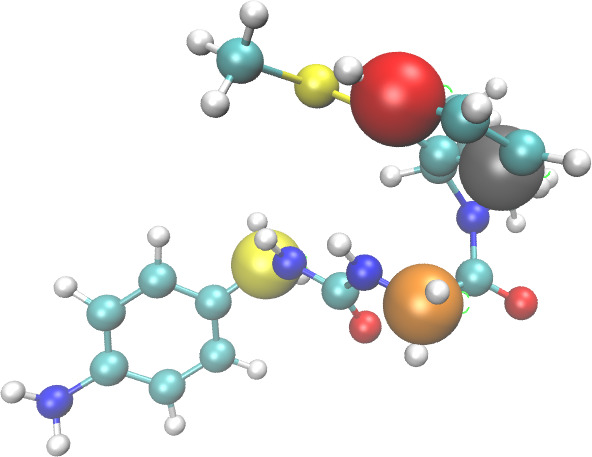} & \\
\end{tabular}
\caption{Illustration of the different values of $\phi$ during lambda-ABF simulations}
\label{fig:phi}
\end{figure}

\begin{figure}[!htb]
    \centering
    \begin{minipage}{.5\textwidth}
        \centering
        \includegraphics[scale=0.5]{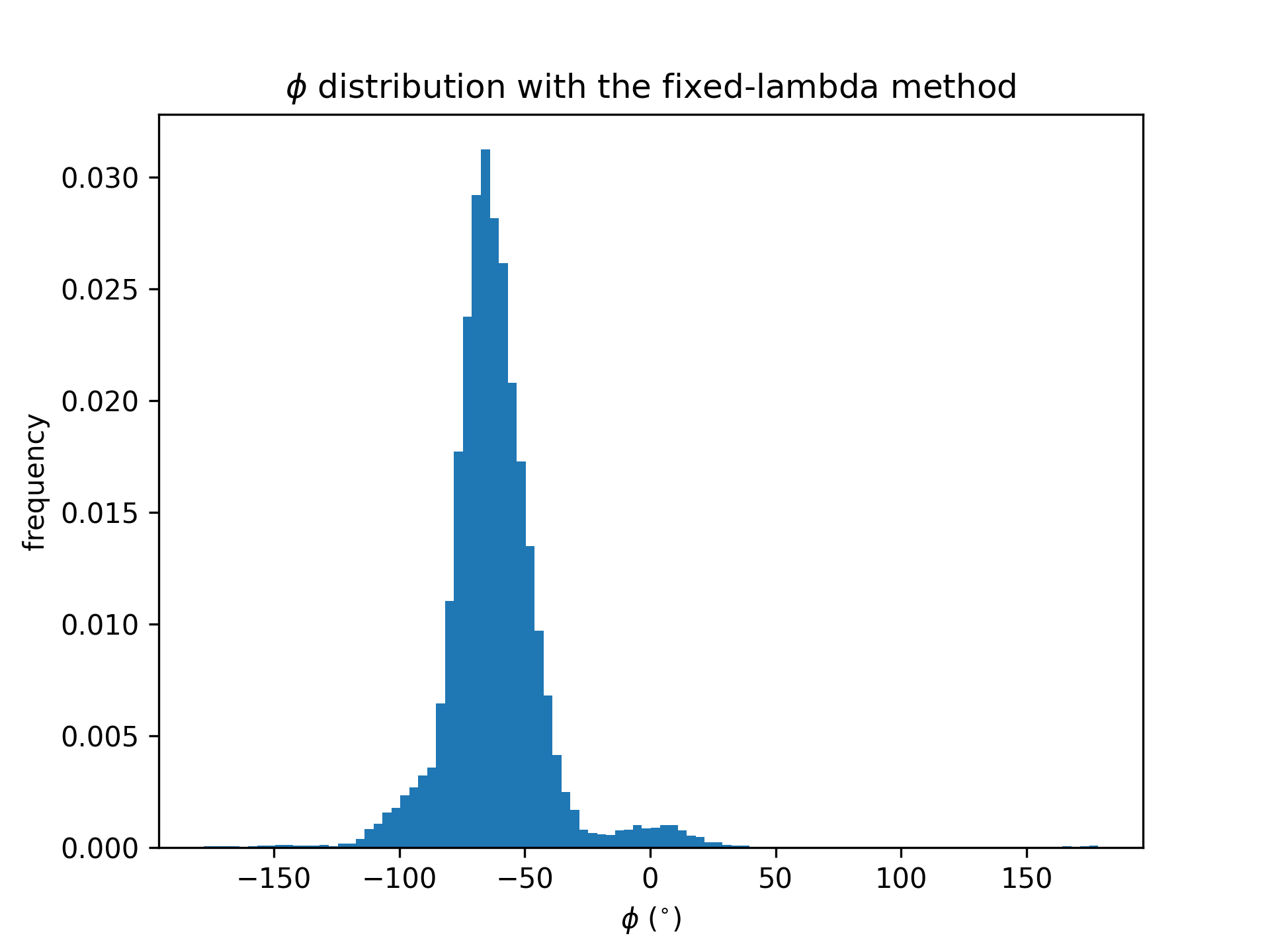}
    \end{minipage}%
    \begin{minipage}{0.5\textwidth}
        \centering
        \includegraphics[scale=0.5]{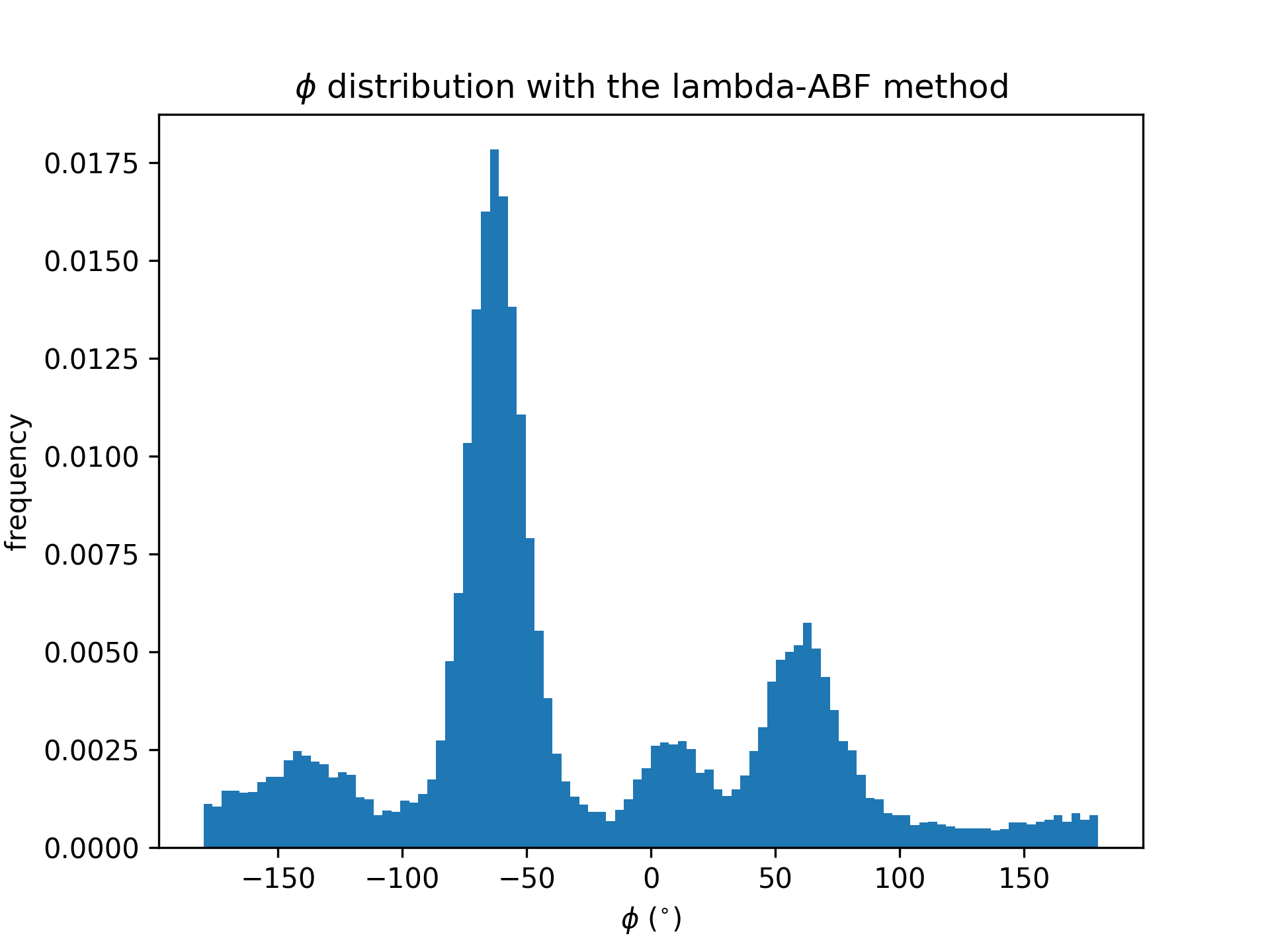}
    \end{minipage}
    \caption{distribution of the diedral angle $\phi$ along the simulation between the fixed-lambda and the lambda-ABF methods.}
    \label{fig:distrib-phi-2}
\end{figure}

The diedral angle $\phi$ that is monitored in section 4.2.4 of the paper involves atoms that are represented in figure \ref{fig:phi}, that shows representatives frames of the $\phi$ distribution sampled by lambda-ABF as shown in figure \ref{fig:distrib-phi-2}. The upper left one represents the reference conformation of the second binding mode, associated to $\phi=-65~^{\circ}$, the second one is representative of the first peak of the $\phi$ distribution, with $\phi=-140~^{\circ}$, the third one with $\phi=-62~^{\circ}$, the fourth one with $\phi=9~^{\circ}$, the fifth one with $\phi=63~^{\circ}$.

\begin{figure}[!htb]
    \centering
    \begin{minipage}{.5\textwidth}
        \centering
        \includegraphics[scale=0.5]{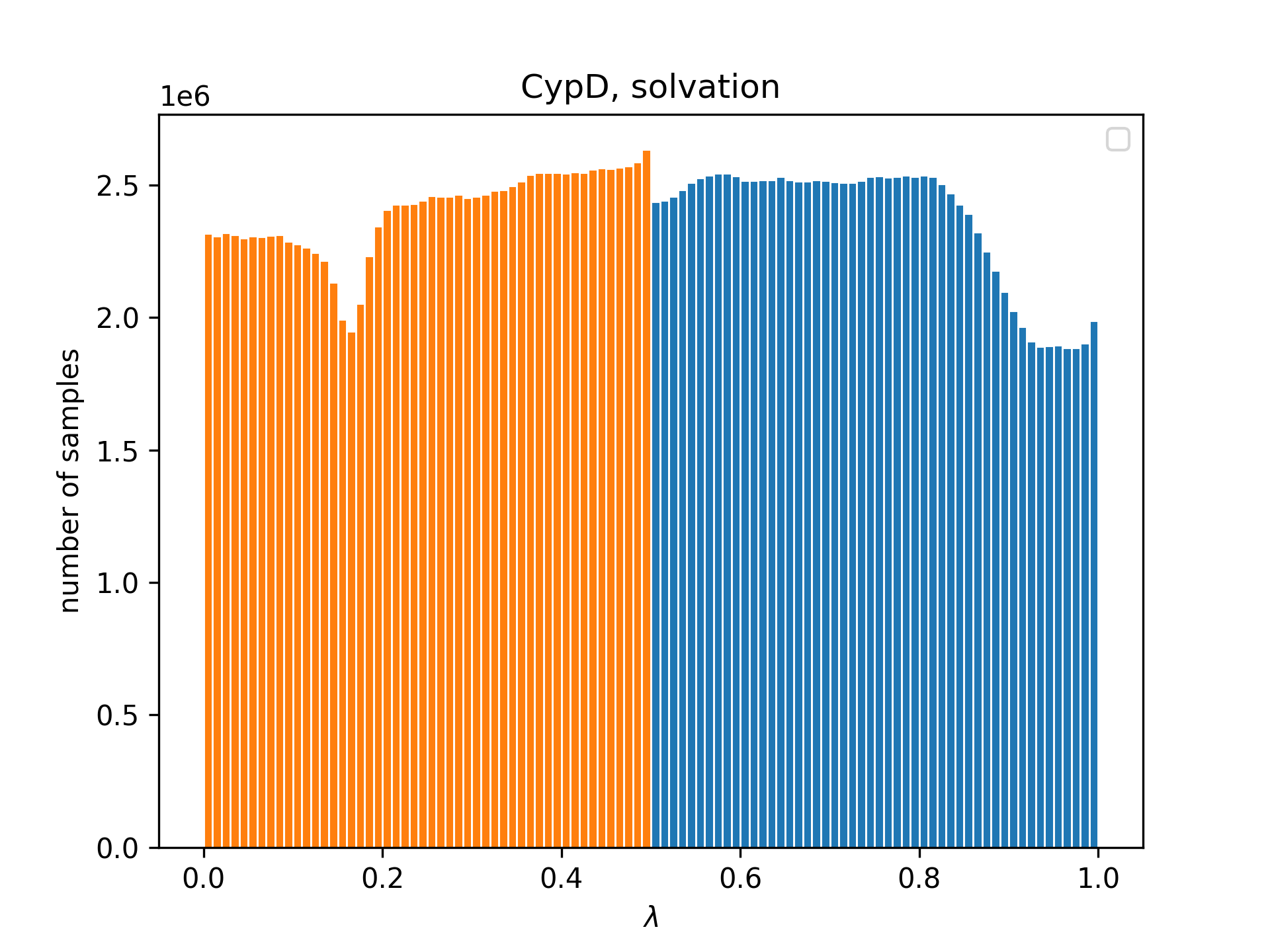}
    \end{minipage}%
    \begin{minipage}{0.5\textwidth}
        \centering
        \includegraphics[scale=0.5]{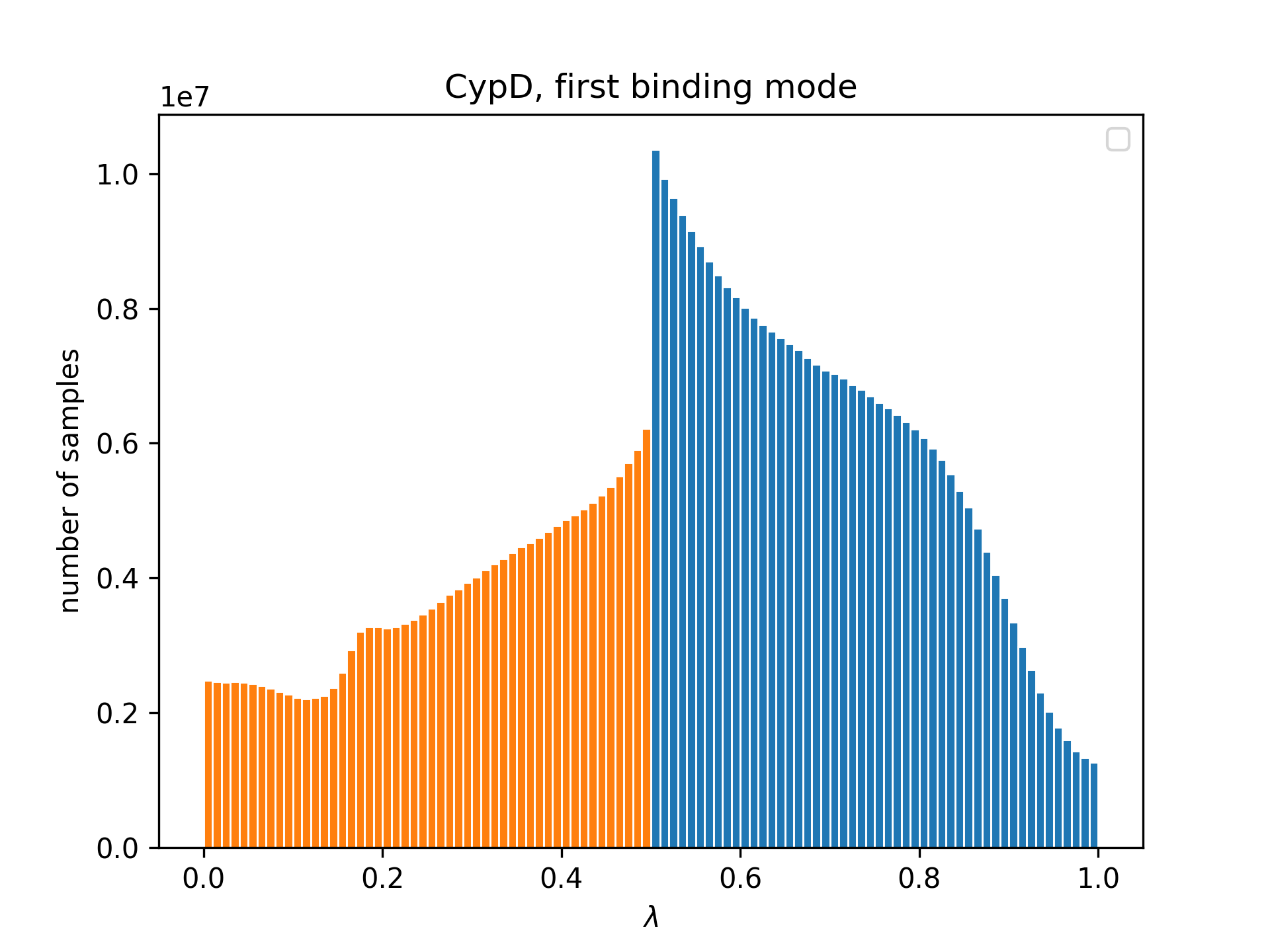}
    \end{minipage}
    \\
    \begin{minipage}{0.5\textwidth}
        \centering
        \includegraphics[scale=0.5]{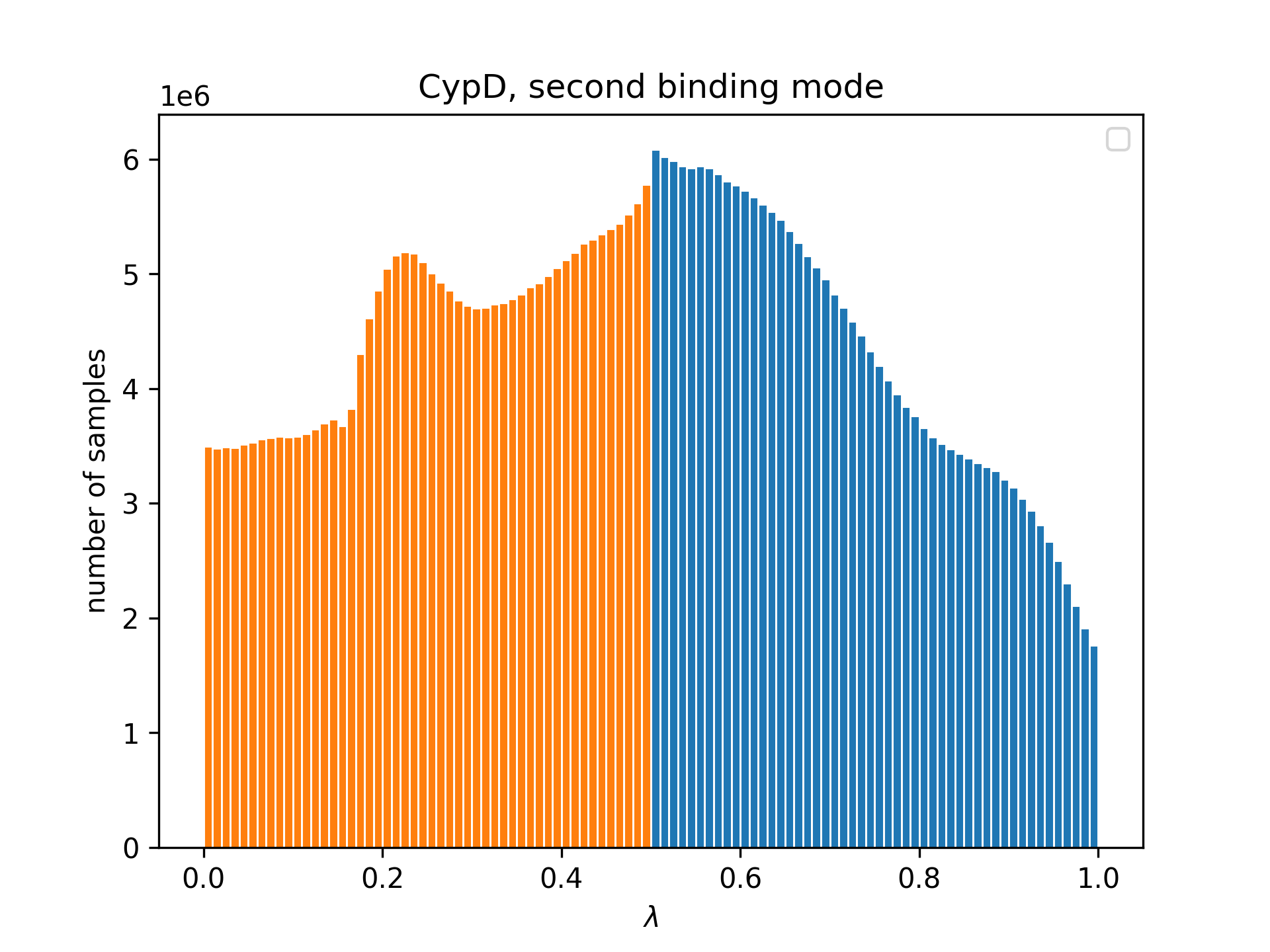}
    \end{minipage}
\caption{Histograms of the different lambda values sampled during the lambda-ABF simulations for the solvation and complexation legs of the CypD ligand}
    \label{fig:count-cypD}
\end{figure}

\subsection{Lysozyme-phenol complex}

\subsubsection{System preparation}

Simulations were started from the crystallographic structure of the L99A/M102H mutated lysozyme in complex with phenol (PDB 4I7L).
The simulation system was prepared using CHARMM-GUI\cite{Jo2008, Lee2016} using a truncated lysozyme (PDB 4I7L, residues 3 to 157) and solvated with TIP3P water~\cite{Jorgensen1983} and 0.15~M NaCl.
The complex was solvated in an elongated periodic box matching the aspect ratio of the protein. In simulations, rotational diffusion of the protein was avoided as explained below to keep the protein aligned with the elongated box.

\subsubsection{General MD simulation parameters}

NAMD simulations were performed using a modified version of NAMD3\cite{phillips2020scalable}, using a constant temperature of 300~K and a constant pressure of 1~bar, controlled by a Langevin thermostat and a Langevin piston algorithm with isotropic cell fluctuations~\cite{feller1995constant,Phillips2005}.
Long-range electrostatics were treated with particle mesh Ewald (PME) \cite{darden1993particle}.
The timestep was set to 2~fs, with SHAKE/RATTLE constraints on all bonds to hydrogen atoms.
The CHARMM36m force field for proteins \cite{Huang2016} was used, together with the CHARMM variant of the TIP3P water model \cite{Jorgensen1983}.

Rotational diffusion of the protein was avoided by imposing a soft harmonic restraint with force constant $10^4~$kcal/mol on its orientation quaternion, using the Colvars library.~\cite{Fiorin2013}
This coordinate robustly represents the orientation of the protein, and is orthogonal to the protein's translational as well as internal degrees of freedom.

\subsubsection{Alchemical transformation protocol}

Alchemical free energy calculations in NAMD involved restraints on the symmetry-adapted DBC coordinate.~\cite{ebrahimi2022symmetry}
A symmetry permutation of reference atoms of phenol was accounted for using the \texttt{atomPermutation} keyword, as shown in the coordinate definition listed below:

\begin{verbatim}
colvar {
    name DBC

    rmsd {
        # Reference coordinates (for ligand RMSD computation)
        refPositionsFile alchemy_site.pdb

        atomPermutation 1 5 3 9 7 11 12
        atoms {
            atomNumbers 1 3 5 7 9 11 12

            centerReference  yes
            rotateReference  yes
            fittingGroup {
                atomNumbers 1207 1315 1370 1386 1556 1566 1599 1616 1730 1827
            }
            # Reference coordinates for binding site atoms
            refPositionsFile alchemy_site.pdb
        }
    }
}
\end{verbatim}

Only nonbonded interactions between the ligand and its environment were perturbed (\texttt{alchDecouple on}).
The $\lambda$-dependent potential energy function was set to fully decouple electrostatics over the first half of the transformation, and Lennard-Jones terms over the full range.
The corresponding NAMD options are:
\begin{verbatim}
alchElecLambdaStart     0.5
alchVdwLambdaEnd        1.0
\end{verbatim}
Lennard-Jones potentials were perturbed through a ``separation-shifted'' soft-core pathway, with a shift parameter of 5~\AA$^2$ (\texttt{alchVdwShiftCoeff 5.0}).~\cite{zacharias1994separation}

For fixed-$\lambda$ simulations, a series of windows were run sequentially using the \texttt{runFEP} script provided within the NAMD distribution under \texttt{lib/alch/fep.tcl}.
20 windows with equal $\lambda$ spacing were run using the Interleaved Double-Wide Sampling method using the \texttt{alchLambdaIDWS} keyword as described in NAMD user's guide~\cite{Bernardi2020}.
To study the rate of convergence, four different window durations were used (0.25, 0.5, 0.75, and 1~ns) resulting in total simulated times of 10.5, 21, 31.5, and 42~ns, respectively.
The resulting two-way comparison energies were parsed using libraries alchemlyb and pymbar as used in the SAFEP Python notebook~\cite{santiago2023computing} (\url{https://github.com/BranniganLab/safep}) to obtain BAR~\cite{bennett1976efficient} free energy estimates.
In each case, five independent replicas were run to compute a mean and standard deviation for the estimates.

For lambda-ABF simulations, a modified branch of NAMD3 was used. The dynamic alchemical variable was defined using the following Colvars configuration:
\begin{verbatim}
colvar {
  name l
  extendedLagrangian on
  extendedMass 150000
  extendedLangevinDamping 1000

   lowerBoundary 0
   upperBoundary 1
   reflectingLowerBoundary
   reflectingUpperBoundary
   width 0.01
  alchLambda {
  }
  outputTotalForce
  outputAppliedforce
}
\end{verbatim}

and the ABF bias was defined by:
\begin{verbatim}
abf {
   colvars l
   fullSamples 1000
   outputFreq  50000
   historyFreq 100000
   shared on
   sharedFreq 5000
}
\end{verbatim}

Multiple-walker ABF was run with 4 walkers, for 10~ns each, or a total simulation length of 40~ns. Free energy gradient estimates were shared between walkers every 5000 steps, i.e. 10~ps. As with the fixed-$\lambda$/BAR simulations, five independent replicas were run.

The command used to start NAMD with 4 walkers on two dual-GPU nodes was:
{\small \noindent
\begin{verbatim}
mpirun -np 4 -N 2 $NAMD +ppn 18 +replicas 4 +devicesperreplica 1 +stdout ${base}.%d.log ${base}.namd
\end{verbatim}
}

\begin{figure}[!htb]
    \centering
    \includegraphics[width=0.5\textwidth]{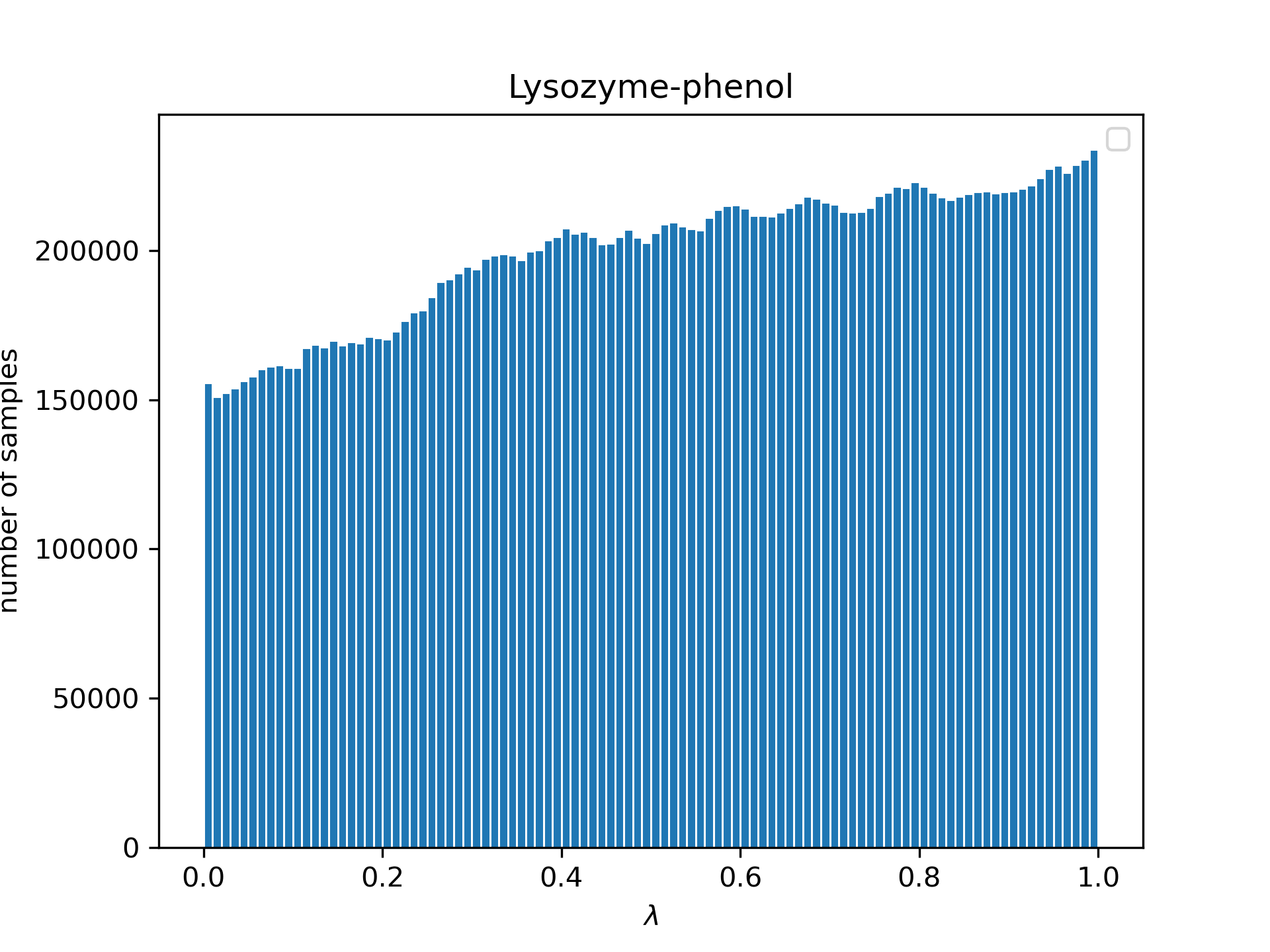}
\caption{Histograms of the different lambda values sampled during the lambda-ABF simulations for the lysozyme-phenol complex.}
    \label{fig:count-lyso}
\end{figure}

\subsubsection{Binding site hydration analysis}

Hydration of the engineered lysozyme pocket was analyzed using VMD.
Two criteria were used: water oxygen atoms within 2.5~\AA{} of phenol heavy atoms were counted as overlapping, while water oxygen atoms within a square radius of 60~\AA$^2$ were considered to be inside the protein interior.

\section{Sensibility analysis of hydration free energies to the mass and friction associated to $\lambda$}
In this section, we report results of the estimation of the hydration free energies of the same systems presented in the main text (water, sodium and potassium ions) with varying mass and friction associated to the alchemical parameter $\lambda$.

In practice, three values are taken for the mass (in the same unit as described in the main text): 15000, 150000 (value used in the maintext) and 1500000.
Additionnally, three values are taken for the friction (in ps$^{-1}$): 100, 1000 (value used in the main text) and 10000.

\begin{table}[h!]
 \center
 \begin{tabular}{|c|c|c|c|}
 \hline
 mass & 1500& 15000&150000 \\
 \hline
 water &5.69 & 5.63&5.60 \\
 \hline
   potassium & 72.34& 72.21&72.3 \\
   \hline
    sodium & 89.53 &89.62 &89.59 \\
 \hline
 \end{tabular}
 \caption{\label{tab:sensanalysismass} Sensitivity analysis of hydration free energies (in kcal/mol) to the mass of $\lambda$. Friction is taken as 1000ps$^{-1}$}
\end{table}

\begin{table}[h!]
 \center
 \begin{tabular}{|c|c|c|c|}
 \hline
 friction & 100& 1000&10000 \\
 \hline
 water &5.64 & 5.63& 5.60 \\
 \hline
   potassium & 72.29&72.21 &72.21 \\
   \hline
    sodium & 89.63& 89.62&89.56 \\
 \hline
 \end{tabular}
 \caption{\label{tab:sensanalysisfriction} Sensibility analysis of hydration free energies (in kcal/mol) to the friction associated to $\lambda$. Mass is taken as 15000.}
\end{table}

\newpage

\bibliographystyle{unsrt}
\bibliography{si}